%% file: superDFT2_paper.tex
\documentclass[11pt]{article}

\usepackage[letterpaper, margin=0.8in]{geometry}

\usepackage{amsmath, amsfonts, amssymb}
\usepackage{slashed}
\usepackage{cite}
\usepackage[vcentermath]{youngtab}
\usepackage{empheq}
\usepackage{booktabs}
\usepackage{mathrsfs}
\usepackage{xcolor}
\usepackage[parsep]{collref}
\usepackage{cancel}
\usepackage[format=hang,labelfont={bf}]{caption}
\usepackage{notes2bib}

\usepackage{hyperref}
\hypersetup{
	unicode,	
	colorlinks,
	citecolor=blue,
	linkcolor=[rgb]{0,.3,.5},
	urlcolor=[rgb]{0,.3,.5}, 
	bookmarksopen=true,
	bookmarksopenlevel=\maxdimen,
        linktocpage,
}

\include{sugracom_v3}
\newcommand{\rep}[1]{\mathbf{#1}}

\DeclareMathOperator{\tr}{tr}
\DeclareMathOperator{\sdet}{sdet}

\newcommand{\rmm}{{\rm m}}

\newcommand{\Lie}{\mathbb L}

\newcommand{\bc}{{\overline c}}
\renewcommand{\bd}{{\overline d}}

\newcommand{\ua}{{\ul a}}
\newcommand{\ub}{{\ul b}}
\newcommand{\uc}{{\ul c}}
\newcommand{\ud}{{\ul d}}
\newcommand{\ue}{{\ul e}}

\newcommand{\hmu}{{\hat\mu}}
\newcommand{\hnu}{{\hat\nu}}

\newcommand{\tA}{{\tiny \rm A}}

\newcommand{\tM}{{\tiny \rm M}}

\newcommand{\drep}[1]{\mathtt{#1}}

\newcommand{\proj}{{\Big\vert}_{\tiny \texttt{proj}}}

\newcommand{\tQ}{\tilde Q}
\newcommand{\wtM}{\widetilde M}
\newcommand{\wtK}{\widetilde K}
\newcommand{\wtL}{\widetilde L}

\newcommand{\tX}{\widetilde X}
\newcommand{\tnabla}{\widetilde\nabla}
\newcommand{\zcV}{\mathring \cV}
\newcommand{\zD}{\mathring D}
\newcommand{\zcD}{\mathring \cD}

\newcommand{\zJ}{\mathring \cJ}

\newcommand{\g}[1]{\mathsf{#1}}

\newcommand{\Clif}{\textrm{Clif}}
\newcommand{\ket}[1]{|#1\rangle}
\newcommand{\bra}[1]{\langle #1|}
\newcommand{\braket}[2]{\langle #1 \vert #2 \rangle}
\newcommand{\ol}{\overline}

\newcommand{\bmu}{{\bar\mu}}

\newcommand{\rba}{{\ol\ra}}
\newcommand{\rbb}{{\ol\rb}}
\newcommand{\rbc}{{\ol\rc}}
\newcommand{\rbd}{{\ol\rd}}

\newcommand{\balpha}{{\overline\alpha}}
\newcommand{\bbeta}{{\overline\beta}}
\newcommand{\bgamma}{{\overline\gamma}}
\newcommand{\bdelta}{{\overline\delta}}

\newcommand{\rA}{{\textrm A}}
\newcommand{\rB}{{\textrm B}}
\newcommand{\rC}{{\textrm C}}

\newcommand{\rbA}{{\ol \rA}}
\newcommand{\rbB}{{\ol \rB}}
\newcommand{\rbC}{{\ol \rC}}

\def\Polacek{Pol\'a\v{c}ek}
\newcommand{\lDirac}{\!\!\stackrel{\leftarrow}{\slashed{\pa}}}
\newcommand{\dbar}{\vert\!\vert}
\newcommand{\Dbar}{\Big{\vert}\! \Big{\vert}}

\newcommand{\cswV}{\widehat{\slashed{\cV}}{}}

\newcommand{\fvac}{\mathfrak{0}}

\newcommand{\talpha}{{\tilde\alpha}}
\newcommand{\tbeta}{{\tilde\beta}}
\newcommand{\tgamma}{{\tilde\gamma}}

\newcommand{\psiGamma}{\beta}

\newcommand{\zLambda}{\mathring\Lambda}

\newcommand{\tra}{{\tilde\ra}}
\newcommand{\trb}{{\tilde\rb}}
\newcommand{\trc}{{\tilde\rc}}
\newcommand{\trd}{{\tilde\rd}}

\newcommand{\zcS}{{\mathring \cS}}

\DeclareMathOperator{\Tr}{Tr}

\makeatletter
\g@addto@macro\bfseries{\boldmath}
\makeatother

\numberwithin{equation}{section}

\begin{document}

\thispagestyle{empty}

%


\bigskip
\bigskip

\vskip 10mm

\begin{center}

{\Large{\bf Type II Double Field Theory in Superspace}}

\end{center}


\vskip 6mm

\begin{center}

{\bf Daniel Butter}

\vspace{1.5ex}

{\em George P. and Cynthia W. Mitchell Institute \\for Fundamental
Physics and Astronomy \\
Texas A\&M University, College Station, TX 77843-4242, USA}\\

\vspace{2ex}
\centerline{\small \texttt{dbutter@gmail.com}}

\end{center}

\vskip0.5cm

\begin{center} {\bf Abstract } \end{center}

\begin{quotation}\noindent
We explore type II supersymmetric double field theory in superspace. The double supervielbein is an element of the orthosymplectic group $\g{OSp}(10,10|64)$, which also governs the structure of generalized superdiffeomorphisms. Unlike bosonic double field theory, the local tangent space must be enhanced from the double Lorentz group in order to eliminate unphysical components of the supervielbein and to define covariant torsion and curvature tensors. This leads to an infinite hierarchy of local tangent space symmetries, which are connected to the super-Maxwell$_\infty$ algebra. A novel feature of type II is the Ramond-Ramond sector, which can be encoded as an orthosymplectic spinor (encoding the complex of super $p$-forms in conventional superspace). Its covariant field strength bispinor itself appears as a piece of the supervielbein. We provide a concise discussion of the superspace Bianchi identities through dimension two and show how to recover the component supersymmetry transformations of type II DFT. In addition, we show how the democratic formulation of type II superspace may be recovered by gauge-fixing.
\end{quotation}


\newpage

\tableofcontents

\newpage

\section{Introduction}

The massless sector of the bosonic string can be described in a duality-covariant way in the language of double field theory (DFT) \cite{Siegel:1993xq,Siegel:1993th, Hull:2009mi,Hull:2009zb,Hohm:2010jy,Hohm:2010pp}. In this approach, the metric, Kalb-Ramond field, and dilaton are encoded in a generalized metric that transforms under generalized diffeomorphisms, which encompass standard diffeomorphisms and abelian $B$-field transformations. The generalized metric nominally depends on twice the usual number of coordinates: the additional coordinates can be understood as dual to winding modes of strings. Different duality frames correspond to different solutions of a section condition that determines which coordinates are the physical ones. 

For type II superstrings, the above constituents describe only the massless part of the NS-NS sector. An additional (bosonic) sector, the massless Ramond-Ramond sector, involves a complex of $p$-form field strengths that describe either the IIA or IIB superstring depending on whether $p$ is even or odd. Finally, the massless R-NS sectors involve the gravitini and dilatini. Since the IIA and IIB superstrings are related by (spacelike) T-duality, their complete massless sectors ought to also be described within the framework of double field theory. Indeed, this was shown by Hohm, Kwak, and Zwiebach for the Ramond-Ramond sector \cite{Hohm:2011zr, Hohm:2011dv}, and by Jeon, Lee, Park, and Suh for the Ramond-Ramond and fermionic sectors as well \cite{Jeon:2012kd, Jeon:2012hp}.\footnote{The generalized metric and Ramond-Ramond sector also emerge as the lowest two levels in the $\g{O}(10,10)$ decomposition of $E_{11}$ \cite{West:2010ev,Rocen:2010bk}. We will return to this point in the conclusion.}

The two approaches \cite{Hohm:2011zr, Hohm:2011dv} and \cite{Jeon:2012kd, Jeon:2012hp} are fascinating in how complementary they are in their treatment of the Ramond-Ramond sector. Hohm, Kwak, and Zwiebach employ the metric formulation of double field theory and encode the Ramond-Ramond sector in a spinor of $\g{O}(D,D)$. Since spinors of $\g{O}(D,D)$ encode $p$-form complexes of $\g{GL}(D)$, this is a very natural assignment and it is inspired by earlier approaches involving torus compactifications \cite{Brace:1998xz, Fukuma:1999jt}. No mention is made of the generalized vielbein of DFT since there is no need for it: as a purely bosonic formulation, one can make do with the generalized metric.

In contrast, the supersymmetric approach of Jeon, Lee, Park, and Suh includes the gravitini and dilatini and fully analyzes their supersymmetry transformations. The Ramond-Ramond sector is described not as a spinor of $\g{O}(D,D)$ but as a bispinor of $\g{O}(D-1,1) \times \g{O}(1,D-1)$; in common parlance, it possesses ``flat'' rather than ``curved'' spinor indices, and this in turn bears relation to early work by Hassan \cite{Hassan:1999bv, Hassan:1999mm}. All of this is quite natural from the perspective of the supersymmetry transformations, of course, because the complex of Ramond-Ramond $p$-forms appears there as a flattened bispinor. (A closely related discussion was given in the context of generalized geometry \cite{Coimbra:2011nw,Coimbra:2012yy}, which is equivalent to double field theory when the section condition is explicitly solved.) Very recently, we have discussed in some detail the connection between the spinor and bispinor formulations of the Ramond-Ramond sector \cite{Butter:2022sfh}, and this permits one to connect the results of \cite{Hohm:2011zr, Hohm:2011dv} and \cite{Jeon:2012kd, Jeon:2012hp} without any gauge-fixing.

The goal of the present paper is to explore type II double field theory in superspace, where supersymmetry is geometrized along with the duality symmetry, in a way that allows us to directly make contact with \cite{Hohm:2011zr, Hohm:2011dv} and \cite{Jeon:2012kd, Jeon:2012hp}. In fact, one of Siegel's early papers on what we now call type I double field theory couched it in a superspace setting \cite{Siegel:1993th}.  We revisited Siegel's work recently \cite{Butter:2021dtu}. Let us briefly highlight some of the crucial features of this superspace approach:
\begin{enumerate}
\item 
The type I supervielbein is an element of the orthosymplectic group $\g{OSp}(10,10|32)$, which is a straightforward supersymmetrization of $\g{O}(10,10)$. However, the local tangent space group is \emph{not} just $\g{O}(9,1)_L \times \g{O}(1,9)_R$. Nor is it the largest natural sub-supergroup, which would be $\g{OSp}(9,1|32)_L \times \g{O}(1,9)_R$. Rather, the left tangent space symmetry is extended from $\g{O}(9,1)_L$ to a proper subgroup $\g{H}_L \subset \g{OSp}(9,1|32)_L$. (Actually, it is only the connected part, including $\g{SO}^+(9,1)$, because we fix the chirality of supersymmetry.) The Lie algebra of this group was explored and it was shown how it is precisely what is needed to eliminate from the type I supervielbein all but the physical fields.

\item In order for a local connection to exist to define the torsion tensor, the local symmetry group $\g{H}_L$ must be extended to a new group $\widehat{\g{H}}_L$ (with the additional generators leaving the supervielbein inert). Notably, the bosonic truncation of this suggested that the right-handed sector $\g{O}(1,9)_R$ should \emph{also} be extended to a new group $\widehat{\g{O}}(1,9)_R$. For example, a new local symmetry appears at dimension 1 that gauges the shift symmetry of the irreducible hook representation of the Lorentz connection. A new connection is then introduced for this local symmetry, new symmetries must be gauged, and so on, leading to a (presumably) infinite hierarchy of additional local gauge symmetries and associated (composite) connections of increasing engineering dimension.

\item The extended group $\widehat{\g{H}}_L$ is dual to the super-Maxwell$_\infty$ algebra recently explored in \cite{Gomis:2017cmt,Gomis:2018xmo}. Using the \Polacek-Siegel framework for gauge symmetries in double field theory \cite{Polacek:2013nla} (see also appendix B of \cite{Butter:2021dtu} for a detailed discussion), this lets one naturally define the torsion and curvature tensors that either vanish or involve only physical components (i.e. the generalized Ricci tensor and scalar in the Riemann tensor). After solving the superspace Bianchi identities, the covariant torsion and curvature tensors lead to the correct supersymmetry transformations of supersymmetric type I DFT \cite{Hohm:2011nu, Jeon:2011sq}. In addition, it appears possible there may exist higher curvatures beyond dimension two, although this point has yet to be explored.

\end{enumerate}

The above results each have clear extensions to type II double field theory, and the first part of this paper will address this. There is past work on type II DFT in superspace by Hatsuda, Kamimura, and Siegel \cite{Hatsuda:2014qqa,Hatsuda:2014aza} and by Cederwall \cite{Cederwall:2016ukd}, elements of which inspire the work here. In the former approach, only the double Lorentz group is employed and a number of torsion constraints are imposed on the supervielbein. This appears to be in contradiction with our results, where the additional connections in $\widehat{\g{H}}_L \times \widehat{\g{H}}_R$ are crucial for constraining the torsion tensors. Presumably the results in \cite{Hatsuda:2014qqa,Hatsuda:2014aza} could be recoverable after further constraining our supervielbein (either by hand or by gauge choice); but as we will see, the extra gauge symmetries are crucial for building invariant torsions and curvatures and for making sense of the truncations to component DFT and conventional type II superspace. 
The work by Cederwall (who also gauged the double Lorentz group) introduced a key idea:  just as the Ramond-Ramond $p$-forms of $\g{GL}(D)$ lift to super $p$-forms of $\g{GL}(D|s)$ in conventional type II superspace (for $D=10$ and $s=32$), Ramond-Ramond spinors of $\g{O}(D,D)$ ought to lift to spinors of $\g{OSp}(D,D|2s)$. Understanding this prescription in detail and how the Ramond-Ramond sector is simultaneously observed in the DFT supervielbein (as argued in 
\cite{Hatsuda:2014qqa,Hatsuda:2014aza,Cederwall:2016ukd}) is one of our main results.

The paper is arranged as follows. Section \ref{S:Supergeometry} provides a concise discussion of the superspace structure of type II DFT from the perspective of the supervielbein, extended connections, torsion constraints, and Bianchi identities. Much of this is directly analogous to the type I situation. We give a complete solution to the Bianchi identities, specifying torsions and curvatures through dimension two.
The new ingredient, the Ramond-Ramond orthosymplectic spinor, will be introduced in section \ref{S:OSpSpinor} where we discuss both curved and flat spinors and the constraints on the Ramond-Ramond spinor field strength. The physical component fields and their supersymmetry transformations will be derived in section \ref{S:DFTcomps}, recovering the results \cite{Jeon:2012hp} of Jeon et al. Then in section \ref{S:TypeIISS}, we will explain how conventional type II superspace can be recovered in a democratic way, encompassing not only IIA and IIB but also their variants IIA$^*$ and IIB$^*$, which arise by timelike T-duality \cite{Hull:1998vg}. We offer some concluding comments and speculate on extensions and open problems in section \ref{S:Conclusion}. Finally, there are two appendices. The first addresses our 10D spinor conventions, while the second is a technical discussion about decomposing the DFT supervielbein.

\section{Supergeometry of double field theory}
\label{S:Supergeometry}

\subsection{Elements of $\g{OSp}(D,D|2s)$ and the supervielbein}
In analogy to conventional $\g{O}(D,D)$ double field theory, let us introduce $\g{OSp}(D,D|2s)$ double field theory with $2D$ bosonic coordinates and $2s$ fermionic ones. This approach was essentially pioneered by Siegel \cite{Siegel:1993th}, who addressed what we now call type I and heterotic double field theory, and reintroduced by Hatsuda, Kamimura, and Siegel \cite{Hatsuda:2014qqa, Hatsuda:2014aza} and Cederwall \cite{Cederwall:2016ukd}. 
Our previous work \cite{Butter:2021dtu} addressed type I DFT ($D=10$, $s=16$) and here we will be concerned with type II ($D=10$, $s=32$). The details are essentially the same so we will be brief. The supercoordinates are collectively denoted $z^\cM$. Superdiffeomorphisms have an $\g{OSp}(D,D|2s) \times\mathbb R^+$ structure, 
and act on a supervector $W^\cM$ of weight $w$ as
\begin{align}
\Lie^{(w)}_\xi W^\cM = \xi^\cN \pa_\cN W^\cM 
    - \cW^\cN \Big(\pa_\cN \xi^\cM - \pa^\cM \xi_\cN (-1)^{nm}\Big)
    + w\, \pa_\cN \xi^\cN\, W^{\cM}\, (-1)^n 
\end{align}
where the factor $(-1)^{nm}$ is a grading, denoting $-1$ if both $\cN$ and $\cM$ are fermionic and $+1$ otherwise. Indices $\cM$ can be raised or lowered with the canonical orthosymplectic element $\eta$ of $\g{OSp}(D,D|2s)$, with the usual NW-SE convention, i.e.
\begin{align}
W^\cM = \eta^{\cM \cN} W_\cN~, \qquad
W_\cM = W^\cN \eta_{\cN \cM}~.
\end{align}
The metric $\eta$ itself is graded symmetric, $\eta_{\cM \cN} = \eta_{\cN \cM} (-1)^{nm}$.
Because of the grading, $\eta_{\cM\cN}$ is not quite the inverse of $\eta^{\cM \cN}$;
instead, one finds $\eta^{\cM \cP} \eta_{\cP\cN} = \delta_\cN{}^\cM  (-1)^{nm}$.
Closure of the algebra of superdiffeomorphisms is ensured by the section condition
\begin{align}
\eta^{\cM \cN} \pa_\cN \otimes \pa_\cM = 0~.
\end{align}

Under the $\g{GL}(D|s) \subset \g{OSp}(D,D|2s)$ subgroup, the coordinates and derivatives decompose as
\begin{align}
\pa_\cM = \Big(\pa_M, \tilde \pa^M\Big)~, \qquad
z_\cM = (\tilde z_M, z^M)~, \qquad
z^\cM = \Big(z^M, \tilde z_M (-)^m \Big)~,\\
\pa_\cM z^\cN = \delta_\cM{}^\cN \quad \implies \quad \pa_M z^N = \delta_M{}^N~, \qquad
\tilde\pa^M \tilde z_N = \delta_N{}^M (-)^{nm}~.
\end{align}
The $\g{OSp}(D,D|2s)$ metric is
\begin{align}\label{E:eta.GLbasis}
\eta^{\cM \cN} =
\begin{pmatrix}
0 & \delta^M{}_N  \\
\delta_M{}^N (-)^{mn} & 0
\end{pmatrix}~, \qquad
\eta_{\cM \cN} =
\begin{pmatrix}
0 & \delta_M{}^N  \\
\delta^M{}_N (-)^{mn} & 0
\end{pmatrix}~.
\end{align}
The section condition becomes $\tilde \pa^M \otimes \pa_M = 0$ and we solve it by taking $\tilde \pa^M = 0$ to recover a conventional $\g{GL}(D|s)$ superspace described by coordinates $z^M$. This coordinate further decomposes into $D$ bosonic coordinates and $s$ fermionic ones, $z^M = (x^m, \theta^\hmu)$. A similar statement applies to its dual winding coordinate, $\tilde z_M = (\tilde x_m, \tilde \theta_\hmu)$. We have denoted the fermionic index with a hat for later convenience: in type II superspace, there are two copies of the fermions and we can further decompose $\theta^\hmu = (\theta^\mu, \theta^\bmu)$.

The supervielbein is naturally taken as a weighted element of $\g{OSp}(D,D|2s) \times \mathbb R^+$. As in the bosonic case, it is more convenient to split the supervielbein into an $\g{OSp}(D,D|2s)$ element, which we denote $\cV_\cM{}^\cA$, and a separate scalar density, the superdilaton $\Phi$, which we take to have weight $1$. These transform respectively as
\begin{subequations}\label{E:SDFT.VPhi.trafos}
\begin{align}
\label{E:SDFT.VPhi.trafos.a}
\delta \cV_\cM{}^\cA 
    &= \xi^\cN \pa_\cN \cV_\cM{}^\cA + \Big(\pa_\cM \xi^\cN - \pa^\cN \xi_\cM (-1)^{nm}\Big) \cV_\cN{}^\cA~, \\
\label{E:SDFT.VPhi.trafos.b}
\delta \Phi 
    &= \xi^\cN \pa_\cN  \Phi
    + \pa_\cN \xi^\cN\,  \Phi\, (-1)^n~.
\end{align}
\end{subequations}
The condition that the supervielbein is a group element amounts to
\begin{align}
\cV_\cA{}^\cM = \eta^{\cM \cN} \cV_\cN{}^\cB \eta_{\cB \cA} \,(-1)^{am}~.
\end{align}

The supervielbein admits a conventional level decomposition, with the $\g{OSp}(D,D|2s)$ generator $T_{\cM \cN}$ decomposed into the $\g{GL}(D|s)$ generator $T_M{}^N$ at level 0 and
nilpotent generators $T_{MN}$ and $T^{MN}$ at levels $\pm1$. Exponentiating each of these gives three independent matrix factors which can be combined in the conventional way\footnote{The field $S$ was denoted $C$ in the context of type I DFT \cite{Butter:2021dtu} and is typically denoted $\beta$ in bosonic DFT.}
\begin{equation}
\begin{aligned}
\cV_\cM{}^\cA &=
\begin{pmatrix}
1 & B \\
0 & 1
\end{pmatrix}
\times
\begin{pmatrix}
E & 0 \\
0 & E^{-T}
\end{pmatrix}
\times
\begin{pmatrix}
1 & 0 \\
S & 1
\end{pmatrix}~, \\[2ex]
B &= B_{MN} (-)^n, \quad
E = E_M{}^A,\quad
E^{-T} = E_A{}^M (-1)^{am+a},\quad
S = S^{AB}~.
\end{aligned}
\label{E:SuperVielbein.BES}
\end{equation}
In this expansion, $E_M{}^A$ is an invertible supermatrix that can be identified as the supervielbein, $B_{MN}$ is the Kalb-Ramond super-two-form, and $S^{AB}$ is an additional scalar superfield, which will turn out to contain the dilatino and Ramond-Ramond bispinor, as we will discuss shortly. Let us presume the section condition to be trivially satisfied with $\tilde \pa^M = 0$. Then under generalized diffeomorphisms \eqref{E:SDFT.VPhi.trafos} with
$\xi_\cM = (\tilde \xi_M, \xi^M)$, these fields transform as
\begin{subequations}
\begin{align}
\delta B_{MN} &= 2\,\pa_{[M} \tilde \xi_{N]} + \xi^P \pa_P B_{MN} + 2\,\pa_{[M|} \xi^P B_{P |N]}~, \\
\delta E_M{}^A &= \xi^N \pa_N E_M{}^A + \pa_M \xi^N E_N{}^A~, \\
\delta S^{AB} &= \xi^M \pa_M S^{AB}
\end{align}
\end{subequations}

This decomposition of the supervielbein is the simplest means of identifying the various type II superfields present, although it is not completely correct: not every orthosymplectic element (or $\g{O}(D,D)$ element for that matter) can be put in this form. The above decomposition really applies to a specific connected component, and this is related to the topology of the supergroup in ways we will discuss in due course. A key point to mention is that the topology of a supergroup is determined by its bosonic part, which in the case of $\g{OSp}(D,D|2s)$ is $\g{O}(D,D) \times \g{Sp}(2s, \mathbb R)$. Since the symplectic groups are connected, the topology of the orthosymplectic group is determined by the split orthogonal group, and this decomposes into four pieces $\g{O}^{(\alpha,\beta)}(D,D)$ with $\alpha=\pm1$ and $\beta=\pm1$ depending on whether there is an orientation reversal in either factor in the maximal compact subgroup $\g{O}(D) \times \g{O}(D)$. This has been discussed at length in \cite{Butter:2022sfh}, building off a discussion in \cite{Jeon:2012kd, Jeon:2012hp}, and we will discuss in section \ref{S:TypeIISS} how to generalize it.\footnote{A slightly different parametrization will be used in section \ref{S:DFTcomps} for component DFT, but we will show how to relate them.}

The superdilaton meanwhile transforms as a scalar density
\begin{align}
\delta \Phi 
    &= \xi^N \pa_N  \Phi
    + \pa_N \xi^N\,  \Phi\, (-1)^n
\end{align}
and one can define the conventional dilaton superfield $\varphi$ by factoring out the superdeterminant of $E_M{}^A$,
\begin{align}
e^{-2 \varphi} = \Phi \times \sdet E_M{}^A~.
\end{align}
Recalling that the conventional dilaton is given in component DFT by $e^{-2 \varphi} = e^{-2d} \times \det e_m{}^a$, one can see that $e^{-2d}$ differs from $\Phi$ by a factor of $\sdet E_M{}^A / \det e_m{}^a$.

\subsection{The local tangent space group of the supervielbein: $\g{H}_L \times \g{H}_R$}
The DFT supervielbein $\cV_\cM{}^\cA$ is acted on by a local symmetry group, whose infinitesimal form is
\begin{align}
\delta\cV_\cM{}^\cA = - \cV_\cM{}^\cB \lambda_\cB{}^\cA
\end{align}
for some parameter obeying $\lambda_{\cB \cA} = - \lambda_{\cA \cB} (-)^{ab}$. This cannot be a generic element of $\g{OSp}(D,D|2s)$, as then the entire DFT supervielbein could be gauged away. 
The simplest possibility, advocated by Hatsuda, Kamimura, and Siegel \cite{Hatsuda:2014qqa,Hatsuda:2014aza} and Cederwall \cite{Cederwall:2016ukd}, is for the local symmetry group to simply be the double Lorentz group. Recall that in the bosonic case, the local tangent space symmetry group is $\g{O}(D-1,1) \times \g{O}(1,D-1)$ and can be defined as a subgroup of $\g{O}(D,D)$ that also leaves invariant not just $\eta_{\ha \hb}$ but also a flat metric $\cH_{\ha \hb}$; these are given by
\begin{align}
\eta_{\ha \hb} = 
\begin{pmatrix}
0 & \delta_a{}^b \\
\delta^a{}_b & 0
\end{pmatrix}~, \qquad
\cH_{\ha \hb} = 
\begin{pmatrix}
\eta_{a b} & 0 \\
0 & \eta^{ab}
\end{pmatrix}~.
\end{align}
Then for the bosonic double vielbein, decomposed as in \eqref{E:SuperVielbein.BES}, one can gauge $S^{ab}$ to zero, leaving behind the diagonal Lorentz subgroup to be the local symmetry group of the vielbein $e_m{}^a$. However, in the superspace framework, this is unsatisfactory because it leaves unphysical components behind in the supervielbein, specifically in whatever is left unfixed in $S^{AB}$.

Here is a good opportunity to elaborate on the tangent space vector indices $\cA$. In the bosonic theory, a double vector $W^\ha$ can be decomposed either toroidally as $W^\ha = (W^a, \tilde W_a)$, or chirally as $W^\ha = (W^\ra, W^\rba)$; these are related by
\begin{align}
W^\ra = \frac{1}{\sqrt 2} (W^a + \eta^{ab} \tilde W_b)~, \qquad
W^\rba = \frac{1}{\sqrt 2} (W^a - \eta^{ab} \tilde W_b)~.
\end{align}
In the chiral decomposition, $\eta_{\ha \hb}$ and $\cH_{\ha \hb}$ are given by\footnote{In our type I DFT paper \cite{Butter:2021dtu}, we took the opposite sign for $\eta_{\ol{\ra\rb}}$. The advantage of the choice made here is that all formulae for the right sector follow from the left sector merely by barring expressions.}
\begin{align}\label{E:bosonic.eta.H.chiral}
\eta_{\ha \hb} = 
\begin{pmatrix}
\eta_{\ra \rb} & 0 \\
0 & \eta_{\ol{\ra\rb}}
\end{pmatrix}~, \qquad
\cH_{\ha \hb} = 
\begin{pmatrix}
\eta_{\ra \rb} & 0 \\
0 & -\eta_{\ol{\ra\rb}}
\end{pmatrix}~, \qquad \eta_{\ol{\ra\rb}} = -\eta_{\ra \rb} = -\eta_{ab}~.
\end{align}
In the superspace case, one naturally assigns the fermionic components of $\cA$ to be spinors of one or the other Lorentz groups. For type I DFT, they carried spinor indices $\alpha$ of the left Lorentz group. For type II, there will be both left and right spinor indices, $\alpha$ and $\balpha$. It will occasionally be useful to group all left indices together. Then the unbarred capital Roman indices $\rA = (\ra, \alpha)$ transform only under the left Lorentz group and the barred capital Roman indices $\rbA = (\rba, \balpha)$ transform only under the right, and we write
\begin{align}
W_\cA &= (W_\rA, W_\rbA) = ( W_\ra, W_\alpha, W^\alpha, \,\,W_\rba, W_\balpha, W^\balpha)~, \eol
W^\cA &= (W^\rA, W^\rbA) = ( W^\ra, W^\alpha, -W_\alpha, \,\,W^\rba, W^\balpha, -W_\balpha)~, \end{align}
and
\begin{align}
\eta_{\cA \cB} = 
\begin{pmatrix}
\eta_{\rA\rB} & 0 \\
0 & \eta_{\rbA \rbB}
\end{pmatrix}
=
\left(\begin{array}{ccc|ccc}
\eta_{\ra\rb} & 0 & 0 
    & 0 & 0 & 0\\
0 & 0 & \delta_\alpha{}^\beta 
    & 0 & 0 & 0\\
0 & -\delta^\alpha{}_\beta & 0
    & 0 & 0 & 0 \\ \hline
0 & 0 & 0
    & \eta_{\ol{\ra \rb}} & 0 & 0\\
0 & 0 & 0
    & 0 & 0 & \delta_\balpha{}^\bbeta \\
0 & 0 & 0
    & 0 & -\delta^\balpha{}_\bbeta & 0
\end{array}\right)~, \eol[2ex]
\eta^{\cA \cB} = 
\begin{pmatrix}
\eta^{\rA\rB} & 0 \\
0 & \eta^{\rbA \rbB}
\end{pmatrix}
=
\left(\begin{array}{ccc|ccc}
\eta^{\ra\rb} & 0 & 0 
    & 0 & 0 & 0\\
0 & 0 & \delta^\alpha{}_\beta 
    & 0 & 0 & 0\\
0 & -\delta_\alpha{}^\beta & 0
    & 0 & 0 & 0 \\ \hline
\phantom{{}^{A^{A^A}}\!\!\!\!\!\!\!\!\!\!} 0 & 0 & 0
    & \eta^{\ol{\ra \rb}} & 0 & 0\\
0 & 0 & 0
    & 0 & 0 & \delta^\balpha{}_\bbeta \\
0 & 0 & 0
    & 0 & -\delta_\balpha{}^\bbeta & 0
\end{array}\right)~.
\end{align}

A natural proposal for the tangent space group might seem to be extend the bosonic case by introducing a flat supermetric $\cH_{\cA \cB}$. This is not really the right approach. An easy way to see this is that in conventional superspace, the supervielbein $E_M{}^A$ is the fundamental geometric object and there is no natural notion of an invertible supermetric. The closest analogue is $G_{MN} = E_M{}^a E_N{}^b \eta_{ab}$; however, (i) this is not invertible, and (ii) does not involve the gravitino one-form $E_M{}^\alpha$ and so it doesn't completely encode the relevant physics. Within double field theory, one could address at least the first issue by constructing an invertible $\cH_{\cA \cB}$ by adding a sign to the right-handed sector of $\eta_{\cA \cB}$; this mirrors the bosonic situation \eqref{E:bosonic.eta.H.chiral} and then one could build projectors $\tfrac{1}{2} (\delta_\cA{}^\cB \pm \cH_\cA{}^\cB)$ onto the left and right-handed sectors.  However, the relevant local symmetry group is \emph{not} the group leaving invariant $\eta$ and $\cH$: this group, $\g{OSp}(D-1,1|s)_L \times \g{OSp}(1,D-1|s)_R $, is too large.

Instead, we argued in \cite{Butter:2021dtu} that the original proposal of Siegel \cite{Siegel:1993th}, inspired by the Hamiltonian description of the superstring worldsheet, was more sensible. Let's review Siegel's proposal for type I DFT. There the spinors are valued in the left Lorentz group and this group is extended with the parameters $\lambda$ subject to the conditions
\begin{align}
\lambda_\alpha{}^\rb = 0~, \qquad 
\lambda_\alpha{}^\beta = \frac{1}{4} \lambda^{\ra\rb} (\gamma_{\ra\rb})_\alpha{}^\beta~.
\end{align}
with $\lambda^{\alpha \beta}$, $\lambda_\ra{}^{\rb}$, and $\lambda_\ra{}^{\beta}$ unconstrained except for the symmetry conditions 
$\lambda^{\alpha \beta} = \lambda^{\beta\alpha}$ 
and
$\lambda^{\ra \rb} = - \lambda^{\rb \ra}$.
The second condition above amounts to the requirement that fermionic orthosymplectic indices transform as spinors under the $\g{SO}(9,1)$ subgroup of $\g{OSp}(9,1|32)$. In addition, as we showed in \cite{Butter:2021dtu}, it is natural to include an additional constraint on the fermionic parameter $\lambda_\ra{}^\beta$,
\begin{align}
(\gamma^\ra)_{\alpha \beta} \lambda_\ra{}^\beta = 0
\end{align}
which eliminates its spin-1/2 part. The upshot is that we have three local symmetries, generated by $\lambda_\ra{}^\rb$, $\lambda_\ra{}^\beta$, and $\lambda^{\alpha\beta}$. This group, which we denoted $\g{H}_L$, is a subgroup (for type I) of $\g{OSp}(9,1|32)$. It is also precisely the right group to eliminate the unphysical parts of the tensor $S^{AB}$ of the supervielbein \eqref{E:SuperVielbein.BES}. For type I, this tensor consists of $S^{ab}$, $S^{a \beta}$, and $S^{\alpha\beta}$, and the $\lambda$ parameters allow one to eliminate all but the spin-1/2 part of $S^{a \beta}$ --- this is the dilatino.

For type II DFT, the obvious proposal is to require the local symmetry group of the supervielbein to be $\g{H}_L \times \g{H}_R$ where $\g{H}_R$ is just a copy of $\g{H}_L$. For type II DFT, the tensor $S^{AB}$ now consists of
\begin{gather*}
S^{\alpha \beta} \qquad S^{\alpha \bbeta} \qquad S^{\ol{\alpha\beta}} \\
S^{a \beta} \qquad S^{a \bbeta} \\
S^{a b}
\end{gather*}
We can eliminate $S^{a b}$, the spin-3/2 pieces of $S^{a\beta}$ and $S^{a \bbeta}$, and $S^{\alpha \beta}$ and $S^{\ol{\alpha\beta}}$. The remaining spin-1/2 pieces of $S^{a \beta}$ and $S^{a \bbeta}$ will become the two dilatini, and $S^{\alpha \bbeta}$ must become the Ramond-Ramond polyform field strength, written as a bispinor. This identifies all the physical fields from a superspace perspective.

\subsection{Generalized fluxes and torsion and extending $\g{H}_L \times \g{H}_R$}
To see how we recover the physical spectrum of type II DFT, we will analyze the possible constraints on the torsion tensor. The analysis will be extremely similar to the type I analysis \cite{Butter:2021dtu}, so we will not go into exhaustive detail.

The generalized fluxes $\cF_{\cA \cB \cC}$ and $\cF_\cA$ are given by
\begin{align}\label{E:genFluxes}
\cF_{\cA \cB \cC} = - 3 \,D_{[\cA} \cV_\cB{}^\cM \, \cV_{\cM \cC]}~, \qquad
\cF_\cA = D_\cA \log \Phi + \pa_\cM \cV_\cA{}^\cM (-)^{a m+m}~.
\end{align}
These are the only scalars (under diffeomorphisms) that can be built purely from the supervielbein 
involving a single derivative, and can be alternatively defined as the generalized
Lie derivative of the supervielbein and superdilaton with respect to the supervielbein,
\begin{align}\label{E:genFluxes.LieDef}
\Lie_{\cV_\cA} \cV_\cB{}^\cM = -\cF_{\cA \cB}{}^\cC \cV_\cC{}^\cM~, \qquad
\Lie_{\cV_\cA} \Phi = \cF_{\cA} \Phi~.
\end{align}
The flattened derivatives $D_\cA := \cV_\cA{}^\cM \pa_\cM$ here obey
\begin{align}\label{E:flatDalgebra}
[D_\cA, D_\cB] = -\cF_{\cA \cB}{}^\cC D_\cC~, \qquad
D^\cA D_\cA = - \cF^\cA D_\cA~.
\end{align}
These flux tensors in turn obey the following Bianchi identities:
\begin{subequations}
\begin{align}
4 \,D_{[\cA} \cF_{\cB \cC \cD]} &= -3 \cF_{[\cA \cB|}{}^{\cE} \cF_{\cE |\cC \cD]}~, \\
2 D_{[\cA} \cF_{\cB]} &= -\cF_{\cA \cB}{}^\cC \cF_\cC - D^\cC \cF_{\cC \cA \cB}~, \\
D^\cA \cF_{\cA} &= -\frac{1}{2} \cF^\cA \cF_\cA - \frac{1}{12} \cF^{\cA \cB \cC} \cF_{\cC \cB \cA}~.
\end{align}
\end{subequations}
We give the various components of the flux tensors in Table \ref{T:Fluxes}, organized by engineering dimension. We use hatted indices $\ha = (\ra, \rba)$ and $\halpha = (\alpha,\balpha)$ to denote both left and right-handed vector and spinorial indices collectively. 

The basic definition of engineering dimension is to assign
dimensions 1/2, 1, and 3/2 to $D_\halpha$, $D_\ha$, and $D^\halpha$, respectively,
and identically for $\pa_\hmu$, $\pa_\hm$, and $\pa^\hmu$. In this way, for example,
$\cV_\ha{}^\hm$ has vanishing engineering dimension. 
This rule implies that a flux $\cF_{\cA \cB \cC}$ has dimension
$-2 + \Delta(\cA) + \Delta(\cB) + \Delta(\cC)$ where $\Delta(\cA)$ denotes the dimension of
$D_\cA$, consistent with \eqref{E:flatDalgebra}.\footnote{The factor of $-2$ arises because the flux $\cF_{\cA \cB}{}^\cC$
has dimension $\Delta(\cA) + \Delta(\cB) - \Delta(\cC)$ and then lowering the $\cC$ index with
$\eta$ exchanges $\Delta(\cC)$ with $2 - \Delta(\cC)$.} Similarly, $\cF_\cA$ has 
dimension $\Delta(\cA)$.

\begin{table}[t]
\centering
\renewcommand{\arraystretch}{1.5}
\begin{tabular}{c|l}
\toprule
dimension &  fluxes \\ 
\hline
$-\frac{1}{2}$ & $\cF_{\halpha \hbeta \hgamma} = \cT_{\halpha\hbeta\hgamma}$ \\ 
\hline
$0$ & $\cF_{\halpha \hbeta \hc} = \cT_{\halpha \hbeta \hc} $ \\
\hline
$\frac{1}{2}$
    & $\cF_{\halpha \hb \hc}$, $\cF_{\halpha \hbeta}{}^\hgamma$, $\cF_\halpha$ \\
\hline
$1$
    & $\cF_{\ha \hb \hc}$, $\cF_{\ha \hbeta}{}^\hgamma$, $\cF_{\ha}$ \\
\hline
$\tfrac{3}{2}$
    & $\cF_{\ha \hb}{}^\hgamma$, $\cF_{\halpha}{}^{\hbeta \hgamma}$, $\cF^\halpha$ \\
\hline
$2$
    & $\cF_{\ha}{}^{\hbeta \hgamma}$ \\
\hline
$\tfrac{5}{2}$
    & $\cF^{\halpha \hbeta \hgamma}$ \\
\bottomrule
\end{tabular}
\captionsetup{width=0.6\textwidth}
\caption{Generalized fluxes in type II super-DFT}
\label{T:Fluxes}
\end{table}

The fluxes themselves are not invariant under the local symmetry group $\g{H}_L \times \g{H}_R$. This group leaves the superdilaton invariant and acts on the supervielbein and fluxes as
\begin{align}
\delta \cV_\cA{}^\cM = \lambda_\cA{}^\cB \cV_{\cB}{}^\cM~, \quad
\delta \cF_{\cA \cB \cC} &= - 3 D_{[\cA} \lambda_{\cB \cC]} + 3 \lambda_{[\cA}{}^\cD \cF_{\cD| \cB \cC]}~, \qquad
\delta \cF_\cA = -D^\cB \lambda_{\cB \cA} - \cF^\cB \lambda_{\cB \cA}~.
\end{align}
In order to build invariant torsions, we introduce the $\g{H}_L\times\g{H}_R$ connection 
$\Omega_{\cA \cB \cC} = (\Omega_{\cA\, \rB \rC}, \Omega_{\cA \, \rbB \rbC})$ with non-vanishing components
\begin{align}\label{E:Omega.nonvanishing}
\Omega_{\cA\, \rb \rc}~, \qquad
\Omega_{\cA\, \beta}{}^\gamma = 
    \frac{1}{4} \Omega_{\cA\, \rb \rc} \,(\gamma^{\rb \rc})_\beta{}^\gamma~, \qquad
\Omega_{\cA\, \rb}{}^\gamma~, \qquad
\Omega_{\cA\,}{}^{\beta\gamma}~,\eol
\Omega_{\cA\, \ol{\rb \rc}}~, \qquad
\Omega_{\cA\, \bbeta}{}^\bgamma = 
    \frac{1}{4} \Omega_{\cA\, \ol{\rb \rc}} \,(\gamma^{\ol{\rb\rc}})_\bbeta{}^\bgamma~, \qquad
\Omega_{\cA\, \rbb}{}^\bgamma~, \qquad
\Omega_{\cA\,}{}^{\ol{\beta \gamma}}~.
\end{align}
The invariant torsions are then given by
\begin{alignat}{2}
\label{E:Torsion}
\cT_{\cA \cB \cC} &:= - 3 \,\nabla_{[\cA} \cV_\cB{}^\cM \, \cV_{\cM \cC]}  & 
    &= \cF_{\cA \cB \cC} + 3\, \Omega_{[\cA \cB \cC]}~, \\
\label{E:DilTorsion}
\cT_\cA &:= \nabla_\cA \log \Phi + \cD_\cM \cV_\cA{}^\cM & 
    &= \cF_\cA + \Omega^\cB{}_{\cB \cA}~.
\end{alignat}
These are the covariantizations of the generalized fluxes \eqref{E:genFluxes} and
can be defined as the \emph{covariant} generalized Lie derivatives of the supervielbein and superdilaton,
similarly to \eqref{E:genFluxes.LieDef}.
Note that the dimension -1/2 and dimension 0 torsion tensors coincide with the fluxes, because the $\Omega$ connections are dimension +1/2 and higher. (We assign dimension to $\Omega$ so that the torsion and flux dimensions match.)

A conventional $\g{H}_L \times \g{H}_R$ connection would transform as
\begin{align}\label{E:dOmega.expected}
\delta \Omega_{\cM \, \cA\cB}
    \stackrel{?}{=} \Lie_\xi \Omega_\cM{}_{\cA\cB}
    + \pa_\cM \lambda_{\cA \cB}
    - \Omega_{\cM \, \cA}{}^\cC \lambda_{\cC \cB}
    + \Omega_{\cM \, \cB}{}^\cC \lambda_{\cC \cA} (-)^{ab}~.
\end{align}
However, $\Omega$ is actually going to transform a bit differently and this is related to an enhancement of the local symmetry group. This comes about for two reasons: (i) certain components of $\Omega$ are absent in the torsion, implying an enhanced symmetry that shifts these components; and (ii) the supersymmetry constraints we wish to impose on the torsion require modifications of the $\Omega$ transformations. The upshot is that the local symmetry group is enhanced from $\g{H}_L \times \g{H}_R$ to $\widehat{\g{H}}_L \times \widehat{\g{H}}_R$, but only the subgroup $\g{H}_L \times \g{H}_R$ actually acts on the supervielbein.

In our previous work on type I DFT \cite{Butter:2021dtu}, we defined the group $\widehat{\g{H}}_L$ by relating it to the so-called super-Maxwell$_\infty$ algebra, using the framework of \Polacek{} and Siegel \cite{Polacek:2013nla} where one doubles not just spacetime but also the local gauge symmetries (e.g. the Lorentz group). For a detailed discussion of this framework, see appendix B of \cite{Butter:2021dtu}. In the interests of being pedagogical, we will first sketch here why such an extended group is necessary by reviewing how constraints, both physical and conventional, may be imposed on the torsion.

Let us begin with an observation. All components of torsions and curvatures are covariant objects. However, the only covariant objects in type II double field theory, at least at the component level, are found at dimension 3/2 and dimension 2: these are the generalized gravitino curvature and the generalized Riemann tensor, and they are built out of derivatives of the physical fields (e.g. double vielbein, gravitini, dilatini, dilaton) that lie at lower dimensions. This means that in order to recover component double field theory, all components of torsions and curvatures through dimension 1 must vanish. Let us describe how this comes about by analyzing the constraints we can impose on the various torsion tensors.

Using the various components of the spin connection $\Omega$, it is possible to impose constraints on the torsion tensor:
\begin{alignat}{2}
\text{fixing   }  &\Omega_{\cA \, \rb \rc}
    &\quad \implies \quad 
    \cT_{\halpha \rb \rc} &= \cT_{\ra \rb \rc} = \cT_{\ra \rb}{}^\gamma = 0~, \quad 
    \cT_\ra = 0~,\\
\text{fixing   }  &\Omega_{\cA \, \rb}{}^{\gamma}
    &\quad \implies \quad 
    \cT_{\halpha \rb}{}^\gamma &= \frac{1}{10} \cX_{\halpha, \beta} (\gamma_\rb)^{\beta \gamma}~, \quad
    \cT_{\rba \rb}{}^\gamma = \frac{1}{10} \cX_{\rba \beta} (\gamma_\rb)^{\beta \gamma}~, \quad
    \cT^{\balpha}{}_\rb{}^\gamma = \frac{1}{10} \cX^\balpha{}_{\beta} (\gamma_\rb)^{\beta \gamma}~, \\
\text{fixing   }  &\Omega_{\cA}{}^{\beta \gamma}
    &\quad \implies \quad 
    \cT_{\halpha}{}^{\beta \gamma} &= \cT_{\ha}{}^{\beta \gamma} = \cT^{\halpha \beta \gamma} = 0
\end{alignat}
and similarly for their barred versions. Because $\Omega_{\cA \rb}{}^\gamma$ is $\gamma$-traceless, only certain representations may be eliminated.

The remaining torsion tensors may be organized by dimension. At dimensions -1/2 and dimension 0, no $\Omega$ connections appear and so the torsion tensors can be identified with the flux tensors:
\begin{gather}
\label{E:BasicTorsionConstraint}
\cT_{\halpha \hbeta \hgamma} = 0~, \qquad
\cT_{\alpha \beta \, \rc} = k\,(\gamma_\rc)_{\alpha \beta}~, \qquad
\cT_{\ol{\alpha \beta}\, \ol\rc} = k\, (\bar\gamma_\rbc)_{\ol{\alpha\beta}}~, \qquad
\cT_{\alpha \bbeta\, \hc} = \cT_{\alpha \beta\, \rbc}
= \cT_{\ol{\alpha \beta}\, \rc} = 0~.
\end{gather}
The constant $k$ fixes the normalization of supersymmetry and is imaginary for a Majorana representation of the $\gamma$-matrices. We leave it unspecified, because then we can more easily compare against results in other papers with different conventions. Finally, we fix $\cT_\halpha$ to vanish: this defines $\Phi$ as the superdilaton uniquely. We summarize the various torsion tensors and constraints imposed in Table \ref{T:TConstraints}. 

\begin{table}[t]
\centering
\renewcommand{\arraystretch}{1.5}
\begin{tabular}{c|c|c|c}
\toprule 
dimension & conventional constraints & physical constraints & remaining torsions \\ \hline
$-\frac{1}{2}$ 
    & $-$ 
    & $\cT_{\halpha\hbeta\hgamma} = 0$ 
    & $-$ 
    \\ \hline
$0$ 
    & $-$
    & $\cT_{\alpha \beta \,\rc} = k \, (\gamma_\rc)_{\alpha\beta}$ \,\,
        $\cT_{\alpha \beta \,\rbc} = 0$ 
    & $-$ 
    \\
& 
    & $\cT_{\ol{\alpha \beta}\, \rbc} = k\, (\gamma_\rbc)_{\ol{\alpha\beta}}$ \,\,
        $\cT_{\ol{\alpha \beta} \,\rc} = 0$ \\
& 
    & $\cT_{\alpha \bbeta \,\hc} = 0$ \\
\hline
$\frac{1}{2}$
    & $\cT_{\halpha \rb \rc} = \cT_{\halpha \overline{\rb\rc}} = 0$
    & $\cT_\halpha = 0$ 
    & $\cT_{\halpha \rb \rbc}$, $\cT_{\halpha \hbeta}{}^\hgamma$\\
\hline
$1$ 
    & $\cT_{\ha \hb \hc} = \cT_\ha = 0$ 
        & $-$
        & $\cT_{\halpha \rbb}{}^\gamma$,  $\cT_{\halpha \rb}{}^\bgamma$ \\
    & $\cT_{\halpha \rb}{}^\gamma = \tfrac{1}{10}\,\cX_{\halpha, \beta} (\gamma_\rb)^{\beta\gamma} $ 
        & 
        & $\cX_{\halpha, \hbeta}$ \\ 
    & $\cT_{\halpha \rbb}{}^\bgamma = \tfrac{1}{10}\,\cX_{\halpha, \bbeta} (\gamma_\rbb)^{\ol{\beta\gamma}}$ & 
\\ \hline
$\tfrac{3}{2}$
    & $ \cT_{\ra \rb}{}^\hgamma = \cT_{\overline{\ra\rb}}{}^\hgamma = 0$ 
        & $-$
        & $\cT^\halpha$, $\cT_\halpha{}^{\beta \bgamma}$ \\
    & $\cT_{\rba \rb}{}^\gamma = \tfrac{1}{10}\, \cX_{\rba \beta} (\gamma_\rb)^{\gamma \beta} $
        & 
        & $\cX_{\rba \, \beta}$, $\cX_{\ra\, \bbeta}$\\
    & $\cT_{\ra \rbb}{}^\bgamma = \tfrac{1}{10}\, \cX_{\ra \bbeta} (\gamma_\rbb)^{\ol{\gamma \beta}} $
        & 
        & \\
    & $\cT_{\halpha}{}^{\beta \gamma} = \cT_{\halpha}{}^{\ol{\beta\gamma}} = 0 $
        &
        & \\ \hline
$2$ 
    & $\cT_{\ha}{}^{\beta \gamma} = \cT_{\ha}{}^{\ol{\beta\gamma}} = 0$ 
        & $-$ 
        & $\cX^\balpha{}_\beta$, $\cX^\alpha{}_\bbeta$\\
    & $\cT^\balpha{}_\rb{}^{\gamma} = \frac{1}{10} \cX^\balpha{}_\beta (\gamma_\rb)^{\beta \gamma } $
        &
        & \\
    & $\cT^\alpha{}_\rbb{}^{\bgamma} = \frac{1}{10} \cX^\alpha{}_\bbeta (\gamma_\rbb)^{\ol{\beta\gamma}} $
        &
        & \\ \hline
$\tfrac{5}{2}$ 
    & $\cT^{\halpha \hbeta \hgamma} = 0$ 
    & $-$ 
    & $-$ \\
\bottomrule
\end{tabular}
\captionsetup{width=0.8\textwidth}
\caption{Conventional and physical constraints on torsion.
Conventional constraints arise from a specific choice of $\Omega_{\cA \cB \cC}$. The remaining torsions vanish upon solving the Bianchi identities. Engineering dimensions follow from the
same considerations as for the fluxes.}
\label{T:TConstraints}
\end{table}

In using the $\Omega$ connections to eliminate various torsion tensors, it happens that the putative transformation rule \eqref{E:dOmega.expected} receives corrections. This happens for two reasons. First, the naive action of $\g{H}_L \times \g{H}_L$ would rotate $\cT_{\halpha \hb \hc}$ into $\cT_{\halpha \hbeta \hc}$, but this is contradicted by the former vanishing with the latter non-vanishing (and depending on the constant $k$). So there must appear $k$-dependent corrections to the $\g{H}_L \times \g{H}_R$ transformations:
\begin{subequations}
\label{E:Omega.lambda}
\begin{align}
\label{E:Omega.lambda.a}
\Delta_\lambda \Omega_{\cM \rb \rc}
    &=
    -2 k \, \cV_\cM{}^\alpha (\gamma_{[\rb})_{\alpha\beta} \lambda_{\rc]}{}^\beta
    -\frac{2}{9} k\, \cV_{\cM [\rb} (\gamma_{\rc]})_{\alpha \beta} \lambda^{\alpha \beta}~, \\
\label{E:Omega.lambda.b}
\Delta_\lambda \Omega_{\cM\, \rb}{}^\beta
    &= 
    k \cV_\cM{}^\alpha \Big((\gamma_\rb)_{\alpha \gamma} \lambda^{\gamma\beta}
    - \frac{1}{18} (\gamma_{\rb \rc})_\alpha{}^\beta(\gamma^\rc)_{\gamma\delta} \lambda^{\gamma\delta}\Big)~.
\end{align}
\end{subequations}
The second complication is that the $\Omega$ connections are not uniquely determined by fixing the torsions. This is well-known in the bosonic sector where there remains unfixed the irreducible hook representation of $\Omega_{\ra\,\rb\rc}$; it turns out similar ambiguities appear in the other connections. It is natural to associate these ambiguities with additional local gauge symmetries, and indeed we must do so, because the algebra of the $\g{H}_L \times \g{H}_R$ transformations only closes on $\Omega$ subject to these new transformations \cite{Butter:2021dtu}. These additional transformations involve parameters $\Lambda_{\ra, \rb}{}^\beta$ and $\Lambda^{\alpha,}{}_\rb{}^\beta$ (which are both $\gamma$-traceless in $\rb \beta$) and
$\Lambda^{\alpha|\beta\gamma}$ which is symmetric in $\beta\gamma$ but with vanishing totally symmetric part:
\begin{subequations}
\begin{align}
\delta_\Lambda \Omega_{\cM\, \rb\rc}
    &= \cV_\cM{}^\ra \Lambda_{\ra|\rb\rc}
    -2 \cV_\cM{}_\alpha \Lambda_{[\rb, \rc]}{}^\alpha~, \\
\delta_\Lambda \Omega_{\cM\, \rb}{}^\beta
    &= \frac{1}{4} \cV_\cM{}^\alpha (\gamma^{\rc \rd})_\alpha{}^\beta \Lambda_{\rb|\rc\rd}
    + \cV_\cM{}^\ra \Lambda_{\ra, \rb}{}^\beta
    + \cV_{\cM \alpha} \Lambda^\alpha{}_\rb{}^\beta~, \\
\delta_\Lambda \Omega_\cM{}^{\beta \gamma} &= 
    \cV_\cM{}^\alpha (\gamma^{\rc\rd})_\alpha{}^{(\beta} \Lambda_{\rc,\rd}{}^{\gamma)}
     + 2 \,\cV_\cM{}^\ra \Lambda^{(\beta}{}_\ra{}^{\gamma)}
     + \cV_{\cM \alpha} \Lambda^{\alpha|\beta\gamma}~.
\end{align}
\end{subequations}

\subsection{Interpreting the torsion constraints}
Before moving on, we want to address the significance of the remaining torsions --- those in the final column in Table \ref{T:TConstraints}. We have already alluded to the fact that the Bianchi identities will tell us that all these torsions vanish. Before getting to that, we should understand just what these objects correspond to at the component level.

For the dimension 1/2 and dimension 1 torsions, they lead to potential contributions to the supersymmetry transformations of 
the vielbein (via $\cT_{\halpha \rb \rbc}$),
the gravitini (via $\cT_{\halpha \rbb}{}^\gamma$, $\cT_{\halpha \rb}{}^\bgamma$, and $\cT_{\halpha \hbeta}{}^\hgamma$) and the dilatini (via $\cX_{\halpha, \hbeta}$). In principle, there could have been new covariant fields into which these fields transform; that there are not will turn out to be a consequence of the Bianchi identities at the superspace level. We emphasize that just because $\cT$ and $\cX$ will turn out to vanish does not mean that there are no supersymmetry transformations; rather, SUSY transformations will arise from the complicated orthosymplectic structure of the supervielbein.

For dimension 3/2, one can interpret $\cT_\halpha{}^{\beta \bgamma}$ similarly as a contribution of new covariant fields to the SUSY transformation of the Ramond-Ramond bispinor. The fact that there is no such new contribution is again a consequence of closure. More interesting are $\cT^\halpha$, $\cX_{\rba \beta}$, $\cX_{\ra \bbeta}$. From their dimension, they have an obvious interpretation as curvatures for the dilatini and gravitini. A linearized analysis --- see section 4.2 of \cite{Butter:2021dtu} for the type I discussion --- would reveal that after solving the section condition in components, these tensors are given by
\begin{align}
\cT^\alpha =
    (\gamma^\ra)^{\alpha \beta} \,\pa_\ra \rho_\beta
    + \pa^\rbb \Psi_{\rbb}{}^\alpha~, \qquad
\cX_{\rbb \alpha} = 
    \pa_\rbb \rho_\alpha - 
    (\gamma^\ra)_{\alpha \beta} \,\pa_\ra \Psi_\rbb{}^\beta
\end{align}
where $\rho_\alpha$ is the dilatino and $\Psi_{\rbb}{}^\alpha$ is the gravitino.
There are two important facts about these quantities. The first is that they are the equations of motion (in the linearized theory) for the dilatini and gravitini respectively; the fact that they vanish in superspace is therefore consistent. The second fact is that when the derivatives are covariantized with the double Lorentz connection, \emph{these are the only invariant curvature tensors that one can construct}. For example, $\cD_{[\rba} \Psi_{\rbb]}{}^\beta$ is Lorentz covariant but it involves the irreducible hook representation of the component Lorentz connection $\omega_{\rba \,\ol{\rb\rc}}$; this means it is not \emph{truly} a covariant object.

At dimension 2, the only quantities we encounter are $\cX^\balpha{}_\beta$ and its barred version.
(The generalized Ricci tensor and curvature scalar are found elsewhere in the generalized Riemann curvature.) A linearized analysis would reveal that $\cX^\balpha{}_\beta$ is nothing but the Dirac operator on the Ramond-Ramond bispinor,
\begin{align}
\cX^\balpha{}_\beta \propto (\gamma^\rb)_{\beta \alpha} \pa_\rb S^{\alpha \balpha}~.
\end{align}
which is the linearized equation of motion of the Ramond-Ramond sector. 

The point we wish to drive home is that the majority of the torsion tensors vanish purely for conventional reasons --- some $\Omega$ is being fixed --- and involve no dynamical information. It was crucial here that the $\Omega$ connections be extended from the double Lorentz group to $\g{H}_L \times \g{H}_R$ so that conventional constraints could be imposed on those torsion components (see Table \ref{T:TConstraints}) that have no component interpretation. For those not constrained in this way, there is physically meaningful data --- a supersymmetry transformation or an equation of motion --- associated with them. This is one reason to believe the results of \cite{Hatsuda:2014qqa,Hatsuda:2014aza} should be understandable only after a significant gauge-fixing.

\subsection{The extended gauge group $\widehat{\g{H}}_L \times \widehat{\g{H}}_R$, connections, and curvatures}

\begin{table}[t]
\centering
\begin{tabular}{c|c|c|c|c}
\toprule
dimension & generator & constraint & dual generator & dual dimension \\ \hline
1    
& $P_\ra$ 
& $-$ 
& $P^\ra$ 
& 1 \\[1ex] \hline
$1/2$ & 
$Q_\alpha$& 
$-$ & 
$\tQ^\alpha$ $\phantom{\Big\vert}$ &
$3/2$ \\[1ex] \hline
$0$ & 
$M_{\ra\rb}$  & 
antisymmetric & 
$\wtM^{\ra\rb}$ $\phantom{\Big\vert}$ & 
$2$ \\[1ex] \hline
$-1/2$ & 
$M_{\alpha \rb}$  & 
$\gamma$-traceless  & 
$\wtM^{\ra\beta}$ $\phantom{\Big\vert}$  & 
$5/2$  \\[1ex] \hline
$-1$ & 
$M_{\alpha \beta}$ &
symmetric   & 
$\wtM^{\alpha \beta}$ $\phantom{\Big\vert}$ & 
$3$ \\
& 
$K_{\ra|\rb\rc}$ &  
hook irrep & 
$\wtK^{\ra|\rb\rc}$  & 
\\[1ex] \hline
$-3/2$ & 
$K_{\ra,\rb\beta}$  & 
$\gamma$-traceless on $\rb\beta$ & 
$\wtK^{\ra,\rb\beta}$ $\phantom{\Big\vert}$  &  
$7/2$ \\[1ex] \hline
$-2$ & 
$K_{\alpha,\rb\beta}$  & 
$\gamma$-traceless on $\rb\beta$ & 
$\wtK^{\alpha, \rb \beta}$ $\phantom{\Big\vert}$  &  
$4$ \\
& 
$L_{\ra\rb,\rc\rd}$ &
pairwise antisymmetric & 
$\wtL^{\ra\rb,\rc\rd}$ $\phantom{\Big\vert}$ &  \\
& $L_{\ra|\rb|\rc\rd}$ &
$\drep{21000}$ irrep &
$\wtL^{\ra|\rb|\rc\rd}$ $\phantom{\Big\vert}$ & \\[1ex]\hline
$\vdots$ & $\vdots$ & $\vdots$ & $\vdots$ & $\vdots$ \\
\bottomrule
\end{tabular}
\captionsetup{width=0.8\textwidth}
\caption{Generators of $\widehat{\g{H}}_L$ and their duals. 
The positive dimension generators make up the super-Maxwell$_\infty$ algebra.}
\label{T:GeneratorsHL}
\end{table}

Let's return to the discussion of the $\Omega$ connections. Already we have discussed how constraining the torsion tensors leads to an extension of the local symmetry group to $\widehat{\g{H}}_L \times \widehat{\g{H}}_R$. We have elaborated in \cite{Butter:2021dtu}, for the case of type I DFT, how this group can be seen to come about by considering successive commutators of the corresponding generators, and gave a proposal for the connection to the super-Maxwell$_\infty$ algebra. For type II DFT, the right-handed sector is just a copy of the left-handed sector, so we may bring over all of our results. 

We summarize the generators of the left-handed sector in Table \ref{T:GeneratorsHL}. A simple way of understanding this set is that the tilde generators $\tQ^\alpha$, $\wtM^{\ra\rb}$ and so on arise as a free Lie algebraic extension of the super-Poincar\'e algebra of $Q_\alpha$ and $P_\ra$. In the case of a Yang-Mills superalgebra (a possible extension of super-Poincar\'e that highlights the various possible structures), $\tQ^\alpha$ can be interpreted as the spinorial gaugino superfield, $\wtM^{\ra\rb}$ is the bosonic field strength, and so on, with higher dimension generators corresponding to covariant derivatives of these objects and commutators thereof. Their Lorentz representations and engineering dimensions follow from the free Lie algebra construction. The dual generators lying at non-positive dimensions are implied by extending the free Lie algebra to a double superalgebra; this requires an invariant bilinear form $\eta$ pairing generators of dimension $\Delta$ and $2-\Delta$ (such as $P_\ra$ with itself, $Q_\alpha$ with $\tQ^\alpha$, and so on).
Further details may be found in \cite{Butter:2021dtu}.

Since we will only be working to dimension 2 in curvatures, only a few of these generators will be relevant for us. The generators corresponding to the $\lambda_{\cA \cB}$ parameters we denote $M_{\cA \cB}$. They are normalized to recover the $\g{OSp}(9,1|32)$ transformation
\begin{align}\label{E:OSpAction}
\frac{1}{2} [\lambda^{\rB\rC} M_{\rC\rB}, P_\rA] = \lambda_\rA{}^\rB P_\rB + \cdots
\end{align}
where the elided terms correspond to deformations arising from the background
$k$-dependent SUSY algebra. The Lorentz generator acts in the usual way
\begin{align}
[M_{\rc \rb}, P_\ra] = \eta_{\rb\ra} P_\rc - \eta_{\rc \ra} P_\rb~, \quad
[M_{\rc \rb}, Q_\alpha] = -\tfrac{1}{2} (\gamma_{\rc \rb})_\alpha{}^\beta Q_\beta~, \quad
[M_{\rc \rb}, \tQ^\alpha] = -\tfrac{1}{2} (\gamma_{\rc \rb})^\alpha{}_\beta \tQ^\beta
\end{align}
while the other $M$ generators act as
\begin{subequations}
\label{E:Malphab.P}
\begin{align}
\label{E:Malphab.P.a}
\{M_{\beta \rb}, Q_\alpha\} &= k \, (\gamma^c)_{\beta \alpha} M_{\rb\rc} 
    - \tfrac{1}{10} k (\gamma_\rb \gamma^{\rc\rd})_{\beta \alpha} M_{\rc\rd}~, \\
[M_{\beta \rb}, P_\ra] &= \eta_{\rb \ra} Q_{\beta} 
    - \tfrac{1}{10} (\gamma_{\rb} \gamma_\ra)_\beta{}^\gamma Q_\gamma~, \\
\{M_{\beta \rb}, \tQ^\alpha\} &= -\delta_\beta{}^\alpha P_\rb
    + \tfrac{1}{10} (\gamma_\rb \gamma^\rc)_\beta{}^\alpha P_\rc~,
\end{align}
\end{subequations}
and
\begin{subequations}
\label{E:Malphabeta.P}
\begin{align}
[M_{\gamma \beta}, Q_\alpha] &= 
    -2 k (\gamma^{\rb})_{\alpha (\beta} M_{\gamma) \rb}
    + \tfrac{1}{9} k (\gamma^{\rb})_{\gamma \beta} M_{\alpha \rb}~, \\
[M_{\gamma \beta}, P_\ra] &= \tfrac{2}{9} k (\gamma^\rb)_{\gamma \beta} M_{\rb\ra}~, \\
[M_{\gamma \beta}, \tQ^\alpha] &= 2 \,Q_{(\gamma} \,\delta_{\beta)}{}^\alpha~.
\end{align}
\end{subequations}
The additional generators responsible for the shift symmetries of $\Omega$, are denoted $K_{\ra |\rb\rc}$, $K_{\ra, \rb\beta}$, and $K_{\gamma, \rb\beta}$. Some of their (anti)commutators are given in \cite{Butter:2021dtu}, but we do not reproduce them here.

We adopt the \Polacek-Siegel framework for local gauge symmetries in double field theory \cite{Polacek:2013nla} (see also appendix B of \cite{Butter:2021dtu}).
Superspace double field theory can be built on a rigid double super-Poincar\'e algebra involving the generators $Q_\alpha$, $P_\ra$, and $\tilde Q^\alpha$ in the left-handed sector and $Q_\balpha$, $P_\rba$, and $\tilde Q^\balpha$ in the right.
They obey the algebra
\begin{align}
\{Q_\alpha, Q_\beta\} = -k \, (\gamma^\rc)_{\alpha \beta}\, P_\rc~, \qquad
[Q_\alpha, P_\rb] = -k \, (\gamma_\rb)_{\alpha \gamma} \tQ^\gamma~,
\end{align}
with other (anti-)commutators vanishing.  (The right-handed sector follows by adding bars to all indices.) These relations can be collectively written
\begin{align}
[P_\cA, P_\cB] = -f_{\cA \cB}{}^{\cC} P_\cC
\end{align}
where the structure constants with a lowered index, $f_{\cA \cB \cC} = f_{\cA \cB}{}^{\cD} \eta_{\cD \cC}$, are (graded) totally antisymmetric. This rigid algebra is then augmented by additional generators corresponding to $\widehat{\g{H}}_L \times \widehat{\g{H}}_R$. The latter we denote $X_\ua$, where the graded index $\ua$ has no relation to a vector index, and corresponds to an infinite set of generators lying in increasingly complicated Lorentz representations. The full extended algebra schematically reads
\begin{subequations}\label{E:RigidPSAlgebra}
\begin{align}
[P_\cA, P_\cB] &= -f_{\cA \cB}{}^\cC P_\cC - \tX^\uc  f_{\uc \cA \cB}~, \\
[X_\ua, P_\cB] &= -f_{\ua \cB}{}^\cC P_\cC - f_{\ua \cB}{}^\uc \,X_\uc~, \\
[X_\ua, X_\ub] &= - f_{\ua \ub}{}^\uc X_\uc~, \\[2ex]
[\tX^\ua, \tX^\ub] &= - \tX^\uc \, f_\uc{}^{\ua \ub}~, \\
[P_\cA, \tX^\ub] &= 
    - \tX^\uc f_{\uc\cA}{}^\ub~, \\
[X_\ua, \tX^\ub] &= -\tX^\uc\, f_{\uc\ua}{}^\ub
    - f_{\ua}{}^{\ub}{}^\cC P_\cC
    - f_\ua{}^{\ub \uc} X_\uc~.
\end{align}
\end{subequations}
The generators $X_\ua$ comprise Lorentz transformations $M_{\ra\rb}$, the additional tangent space transformations acting on the supervielbein, which we denote $M_{\beta \ra}$ and $M_{\alpha\beta}$, the new shift symmetries of $\Omega$ involving the $\Lambda$ parameters (whose generators we denote by $K$), and a higher tower of generators that arise when we attempt to build curvatures for $\Omega$. It is important that $X_\ua$ furnish a closed algebra --- this generates $\widehat{\g{H}}_L \times \widehat{\g{H}}_R$.
The generators $\tilde X^\ua$ are additional dual generators that are paired with $X_\ua$ via a non-degenerate $\eta$ metric, in the same manner that $P_a$ is paired with itself and $Q_\alpha$ is paired with $\tilde Q^\alpha$.

A more compact form of the above relations is
\begin{align}\label{E:RigidPSAlgebra.Compact}
[X_{\widehat\cA}, X_{\widehat \cB}] = -f_{\widehat \cA \widehat \cB}{}^{\widehat \cC} X_{\widehat \cC}
\end{align}
for $X_{\widehat\cA} = (X_\ua, P_\cA, \tX^\ua)$ and where
$f_{\widehat \cA \widehat \cB \widehat \cC} = 
    f_{\widehat \cA \widehat \cB}{}^{\widehat \cD} \eta_{\widehat \cD \widehat \cC}$
is totally antisymmetric with $\eta_{\widehat \cA \widehat \cB}$ given by
\begin{align}
\eta_{\widehat \cA \widehat \cB} =
\begin{pmatrix}
0 & 0 & \delta_\ua{}^\ub \\
0 & \eta_{\cA \cB} & 0 \\
(-1)^{\ua \ub} \,\delta_\ub{}^\ua  & 0 & 0
\end{pmatrix}~.
\end{align}
The requirement of a non-degenerate $\eta$ is one reason to introduce the dual generators $\tilde X^\ua$.

Now we want to gauge this formal algebra by introducing a supervielbein, connections, and so on. Following the discussion in appendix B of \cite{Butter:2021dtu}, we introduce the superdilaton $\Phi$, the supervielbein $\cV_\cM{}^\cA$, connections $H_\cM{}^\ua$, and an additional graded antisymmetric superfield $P^{\ua \ub}$. These transform under diffeomorphisms and gauge transformations (with parameter $\Lambda^\ua$) as\footnote{For purposes of legibility, we have suppressed gradings in these and subsequent expressions.}
\begin{subequations}\label{E:PSTrafos}
\begin{align}
\label{E:PSTrafos.Phi}
\delta \Phi &= \Lie_\xi \Phi~, \\
\label{E:PSTrafos.V}
\delta \cV_\cM{}^\cA &= \Lie_\xi \cV_\cM{}^\cA + \cV_\cM{}^\cB \Lambda^\uc f_{\uc \cB}{}^\cA~, \\
\label{E:PSTrafos.H}
\delta H_\cM{}^\ua
    &= \Lie_\xi H_\cM{}^\ua
    + \pa_\cM \Lambda^\ua
    + H_\cM{}^\ub \Lambda^\uc f_{\uc \ub}{}^\ua
    + \cV_\cM{}^\cB \Lambda^\uc f_{\uc \cB}{}^\ua~, \\
\label{E:PSTrafos.P}
\delta P^{\ua \ub}
    &= \xi^\cM \pa_\cM P^{\ua \ub}
    - \Lambda^\uc f_{\uc}{}^{\ua \ub}
    - 2 \,\Lambda^\uc P^{\ud [\ua} f_{\uc \ud}{}^{\ub]}
    - H^{\cM [\ua} \pa_\cM \Lambda^{\ub]}
    - \Lambda^\uc H^{\cD [\ua} f_{\cD \uc}{}^{\ub]}~.
\end{align}
\end{subequations}
The connection $H_\cM{}^\ua$ generalizes $\Omega_\cM{}^{\cA \cB}$ and the field $P^{\ua \ub}$
generalizes the one introduced by \Polacek{} and Siegel \cite{Polacek:2013nla}. One can check that the algebra of $\widehat{\g{H}}_L \times \widehat{\g{H}}_R$ closes on these fields with
\begin{align}
[\delta_{\Lambda_1}, \delta_{\Lambda_2}] = \delta_{\Lambda_{12}} \quad\text{for}\quad
    \Lambda_{12}{}^\ua = \Lambda_1{}^\ub \Lambda_2{}^\uc f_{\uc \ub}{}^\ua
\end{align}

With these ingredients, we can construct covariant derivatives
\begin{align}
\nabla_\cA &= \cV_\cA{}^\cM \pa_\cM - H_\cA{}^{\ub} X_\ub~, \qquad
\tnabla^\ua = H^{\cM \ua} \pa_\cM + (P^{\ua \ub} - \tfrac{1}{2} H^{\cM \ua} H_\cM{}^{\ub} ) X_\ub~.
\end{align}
These correspond to the curved extensions of $P_\cA$ and $\tX^\ua$.
Their algebra can again be written \eqref{E:RigidPSAlgebra.Compact},
but with some of the components of $f$ now becoming structure functions.
These consist of four curvatures
$\cT_{\cA \cB \cC}$, 
$\cR_{\cA \cB}{}^{\uc}$,
$\cR_\cA{}^{\ub \uc}$, and
$\cR^{\ua \ub \uc}$, which appear in the curved algebra as
\begin{subequations}\label{E:CurvedPSAlgebra}
\begin{align}
[\nabla_\cA, \nabla_\cB] &= 
    - \cT_{\cA \cB}{}^\cC \nabla_\cC
    - \cR_{\cA \cB}{}^\uc X_\uc
    - \tnabla^\uc\, f_{\uc \cA \cB} 
    ~, \\
[X_\ua, \nabla_\cB] &= -f_{\ua \cB}{}^\cC \nabla_\cC - f_{\ua \cB}{}^\uc \,X_\uc~, \\
[X_\ua, X_\ub] &= - f_{\ua \ub}{}^\uc X_\uc~, \\[2ex]
[\tnabla^\ua, \tnabla^\ub] &= 
    - \tnabla^\uc \, f_\uc{}^{\ua \ub}
    - \cR^{\ua \ub \cC} \nabla_\cC
    - \cR^{\ua \ub \uc} X_\uc
    ~, \\
[\nabla_\cA, \tnabla^\ub] &=
    - \tnabla^\uc \,f_\uc{}_\cA{}^\ub
    - \cR_\cA{}^\ub{}^\cC \,\nabla_\cC 
    - \cR_\cA{}^\ub{}^{\uc} X_\uc~, \\
[X_\ua, \tnabla^\ub] &= 
    - \tnabla^\uc \,f_\uc{}_\ua{}^{\ub}
    - f_\ua{}^\ub{}^\cC \nabla_\cC
    - f_\ua{}^\ub{}^\uc X_\uc~.
\end{align}
\end{subequations}
The torsion tensor is
\begin{align}
\cT_{\cC\cB\cA} &= -3 \nabla_{[\cC} \cV_\cB{}^\cM \cV_{\cM \cA]}~, \qquad
\nabla_\cC \cV_\cB{}^\cM := D_\cC \cV_\cB{}^\cM + H_\cC{}^\ud f_{\ud \cB}{}^\cA \cV_\cA{}^\cM~.
\end{align}
The curvature tensor $\cR_{\cC \cB}{}^\ua$ is
\begin{align}
\cR_{\cC \cB}{}^\ua &= 
    2\, D_{[\cC} H_{\cB]}{}^\ua
    + \cF_{\cC \cB}{}^\cD H_\cD{}^\ua
    - H_\cB{}^\ub H_\cC{}^\uc f_{\uc \ub}{}^\ua
    - 2\, H_{[\cC}{}^{\ud} f_{\ud \cB]}{}^\ua
    - f_{\cC \cB \,\ud} \, \Big(P^{\ud \,\ua} + \tfrac{1}{2} H^{\cF \ud} H_\cF{}^\ua\Big)~.
\end{align}
The other curvature tensors, particular to double field theory, are $\cR_\cC{}^{\ub \ua}$,
which is the covariantized derivative of the \Polacek-Siegel field,
\begin{align}
\cR_\cC{}^{\ub \ua}
    &= \bigg[- D_\cC P^{\ub \ua}
    - 2 \,H_\cC{}^{\uc} P^{\ub \ud} f_{\ud \uc}{}^\ua
    - 2 \,P^{\ub \ud} f_{\ud \cC}{}^\ua
    + H^{\cD \ub} D_\cC H_\cD{}^\ua
    - 2 \,H^{\cD \ub} D_\cD H_\cC{}^\ua
    - H_\cC{}^\uc f_\uc{}^{\ub \ua}
    \eol & \quad
    + \cF_\cC{}^{\cB\cA} H_\cB{}^\ub H_\cA{}^\ua
    - H^{\cD \,\ub} H_\cD{}^\ud \Big(
        f_{\cC \ud}{}^\ua+ H_\cC{}^{\uc} f_{\uc\ud}{}^\ua
    \Big) \bigg]_{[\ub \ua]}
\end{align}
and the curvature $\cR^{\uc \ub \ua}$,
\begin{align}
\cR^{\ul{cba}}
    &= 3 \times \bigg[
    P^{\uc \ud} f_\ud{}^{\ub \ua}
    - H^{\cC \uc} D_\cC P^{\ub \ua}
    + H^{\cC \uc} H^{\cB \ub} D_{\cC} H_{\cB}{}^{\ua}
    + \frac{1}{3} H^{\cC \uc} H^{\cB \ub} H^{\cA \ua} \cF_{\cC \cB \cA} 
    \eol & \qquad \qquad
    + P^{\uc \ud} P^{\ub \ue} f_{\ue \ud}{}^\ua
    - H^{\cD \uc} H_\cD{}^{\ud} \Big(
        \frac{1}{2} f_\ud{}^{\ub \ua}
        + P^{\ub \ue} f_{\ue \ud}{}^\ua
        + \frac{1}{4} H^{\cE \ub} H_\cE{}^{\ue} f_{\ud \ue}{}^\ua
    \Big)
    \bigg]_{[\uc \ub \ua]}~,
\end{align}
which is essentially the covariantized version of $\tilde\nabla^{[\uc} P^{\ub \ua]}$.

Finally, we include the superdilaton $\Phi$, a scalar density under diffeomorphisms and invariant under the local symmetry group. With it, one can construct dilaton-dependent torsion and curvatures
\begin{subequations}
\begin{align}
\cT_\cA &= \nabla_\cA \log\Phi + \nabla_\cM \cV_\cA{}^\cM ~, \\
\cR^\ua &= 
    D^\cB H_\cB{}^\ua + \cF^\cB H_\cB{}^\ua
    - H^{\cB \uc} f_{\uc \cB}{}^\ua
    + P^{\ub \uc} f_{\uc \ub}{}^{\ua}~.
\end{align}
\end{subequations}
where $\cF_\cA = D_\cA \log\Phi + \pa_\cM \cV_\cA{}^\cM$. 

The two sets of torsions and curvatures obey a set of Bianchi identities. The ones
most relevant to us are
\begin{align}
0 &= 4 \,\nabla_{[\cA} \cT_{\cB \cC \cD]}
    + 3 \,\cT_{[\cA \cB|}{}^\cE \cT_{\cE| \cC \cD] }
    + 6 \,\cR_{[\cA \cB|}{}^\ue f_{\ue |\cC \cD]}~, \\
0 &= 3 \,\nabla_{[\cA} \cR_{\cB \cC]}{}^\ud
    - \nabla^\ud \cT_{\cA \cB \cC}
    + 3 \,\cT_{[\cA \cB|}{}^\cE \cR_{\cE |\cC]}{}^\ud
    + 3 \,\cR_{[\cA \cB|}{}^\ue f_{\ue |\cC]}{}^\ud
    + 3 \,\cR_{[\cA|}{}^{\ud \ue} f_{\ue |\cB \cC]}
\end{align}
and
\begin{align}
0 &= 2 \nabla_{[\cA} \cT_{\cB]} 
    + \cT_{\cA \cB}{}^\cC \cT_\cC
    + \nabla^\cC \cT_{\cC \cA \cB}
    + 2\,\cR_{[\cA|}{}^{\cD\uc} f_{\uc \cD |\cB]}
    + \cR^\uc f_{\uc \cA \cB}~.
\end{align}

In terms of these formulae, we can define the following $\g{H}_L \times \g{H}_R$ curvatures. In terms of the naive $\Omega$ curvatures,
\begin{align}
R(\Omega)_{\cA \cB \cC \cD} &:= 
    2 \,D_{[\cA} \Omega_{\cB] \cC \cD}
    - 2 \,\Omega_{[\cA |\cC}{}^{\cE} \Omega_{|\cB] \cE \cD}
    + \cF_{\cA \cB}{}^\cE \Omega_{\cE \cC \cD}
    + \frac{1}{2} \Omega^\cE{}_{\cA \cB} \,\Omega_{\cE \cC \cD}~,
\end{align}
the $\cR_{\cA \cB \rc \rd}$ are given (through dimension two) by
\begin{subequations}
\begin{align}
\cR_{\alpha \beta \, \rc \rd} &= R(\Omega)_{\alpha \beta \,\rc \rd} 
    + 4 \,k(\gamma_{[\rc})_{\gamma (\beta} \Omega_{\alpha) \rd]}{}^\gamma~, \\
\cR_{\alpha \bbeta \, \rc \rd} &= R(\Omega)_{\alpha \bbeta \,\rc \rd} 
    + 2\,k(\gamma_{[\rc})_{\gamma \bbeta} \Omega_{\alpha \rd]}{}^\gamma~, \\
\cR_{\ol{\alpha \beta} \, \rc \rd} &= R(\Omega)_{\ol{\alpha \beta} \,\rc \rd} ~, \\[3ex]
\cR_{\alpha \rb \,\rc \rd} &= R(\Omega)_{\alpha \rb \,\rc \rd} 
    + \tfrac{2}{9} k \,\eta_{\rb [\rc} (\gamma_{\rd]})_{\gamma \delta} \Omega_{\alpha}{}^{\gamma \delta}
    - 2 k\,(\gamma_{[\rc|})_{\alpha \gamma} \,\Omega_{\rb|\rd]}{}^\gamma
    - H_{\alpha\, \rb | \rc\rd} \\
\cR_{\alpha \rbb \,\rc \rd} &= R(\Omega)_{\alpha \rbb \,\rc \rd} 
    - 2 k\,(\gamma_{[\rc})_{\alpha \gamma} \,\Omega_{\rbb \,\rd]}{}^\gamma \\
\cR_{\balpha \rb \,\rc \rd} &= R(\Omega)_{\balpha \rb \,\rc \rd} 
    + \tfrac{2}{9} k \,\eta_{\rb [\rc} (\gamma_{\rd]})_{\gamma \delta} \Omega_{\balpha}{}^{\gamma \delta}
    - H_{\balpha\, \rb | \rc\rd} \\
\cR_{\ol{\alpha \rb} \,\rc \rd} &= R(\Omega)_{\ol{\alpha \rb} \,\rc \rd} \\[3ex]
\cR_{\ra \rb \,\rc \rd} &= R(\Omega)_{\ra \rb \,\rc \rd} 
    + \tfrac{4}{9} k\, \Omega_{[\ra}{}^{\alpha \beta}\, 
    \eta_{\rb] [\rc} (\gamma_{\rd]})_{\alpha\beta}
    - 2\,H_{[\ra, \rb]|\rc\rd}
    + P_{\ra \rb\, \rc \rd} \\
\cR_{\ra \rbb \,\rc \rd} &=
    R(\Omega)_{\ra \rbb \,\rc \rd}
    - \tfrac{2}{9} k\,\eta_{\ra [\rc} (\gamma_{\rd]})_{\alpha \beta} \Omega_{\rbb}{}^{\alpha \beta}
    + H_{\rbb \, \ra|\rc\rd} \\
\cR_{\overline{\ra \rb} \,\rc \rd} &= R(\Omega)_{\overline{\ra\rb} \,\rc\rd}
    + P_{\overline{\ra \rb} \,\rc\rd}\\[3ex]
\cR_\alpha{}^\beta{}_{\rc\rd} &=
    R(\Omega)_\alpha{}^\beta{}_{\rc\rd} 
    + 2\,H_{\alpha\, [\rc,\rd]}{}^\beta
    + \tfrac{1}{4} (\gamma^{\ra\rb})_\alpha{}^\beta P_{\ra\rb\,\rc\rd}~, \\
\cR_\alpha{}^\bbeta{}_{\rc\rd} &=
    R(\Omega)_\alpha{}^\bbeta{}_{\rc\rd} ~, \\
\cR_\balpha{}^\beta{}_{\rc\rd} &=
    R(\Omega)_\balpha{}^\beta{}_{\rc\rd} 
    + 2\,H_{\balpha\, [\rc,\rd]}{}^\beta~, \\
\cR_\balpha{}^\bbeta{}_{\rc\rd} &=
    R(\Omega)_\balpha{}^\bbeta{}_{\rc\rd} 
    + \tfrac{1}{4} (\gamma^{\ol{\ra\rb}})_\balpha{}^\bbeta P_{\ol{\ra\rb}\,\rc\rd}~.
\end{align}
\end{subequations}
There are additional components of $\cR_{\cA \cB \rc \rd}$, namely
$\cR^\halpha{}_{\hb\,\rc \rd}$ and
$\cR^{\halpha\hbeta}{}_{\rc \rd}$, at dimension 5/2 and dimension 3, but we will not need them here. The dilatonic curvature $\cR_{\ra \rb}$ also lies at dimension 2; it is given by
\begin{align}
\cR_{\ra \rb} &= 
    (D^\cC + \cF^\cC) \Omega_{\cC \ra \rb}
    - H^\rc{}_{\rc|\ra\rb}
    + \frac{2}{9} k \, \Omega_{[\ra}{}^{\gamma \beta} (\gamma_{\rb]})_{\beta\gamma}
    - 2 P^\rc{}_{[\ra, \rb] \rc}~.
\end{align}

The additional contributions from the $H$ connections and the \Polacek-Siegel field permit a number of constraints to be imposed on these curvatures (similarly for their barred versions):
\begin{subequations}
\label{E:HP.Constraints}
\begin{alignat}{2}
\label{E:HP.Constraints.a}
H_{\halpha\, \rb|\rc\rd} \quad &\implies &\quad \cR_{\halpha \rb\,\rc\rd} \,\Big\vert_{\rb |\rc\rd} &= 0~, \\
\label{E:HP.Constraints.b}
H_{\rba \, \rb|\rc\rd} \quad &\implies &\quad \cR_{\rba \rb \,\rc \rd} \Big\vert_{\rb|\rc\rd} &= 0~, \\
\label{E:HP.Constraints.c}
P_{\ra \rb\, \overline{\rc\rd}} \quad &\implies &\quad \cR_{\ra \rb \,\overline{\rc\rd}} &= \cR_{\overline{\rc\rd} \,\ra \rb}~, \\
\label{E:HP.Constraints.d}
H_{\ra,\rb|\rc\rd}~,\,\, P_{\ra\rb\, \rc\rd} \quad &\implies &\quad
\cR_{\ra \rb\, \rc \rd} &= \tfrac{1}{45} \eta_{\ra [\rc} \eta_{\rd] \rb}\, \cR + \cR_{[\ra\rb\, \rc\rd]}~,
\end{alignat}
\end{subequations}
where $\vert_{\rb|\rc\rd}$ denotes projection to the irreducible hook representation.

There are a few additional curvature tensors through dimension two. First, the curvatures
\begin{subequations}
\begin{align}
\cR_{\alpha \beta \,\rc}{}^\gamma &= R(\Omega)_{\alpha \beta \,\rc}{}^\gamma
    - 2 k (\gamma_\rc)_{\delta (\beta } \Omega_{\alpha)}{}^{\delta \gamma}
    + \tfrac{1}{9} k\,
        (\gamma_{\rc \rd})_{(\beta}{}^\gamma  \Omega_{\alpha)}{}^{\delta \epsilon}\,
        (\gamma^\rd)_{\delta \epsilon}
    - \tfrac{1}{2} (\gamma^{\rd\re})_{(\beta}{}^\gamma\, H_{\alpha)\, \rc|\rd\re}~, \\
\cR_{\alpha \bbeta \,\rc}{}^\gamma &= R(\Omega)_{\alpha \bbeta \,\rc}{}^\gamma
    - k (\gamma_\rc)_{\delta \alpha} \Omega_{\bbeta}{}^{\delta \gamma}
    + \tfrac{1}{18} k\,
        (\gamma_{\rc \rd})_{\alpha}{}^\gamma  \Omega_{\bbeta}{}^{\delta \epsilon}\,
        (\gamma^\rd)_{\delta \epsilon}
    - \tfrac{1}{4} (\gamma^{\rd\re})_{\alpha}{}^\gamma\, H_{\bbeta\, \rc|\rd\re}~, \\
\cR_{\ol{\alpha \beta} \,\rc}{}^\gamma &= R(\Omega)_{\ol{\alpha \beta} \,\rc}{}^\gamma~, \\[2ex]
\cR_{\alpha \rb \,\rc}{}^\gamma &=
    R(\Omega)_{\alpha \rb \,\rc}{}^\gamma
    + k\, (\gamma_\rc)_{\alpha \delta} \Omega_{\rb}{}^{\delta \gamma}
    - \tfrac{1}{18} k\, (\gamma_{\rc \rd})_{\alpha}{}^\gamma \, (\gamma^\rd)_{\delta \epsilon} \Omega_\rb{}^{\delta \epsilon}
    + \tfrac{1}{4} (\gamma^{\re\rf})_\alpha{}^\gamma\, H_{\rb \,\,\rc|\re\rf}
    - H_{\alpha\, \rb ,\rc}{}^\gamma~, \\
\cR_{\alpha \rbb \,\rc}{}^\gamma &=
    R(\Omega)_{\alpha \rbb \,\rc}{}^\gamma
    + k \,(\gamma_\rc)_{\alpha \delta} \Omega_\rbb{}^{\delta \gamma}
    - \tfrac{1}{18} k\, (\gamma_{\rc \rd})_{\alpha}{}^\gamma \, (\gamma^\rd)_{\delta \epsilon} \Omega_\rbb{}^{\delta \epsilon}
    + \tfrac{1}{4} (\gamma^{\re\rf})_\alpha{}^\gamma \, H_{\rbb \,\rc|\re\rf}~, \\
\cR_{\balpha \rb \,\rc}{}^\gamma &=
    R(\Omega)_{\balpha \rb \,\rc}{}^\gamma
    - H_{\balpha\, \rb ,\rc}{}^\gamma~, \\
\cR_{\ol{\alpha \rb} \,\rc}{}^\gamma &=
    R(\Omega)_{\ol{\alpha\rb} \,\rc}{}^\gamma~,
\end{align}
\end{subequations}
and
\begin{subequations}
\begin{align}
\cR_{\alpha \beta}{}^{\gamma \delta} &= 
    R(\Omega)_{\alpha \beta}{}^{\gamma \delta}
    - 2 \, H_{(\alpha\, \rb, \rc}{}^{(\gamma} (\gamma^{\rb\rc})_{\beta)}{}^{\delta)}~, \\
\cR_{\alpha \bbeta}{}^{\gamma \delta} &= 
    R(\Omega)_{\alpha \bbeta}{}^{\gamma \delta}
    - H_{\bbeta\, \rb, \rc}{}^{(\gamma} (\gamma^{\rb\rc})_{\alpha}{}^{\delta)}~, \\
\cR_{\ol{\alpha \beta}}{}^{\gamma \delta} &= 
    R(\Omega)_{\ol{\alpha \beta}}{}^{\gamma \delta}
\end{align}
\end{subequations}
(along with their barred versions) correspond to the covariantizations of the remaining dimension $\leq$ 2 pieces of $R(\Omega)_{\cA \cB \cC \cD}$. Among these curvatures, the only constraint we can impose is
\begin{align}
\label{E:HP.Constraints.e}
H_{\halpha \,\rb,\rc}{}^\gamma \quad \implies \quad 
    \cR_{\halpha \rb\, \rc}{}^\gamma = 0~.
\end{align}
The only remaining curvatures at dimension two are the lowest dimension pieces of
$\cR_{\cA \cB}(K)^{\rc|\rd\re}$:
\begin{subequations}
\begin{align}
\cR_{\alpha \beta}(K)_{\rc |\rd \re} &=
    2 \,\cD_{(\alpha} H_{\beta)\, \rc |\rd\re}
    + \cF_{\alpha \beta}{}^\cE H_{\cE\, \rc|\rd\re}
    + 4 k \,\Omega_{(\alpha|\,\rc}{}^{\gamma} (\gamma_{\rd})_{\gamma\delta} 
        \Omega_{|\beta)\,\re}{}^{\delta} 
    \eol & \quad
    + 4k\, H_{(\alpha\, \rd, \re}{}^{\gamma} (\gamma_\rc)_{\beta)\gamma}
    - 4k\, H_{(\alpha\, \rc, \rd}{}^{\gamma} (\gamma_\re)_{\beta)\gamma}
    \proj    ~,\\[2ex]
\cR_{\alpha \bbeta}(K)_{\rc |\rd \re} &=
    2 \,\cD_{(\alpha} H_{\bbeta)\, \rc |\rd\re}
    + \cF_{\alpha \bbeta}{}^\cE H_{\cE\, \rc|\rd\re}
    + 4 k \,\Omega_{(\alpha|\,\rc}{}^{\gamma} (\gamma_{\rd})_{\gamma\delta} 
        \Omega_{|\bbeta)\,\re}{}^{\delta} 
    \eol & \quad
    + 2k\, H_{\bbeta\, \rd, \re}{}^{\gamma} (\gamma_\rc)_{\alpha)\gamma}
    - 2k\, H_{\bbeta\, \rc, \rd}{}^{\gamma} (\gamma_\re)_{\alpha)\gamma}
    \proj    ~,\\[2ex]
\cR_{\ol{\alpha \beta}}(K)_{\rc |\rd \re} &=
    2 \,\cD_{(\balpha} H_{\bbeta)\, \rc |\rd\re}
    + \cF_{\ol{\alpha \beta}}{}^\cE H_{\cE\, \rc|\rd\re}
    + 4 k \,\Omega_{(\balpha|\,\rc}{}^{\gamma} (\gamma_{\rd})_{\gamma\delta} 
        \Omega_{|\bbeta)\,\re}{}^{\delta} ~.
\end{align}
\end{subequations}
The covariant derivative $\cD$ above carries the double Lorentz connection alone.
We denote these curvatures with a $(K)$ separating the form indices from the indices of the generator $K_{\rc|\rd\re}$, both to distinguish the types of indices and to reduce confusion with $\cR_{\cA \cB \cC \cD}$.

\subsection{Solving Bianchi identities through dimension 2}
\label{E:S.Bianchi}
Now we turn to solving the Bianchi identities. We will restrict our analysis to dimension $\leq 2$. The three Bianchi identities read:
\begin{subequations}
\label{E:BI.Torsion}
\begin{align}
\label{E:BI.Torsion.a}
0 &= \cB_{\cA \cB \cC \cD} \equiv \Big[4 \,\nabla_{\cA} \cT_{\cB \cC \cD}
    + 3 \,\cT_{\cA \cB}{}^\cE \cT_{\cE \cC \cD}
    - 6 \,\cR_{\cA \cB \cC \cD}
    \Big]_{[\cA\cB\cC \cD]}~,\\
\label{E:BI.Torsion.b}
0 &= \cB_{\cA \cB } \equiv \Big[2 \nabla_{\cA} \cT_{\cB} 
    + \cT_{\cA \cB}{}^\cC \cT_\cC
    + \nabla^\cC \cT_{\cC \cA \cB}
    - \cR_{\cA}{}^{\cD}{}_{\cD \cB}
    + \cR_{\cB}{}^{\cD}{}_{\cD \cA}
    - \cR_{\cA \cB}
    \Big]_{[\cA\cB]}~, \\
\label{E:BI.Torsion.c}
0 &= \cB \equiv \nabla^\cA \cT_\cA
    + \tfrac{1}{2} \cT^\cA \cT_\cA
    + \tfrac{1}{12} \cT^{\cA \cB \cC} \cT_{\cC \cB \cA}
    - \tfrac{1}{2} \cR^{\cA \cB}{}_{\cB \cA}
\end{align}
\end{subequations}
In our previous work on type I DFT \cite{Butter:2021dtu}, we analyzed all three Bianchi identities simultaneously. Here we will take a bit of a different approach and focus only on the first set \eqref{E:BI.Torsion.a}, which does not involve the superdilaton. The reason for this is that the double supergeometry emerging there will naturally correspond to generalized type II supergravity \cite{Wulff:2016tju}, where a dilaton is not presumed to exist.
The analysis of \eqref{E:BI.Torsion.a} is nearly identical to the type I discussion, so we will be relatively brief, proceeding by dimension.

\paragraph{Dimension 0.}
The Bianchi identities $\cB_{\halpha \hbeta \hgamma \hdelta}$ are all satisfied trivially
given the dimension -1/2 and dimension 0 constraints.

\paragraph{Dimension 1/2.}
The Bianchi identities at dimension 1/2 read
\begin{subequations}
\begin{alignat}{5}
\cB_{\alpha \beta \gamma \rd} &= 0 &\quad &\implies &\quad
\cT_{(\alpha \beta}{}^{\delta} (\gamma_{\rd})_{\gamma) \delta} &= 0 &\quad &\implies &\quad
\cT_{\alpha \beta}{}^{\gamma} &= X_{(\alpha} \delta_{\beta)}{}^\gamma 
    - \frac{1}{2} (\gamma^\rc)_{\alpha \beta} (\gamma_\rc)^{\gamma \delta} X_\delta ~, \\
\cB_{\alpha \beta \gamma \rbd} &= 0 &\quad &\implies &\quad
(\gamma^\re)_{(\alpha \beta} \cT_{\re \gamma) \rbd} &= 0  &\quad &\implies &\quad
\cT_{\alpha \rb \rbc} &= (\gamma_{\rb})_{\alpha \beta} \cW^\beta{}_\rbc~, \\
\cB_{\alpha \beta \bgamma \rd} &= 0 &\quad &\implies &\quad
\cT_{\bgamma (\alpha}{}^{\delta} (\gamma_{\rd})_{\beta) \delta} &= 0 &\quad &\implies &\quad
\cT_{\balpha \beta}{}^\gamma &= 0~, \\
\cB_{\alpha \beta \bgamma \bd} &= 0 &\quad &\implies &\quad
\cT_{\alpha \beta}{}^{\bdelta} (\gamma_{\rbd})_{\bgamma \bdelta} &= 0 &\quad &\implies &\quad
\cT_{\alpha \beta}{}^\bgamma &= 0~,
\end{alignat}
\end{subequations}
plus their barred versions. The dimension 1/2 torsions that remain even in light of the Bianchi
identities are the fields $X_\alpha$ and $\cW^\beta{}_\rbc$ (corresponding to parts of
$\cT_{\alpha \beta}{}^{\gamma}$ and $\cT_{\alpha \rb \rbc}$). Both can be set to zero by redefining the dilatino and the gravitino superfields. In other words, \emph{this is merely another conventional constraint} --- exactly the DFT analogue of a gravitino redefinition discussed in \cite{Wulff:2016tju}.
The dimension 1/2 torsions then all vanish,
\begin{align}
\cT_{\halpha \hb \hc} = 0~, \qquad
\cT_{\halpha \hbeta}{}^\hgamma = 0~.
\end{align}

\paragraph{Dimension 1.}
Using $\Omega_{[\rc \rb \ra]}$ and $\Omega_{\rbc \rb \ra}$, we can set
$\cT_{\ha \hb \hc} = 0$ as a conventional constraint. Since we are not addressing the superdilaton curvatures yet, we do not set $\cT_\ha = 0$, so $\Omega^\rb{}_{\rb \ra}$ remains unfixed. The other torsion tensors at dimension 1 involve $\cT_{\ha \hbeta}{}^\hgamma$. Recall we can use $\Omega_{\hbeta \ra}{}^\gamma$ to fix
$\cT_{\hbeta \ra}{}^\gamma = \frac{1}{10} (\gamma_\ra)^{\gamma \delta} \cX_{\hbeta, \delta}$ and its barred version.
Then the Bianchi identities $\cB_{\halpha \hbeta \hc \hd} = 0$ read:
\begin{subequations}
\begin{alignat}{3}
\label{E:Bianchi2.1a}
\cB_{\alpha \beta c d} &= 0 &\quad &\implies &\quad 
    \cR_{\alpha \beta \, c d} &= \frac{2k}{5} (\gamma_{cd})_{(\alpha}{}^\gamma \cX_{\beta), \gamma}~, \\
\label{E:Bianchi2.1b}
\cB_{\alpha \beta \rc \rbd} &= 0 &\quad &\implies &\quad
\cT_{\bd (\alpha}{}^\gamma (\gamma_\rc)_{\alpha) \gamma} &= 0 \quad \implies \quad
\cT_{\bc \beta}{}^\alpha = 0~, \\
\label{E:Bianchi2.1c}
\cB_{\alpha \beta \ol{\rc\rd}} &= 0 &\quad &\implies &\quad 
\cR_{\alpha \beta \,\ol{\rc\rd}} &= 0 ~, \\
\label{E:Bianchi2.1d}
\cB_{\alpha \bbeta \rc \rd} &= 0 &\quad &\implies &\quad 
    \cR_{\alpha \bbeta \, \rc \rd} &= \frac{k}{5} (\gamma_{\rc\rd})_{\alpha}{}^{\gamma} \cX_{\bbeta, \gamma}~, \\
\label{E:Bianchi2.1e}
\cB_{\alpha \bbeta \rc \rbd} &=0 &\quad &\implies &\quad
    \cT_{\rc \alpha}{}^\bgamma (\gamma_{\rbd})_{\bgamma \bbeta}
    &= \cT_{\rd \bbeta}{}^\gamma (\gamma_{\rc})_{\gamma \beta}
    \quad \implies \quad
    \cT_{\rc \beta}{}^\balpha = \frac{1}{10} (\gamma_\rc)_{\beta \alpha} \cX^{\alpha \balpha} \eol
&\phantom{=} &\quad &\phantom{\implies} &\quad
    &\phantom{= \cT_{\rd \bbeta}{}^\gamma (\gamma_{\rc})_{\gamma \beta}}
    \quad \phantom{\implies} \quad
    \quad\cT_{\rbc \bbeta}{}^\alpha = \frac{1}{10} (\gamma_\rbc)_{\bbeta \balpha} \cX^{\alpha \balpha}
\end{alignat}
\end{subequations}
along with their barred versions. The last factor can be removed by redefining the Ramond-Ramond bispinor $S^{\alpha \bbeta}$; then as another conventional constraint, we fix $\cX^{\alpha \balpha} = 0$ above.

In addition to these, we have $\cB_{\halpha \hbeta \hgamma}{}^\hdelta=0$, which decomposes as
\begin{subequations}
\begin{alignat}{3}
\label{E:Bianchi2.2a}
\cB_{\alpha \beta \gamma}{}^\delta &= 0 &\quad &\implies &\quad 
    \frac{1}{4} \cR_{(\alpha \beta\, c d} (\gamma^{c d})_{\gamma)}{}^\delta &=
        -\frac{k}{10} \cX_{\gamma, \epsilon} 
        (\gamma_\rc)_{\alpha \beta} (\gamma^\rc)^{\epsilon \delta}
        \Big\vert_{(\alpha \beta \gamma)}~, \\
\label{E:Bianchi2.2b}
\cB_{\alpha \beta \bgamma}{}^\delta &= 0 &\quad &\implies &\quad 
    \frac{1}{2} \cR_{(\alpha \bgamma\, c d} (\gamma^{c d})_{\beta)}{}^\delta &= 
        -\frac{k}{10} (\gamma^e)_{\alpha \beta} (\gamma_e)^{\delta \epsilon} \cX_{\bgamma, \epsilon}~, \\
\label{E:Bianchi2.2c}
\cB_{\alpha \bbeta \bgamma}{}^\delta &= 0 &\quad &\implies &\quad 
    0 &= 0~, \\
\label{E:Bianchi2.2d}
\cB_{\balpha \bbeta \bgamma}{}^\delta &= 0 &\quad &\implies &\quad 
    0 &= 0
\end{alignat}
\end{subequations}
Combining \eqref{E:Bianchi2.1d} with \eqref{E:Bianchi2.2b} implies that $\cX_{\balpha, \beta} = 0$.
Combining \eqref{E:Bianchi2.1a} with \eqref{E:Bianchi2.2a} implies that
$\cX_{\alpha, \beta}$ is purely vectorial. We can set this to zero as a conventional
constraint by redefining $\Omega^\rb{}_{\rb \ra}$. In summary, 
\begin{align}
\cT_{\ha \hb \hc} = 0~, \qquad
\cT_{\ha \hbeta}{}^\hgamma = 0
\end{align}
These constraints are consistent with the vanishing of all lower dimensional torsion
tensors except for $\cT_{\halpha \hbeta \hc}$. We also have found that the
curvature tensors at this dimension vanish as well,
\begin{align}
\cR_{\halpha \hbeta\, \rc \rd} = \cR_{\halpha \hbeta\, \ol{\rc \rd}} = 0~.
\end{align}

\paragraph{Dimension 3/2.}
The torsion tensors at dimension 3/2 consist of
$\cT_{\ha \hb}{}^\hgamma$ and $\cT_\halpha{}^{\hbeta \hgamma}$. We can
use $\Omega_\halpha{}^{\beta \gamma}$ and its barred version
to fix $\cT_\halpha{}^{\beta \gamma} = \cT_\halpha{}^{\ol{\beta \gamma}} = 0$.
We can also use $\Omega^{\hgamma}{}_{\ra \rb}$ and its barred version
to fix $\cT_{\ra \rb}{}^{\hgamma} = \cT_{\ol{\ra \rb}}{}^{\hgamma} = 0$.
Finally, we can take $\Omega_{\rba \, \rb}{}^{\gamma}$ and its barred version to fix
\begin{align}
\cT_{\rba \rb}{}^\gamma = \frac{1}{10} (\gamma_\rb)^{\gamma \delta} \cX_{\rba, \gamma}~, \qquad
\cT_{\ra \rbb}{}^\bgamma = \frac{1}{10} (\gamma_\rbb)^{\ol{\gamma \delta}} \cX_{\ra, \bgamma}~. 
\end{align}
The Bianchi identities $\cB_{\halpha \hb \hc \hd} = 0$ lead to
\begin{subequations}
\begin{alignat}{3}
\label{E:Bianchi3.1a}
\cB_{\alpha \ol{\rb\rc\rd}} &= 0 &\quad &\implies &\quad 
    \cR_{\alpha [\ol{\rb\rc\rd}]} &= 0~, \\
\label{E:Bianchi3.1b}
\cB_{\alpha \rb \ol{\rc\rd}} &= 0 &\quad &\implies &\quad 
    \cR_{\alpha \rb \, \ol{\rc\rd}} &= 0~, \\
\label{E:Bianchi3.1c}
\cB_{\alpha \rbb \rc \rd} &= 0 &\quad &\implies &\quad 
    \cR_{\alpha \rbb\, \rc \rd} &= 
        -\frac{k}{5} (\gamma_{\rc\rd})_\alpha{}^\gamma \cX_{\rbb, \gamma}~, \\
\label{E:Bianchi3.1d}
\cB_{\alpha \rb \rc \rd}  &= 0 &\quad &\implies &\quad 
    \cR_{\alpha [\rb \rc \rd]} &= 0
\end{alignat}
\end{subequations}
Next, we have $\cB_{\halpha \hbeta \rc}{}^{\hdelta} = 0$. The first batch is
\begin{subequations}
\begin{alignat}{3}
\label{E:Bianchi3.2a}
\cB_{\alpha \beta \rc}{}^\delta &= 0 &\quad &\implies &\quad 
    \cR_{\alpha \beta \,\rc}{}^\delta &= \frac{1}{2} (\gamma^{\rd\re})_{(\alpha}{}^\delta \cR_{\beta) \rc\, \rd \re}~, \\
\label{E:Bianchi3.2b}
\cB_{\alpha \bbeta \rc}{}^\delta &=0 &\quad &\implies &\quad 
    \cR_{\alpha \bbeta\, \rc}{}^\delta &= \frac{1}{4} (\gamma^{\ra\rb})_{\alpha}{}^\delta \cR_{\bbeta \rc\, \ra\rb}~, \\
\label{E:Bianchi3.2c}
\cB_{\ol{\alpha \beta} \rc}{}^\delta &=0 &\quad &\implies &\quad 
    \cR_{\ol{\alpha \beta}\, \rc}{}^\delta &= 
        \frac{k}{10}\, (\gamma^\rba)_{\ol{\alpha\beta}} (\gamma_\rc)^{\delta \epsilon} \cX_{\rba, \epsilon}~.
\end{alignat}
Each of these expressions is traceless when contracted with $(\gamma^\rc)_{\delta \gamma}$. The first two, combined with \eqref{E:Bianchi3.1a} and \eqref{E:Bianchi3.1d}, imply that only the irreducible hook representations of $\cR_{\halpha\, \rb \,\rc \rd}$ are present, but these are eliminated using the constraint \eqref{E:HP.Constraints.a}. The third equation, being $\gamma$-traceless on the left-hand side but pure trace on the right, is solved only by $\cR_{\ol{\alpha \beta}\, \rc}{}^\delta = \cT_{\rba \rb}{}^\gamma = 0$.
The remaining identities are
\begin{alignat}{3}
\label{E:Bianchi3.2d}
\cB_{\alpha \beta \rbc}{}^\delta &=0 &\quad &\implies &\quad 
    (\gamma^{\ra\rb})_{(\alpha}{}^\delta \cR_{\beta) \rbc \,\ra\rb} &= 
        \frac{k}{5} (\gamma^\ra)_{\alpha \beta} 
        (\gamma_\ra)^{\delta \gamma} \cX_{\rbc, \gamma}~, \\
\label{E:Bianchi3.2e}
\cB_{\alpha \bbeta \rbc}{}^\delta &=0 &\quad &\implies &\quad 
(\gamma_\rbc)_{\ol{\beta \gamma}} \cT_\alpha{}^{\delta \bgamma} &= 0
\quad \implies \quad
\cT_\alpha{}^{\beta \bgamma} = 0~, \\
\label{E:Bianchi3.2f}
\cB_{\ol{\alpha \beta \rc}}{}^\delta &=0 &\quad &\implies &\quad 
    (\gamma_\rc)_{\bgamma (\balpha} \cT_{\bbeta)}{}^{\bgamma \delta} &= 0
    \quad \implies \quad
    \cT_{\balpha}{}^{\beta \bgamma} = 0
\end{alignat}
\end{subequations}
Combining \eqref{E:Bianchi3.1c} and \eqref{E:Bianchi3.2d} tells us $\cX_{\rba, \beta} = 0$.

The upshot is that we have eliminated all torsions and curvatures at dimension 3/2,
\begin{gather}
\cT_{\ha \hb}{}^\hgamma = \cT_{\halpha}{}^{\hbeta \hgamma} = 0~, \qquad
\cR_{\halpha \hb \hc \hd} = \cR_{\halpha \hbeta\, \hc}{}^{\hdelta} = 0~.
\end{gather}

\paragraph{Dimension 2.}
Let's start with $\cB_{\ha \hb \hc \hd}$. As in the bosonic case, this reads
\begin{subequations}
\begin{alignat}{3}
\cB_{\ra\rb\rc\rd} &= 0 
&\quad &\implies &\quad
\cR_{[\ra\rb \, \rc\rd]} &= 0~, \\
\cB_{\ra\rb\rc \rbd} &= 0 
&\quad &\implies &\quad
\cR_{\rbd [\rc \, \ra \rb]} &= 0 ~, \\
\cB_{\ra\rb \overline{\rc\rd}} &= 0 
&\quad &\implies &\quad
\cR_{\ra \rb \, \overline{\rc\rd}} &= -\cR_{\overline{\rc\rd}\, \ra \rb}
\end{alignat}
\end{subequations}
along with their barred versions. Using \eqref{E:HP.Constraints.b} -- \eqref{E:HP.Constraints.d},
we fix
\begin{subequations}
\begin{alignat}{2}
\cR_{\ra \rb\, \rc \rd} &= \tfrac{1}{45} \eta_{\ra [\rc} \eta_{\rd] \rb}\, \cR &\qquad
\cR_{\overline{\ra \rb}\, \overline{\rc \rd}} &= \tfrac{1}{45} \eta_{\rba [\rbc} \eta_{\rbd] \rbb}\, \bar\cR \\
\cR_{\ra \rbb \, \rc \rd} &= \tfrac{2}{9} \,\eta_{\ra [\rc} \cR^\re{}_{\rbb\, \re \rd]}~, &\qquad
\cR_{\ra \rbb \, \overline{\rc\rd}} &= -\tfrac{2}{9} \eta_{\rbb [\rbc} \cR_{\ra \overline{\re}\, \rbd]}{}^{\overline \re}~, \\
\cR_{\overline{\ra \rb}\, \rc\rd} &= 0 ~, & \quad
\cR_{\ra \rb\, \overline{\rc\rd}} &= 0~.
\end{alignat}
\end{subequations}
Recall \eqref{E:HP.Constraints.e} fixes $\cR_{\alpha \rb \,\rc}{}^{\gamma} = 0$.
Next, we use
\begin{subequations}
\begin{alignat}{3}
\cB_{\alpha}{}^\beta{}_{\rc\rd} &= 0  &\quad &\implies &\quad
    \cR_{\alpha}{}^{\beta}{}_{\rc\rd} 
        &= -\frac{1}{4} (\gamma^{\ra\rb})_\alpha{}^\beta \cR_{\rc\rd \, \ra\rb}
        = -\frac{1}{180} (\gamma_{\rc\rd})_\alpha{}^\beta \cR ~,\\
\cB_\alpha{}^\beta{}_{\rc \rbd} &= 0  &\quad &\implies &\quad
    \cR_{\alpha \rbd \,\rc}{}^\beta &= \frac{1}{4} (\gamma^{\ra\rb})_\alpha{}^\beta \cR_{\rc \rbd\, \ra \rb}
    \quad \implies \quad
    \cR_{\alpha \rbd \,\rc}{}^\beta  = \cR_{\rc \rbd\, \ra \rb} = 0~,\\
\cB_\alpha{}^\beta{}_{\ol{\rc \rd}} &= 0   &\quad &\implies &\quad
    \cR_\alpha{}^\beta{}_{\ol{\rc\rd}} &= 
        - \frac{1}{4} (\gamma^{\ra\rb})_\alpha{}^\beta \cR_{\ol{\rc\rd} \, \ra\rb} 
    \quad \implies \quad
    \cR_\alpha{}^\beta{}_{\ol{\rc\rd}} = 0~, \\
\cB_\alpha{}^\bbeta{}_{\rc \rbd} &= 0   &\quad &\implies &\quad
    \cR_{\alpha \rc\, \rbd}{}^\bbeta &= 
        \frac{k}{10} (\gamma_\rc)_{\alpha \gamma} (\gamma_\rbd)^{\bbeta \bdelta} \cX^\gamma{}_\bdelta
        \quad \implies \quad
        \cR_{\alpha \rc\, \rbd}{}^\bbeta = \cX^\gamma{}_\bdelta = 0~, \\
\cB_\alpha{}^\bbeta{}_{\rc\rd} &= 0   &\quad &\implies &\quad
    \cR_\alpha{}^\bbeta{}_{\rc\rd} &= \frac{k}{5}\, (\gamma_{\rc \rd})_\alpha{}^\gamma \cX^\bbeta{}_\gamma
        \quad \implies \quad
        \cR_\alpha{}^\bbeta{}_{\rc\rd} = 0~, \\
\cB_\alpha{}^\bbeta{}_{\ol{\rc\rd}} &= 0   &\quad &\implies &\quad
    \cR_\alpha{}^{\bbeta}{}_{\ol{\rc\rd}} &= -2\,\cR_{\alpha [\rbc \,\rbd]}{}^{\bbeta} 
            \quad \implies \quad
    \cR_\alpha{}^{\bbeta}{}_{\ol{\rc\rd}}  = 0~.
\end{alignat}
\end{subequations}
After that, we can use  $\cB_{\halpha \hbeta}{}^{\hgamma \hdelta}$. These involve
\begin{subequations}
\begin{alignat}{3}
\cB_{\alpha\gamma}{}^{\beta\delta} &= 0  &\quad &\implies &\quad
    \cR_{\alpha \gamma}{}^{\beta \delta} &=
        - \frac{1}{4} (\gamma^{\ra\rb})_{(\alpha}{}^{(\beta} (\gamma^{\rc\rd})_{\gamma)}{}^{\delta)}
            \cR_{\ra \rb\, \rc \rd}
        = - \frac{1}{4} (\gamma^{\ra\rb})_{(\alpha}{}^{(\beta} 
                (\gamma_{\ra\rb})_{\gamma)}{}^{\delta)} \cR~, \\
\cB_{\alpha\gamma}{}^{\beta \bdelta} &= 0  &\quad &\implies &\quad
    0 &= 0~, \\
\cB_{\alpha\gamma}{}^{\bbeta\bdelta} &= 0  &\quad &\implies &\quad
    \cR_{\alpha \gamma}{}^{\bbeta \bdelta} &= 0~, \\
\cB_{\alpha\gamma}{}^{\beta \bdelta} &=0 &\quad &\implies &\quad
    0&= 0~, \\
\cB_{\alpha\bgamma}{}^{\beta \delta} &= 0 &\quad &\implies &\quad
    \cR_{\alpha \bgamma}{}^{\beta \delta} &= 0~.
\end{alignat}
\end{subequations}

The superfield $\cR$ remains unfixed at this stage. To determine it, we need to invoke
the Lorentz curvature Bianchi identity. The dimension 3/2 part of this vanishes
given the conditions already imposed. The non-trivial dimension 2 part reads
\begin{align}
\cR_{\alpha \beta}(K)_{\rc | \rd \re}
    &= 
    \cT_{(\alpha \beta}{}^\rf \cR_{\rf \rc)\,\rd \re}
    - \frac{k}{9} \cR_{\alpha \beta}{}^{\gamma \delta} 
        (\gamma_{[\rd})_{\gamma \delta} \eta_{\re] \rc}
\end{align}
and this is solved by taking
\begin{align}
\cR = 0~, \qquad \cR_{\alpha \beta}(K)_{\rc | \rd \re} = 0~.
\end{align}
The upshot is that all dimension 2 curvatures vanish.

\subsection{Dilatonic torsion Bianchi identities}
The dilatonic torsion Bianchi identities are \eqref{E:BI.Torsion.b}
and \eqref{E:BI.Torsion.c}. These hold when the dilatonic torsion $\cT_\cA$ is
given in terms of a superdilaton $\Phi$. However, we would like to show that as
a consequence of the constraints imposed already on the torsion $\cT_{\cA \cB \cC}$
and curvatures, one can choose $\cT_\cA$ so that these hold \emph{without} supposing
the existence of $\Phi$. This is relevant because the resulting type II supergeometry
must correspond then to that of \cite{Wulff:2016tju}.

The starting point is to suppose $\cT_\halpha = 0$.
Then the dimension 1 Bianchi identities $\cB_{\halpha \hbeta}$ amount to
\begin{align}
\cB_{\alpha \beta} &= \cT_{\alpha \beta}{}^\rc \cT_\rc
    + \cR_{\alpha\gamma}{}_{\beta}{}^\gamma
    + \cR_{\beta\gamma}{}_{\alpha}{}^\gamma
    = k (\gamma^\rc)_{\alpha \beta} \cT_\rc
    ~, \\
\cB_{\alpha \bbeta} &= 
    \cR_{\alpha\bgamma}{}_{\bbeta}{}^\bgamma
    + \cR_{\bbeta\gamma}{}_{\alpha}{}^\gamma
    = 0
\end{align}
and similarly for $\cB_{\ol{\alpha\beta}}$. These vanish if we fix $\cT_\ha = 0$. Remember that we could have chosen this instead as a conventional constraint to fix $\Omega^\rb{}_{\rb \ra}$; then this Bianchi identity would have been responsible for eliminating the purely vectorial part of $\cX_{\alpha, \beta}$ in $\cT_{\beta \ra}{}^\gamma$.

The dimension 3/2 identities are
\begin{align}
\cB_{\alpha \rb} &= - \cT_{\alpha \rb \gamma} \cT^\gamma
    + \cR_{\alpha \rc\, \rb}{}^\rc
    + \cR_{\alpha \gamma\, \rb}{}^\gamma
    - \cR_{\rb \gamma}{}_\alpha{}^\gamma
    = k (\gamma_\rb)_{\alpha \gamma} \cT^\gamma~, \\
\cB_{\alpha \rbb} &= 
    \cR_{\alpha \rbc\, \rbb}{}^\rbc
    + \cR_{\alpha \bgamma\, \rbb}{}^\bgamma
    - \cR_{\rbb \gamma}{}_\alpha{}^\gamma
    = 0~,
\end{align}
and their barred versions. The first vanishes if $\cT^\halpha = 0$ and the second vanishes automatically. So already we have concluded all components of the dilaton torsion vanish,
$\cT_\cA = 0$.
The dimension 2 conditions are quite similar and lead to either
identities or definitions of the dilatonic curvatures. Specifically,
\begin{align}
\cB_{\ra\rb} = -\hat \cR_{\ra \rb} ~, \qquad
\cB_{\ra \rbb} = 0 ~, \qquad
\cB_\alpha{}^\beta = - \frac{1}{4} \hat \cR_{\ra \rb} (\gamma^{\ra \rb})_{\alpha}{}^\beta~, \qquad
\cB_\alpha{}^\bbeta = 0~.
\end{align}
These vanish simply by choosing $\widehat \cR_{\ra \rb} = \widehat \cR_{\ol{\ra \rb}} = 0$.
In like fashion, \eqref{E:BI.Torsion.c} is easily found to satisfy $\cB = 0$.

We have not analyzed torsions and curvatures beyond dimension two, but it seems plausible
that the remaining Bianchi identities $\cB_{\cA \cB}$ at dimensions 5/2 and 3 similarly hold as a consequence of $\cB_{\cA \cB \cC \cD} = 0$. We leave the question of the higher dimension torsions and curvatures to future work.

\section{$\g{OSp}(D,D|2s)$ spinors and the Ramond-Ramond sector}
\label{S:OSpSpinor}
A key difference between type I and type II double field theory is the presence of the Ramond-Ramond sector. In bosonic double field theory, the Ramond-Ramond sector is described either by an $\g{O}(D,D)$ spinor \cite{Hohm:2011zr,Hohm:2011dv} or an
$\g{O}(D-1,1)_L \times \g{O}(1,D-1)_R$ bispinor \cite{Jeon:2012kd, Jeon:2012hp}. These two descriptions can be related using the spinorial vielbein \cite{Butter:2022sfh}. Our goal in this section is to give the superspace lift of these relations. That is, we will describe how the Ramond-Ramond super $p$-forms of type II superspace fit into a spinor of $\g{OSp}(D,D|2s)$, and we will give the prescription for identifying a spinorial supervielbein in this framework. Much of this material follows naturally from the bosonic case, so we will be brief where the analogies are clear. Key initial elements of this discussion were already given by Cederwall some time ago \cite{Cederwall:2016ukd}.

\subsection{The Clifford superalgebra}
A natural starting point for defining $\g{OSp}(D,D|2s)$ spinors is via their associated Clifford superalgebra. We introduce gamma matrices $\Gamma^\cM$ that obey the Clifford superalgebra
\begin{align}
\{ \Gamma^\cM, \Gamma^\cN \} = 2 \,\eta^{\cM \cN}~.
\end{align}
In the standard toroidal basis, 
\begin{align}
\eta^{\cM \cN} =
\begin{pmatrix}
0 & \delta^m{}_n & 0 & 0 \\
\delta_m{}^n & 0 & 0 & 0 \\
0 & 0 & 0 & \delta^\hmu{}_\hnu \\
0 & 0 & -\delta_\hmu{}^\hnu & 0
\end{pmatrix}~, \qquad
\eta_{\cM \cN} =
\begin{pmatrix}
0 & \delta_m{}^n & 0 & 0 \\
\delta^m{}_n & 0 & 0 & 0 \\
0 & 0 & 0 & \delta_\hmu{}^\hnu \\
0 & 0 & -\delta^\hmu{}_\hnu & 0
\end{pmatrix}~, 
\end{align}
where $m=1\cdots D$ and $\hmu = 1 \cdots s$. We include a hat on the spinor index as its flat analogue will be denoted $\halpha = (\alpha, \balpha)$.  The Clifford algebra $\Clif(D,D|2s)$ consists of all products of the $\Gamma$-matrices, combined with the unit element $\mathbf 1$: \begin{align}
\Clif(D,D|2s) = \textrm{span}\, (\{\mathbf 1~, \, \Gamma^\cM~, \, \Gamma^{\cM\cN}~, \, \cdots~, \, \Gamma^{\cM_1 \cdots \cM_{p}}~, \cdots\})~.
\end{align}
This is an infinite dimensional algebra, as the spinor-valued $\Gamma$-matrices are commuting and therefore not nilpotent. It is convenient to decompose it into an infinite set of copies of the standard Clifford algebra $\Clif(D,D)$ tensored with the fermionic gamma matrices, i.e.
\begin{align}
\Clif(D,D|2s) = \sum_{p,q} \Clif(D,D){}_{p,q}
\end{align}
where $\Clif(D,D){}_{p,q}$ consists of all elements of $\Clif(D,D)$ multiplied by
$\Gamma^{\hmu_1 \cdots \hmu_p}{}_{\hnu_1 \cdots \hnu_q}$.

\subsection{Orthosymplectic spinors}
The natural definition of an orthosymplectic spinor follows quite analogously from the bosonic case. In the toroidal basis, we define $\Gamma^M$ and $\Gamma_M$ via
\begin{align}
\Gamma^\cM = (\Gamma^M, \Gamma_M)~,
\end{align}
and take $\psiGamma^M = \frac{1}{\sqrt 2} \Gamma^M$ and $\psiGamma_M = \frac{1}{\sqrt 2} \Gamma_M$ as graded raising and lowering operators. These obey 
\begin{align}
\{\psiGamma^N, \psiGamma_M \} = (-)^{mn} \{\psiGamma_M, \psiGamma^N\} = \delta_M{}^N~.
\end{align}
This is a graded anticommutator, so that $\psiGamma^m$ and $\psiGamma_{m}$ furnish a fermionic oscillator algebra,  and $\psiGamma^\hmu$ and $\psiGamma_\hmu$ a bosonic one, i.e.
\begin{align}
\{\psiGamma_n, \psiGamma^m\} = \{\psiGamma^m, \psiGamma_n\} = \delta_n{}^m~, \qquad
[\psiGamma^\hmu, \psiGamma_\hnu]  = -[\psiGamma_\hnu, \psiGamma^\hmu]  = - \delta_\hnu{}^\hmu~.
\end{align}
In order to choose $\psiGamma^M$ to be raising operators and $\psiGamma_M$ to be lowering operators, we build spinors by acting with $\psiGamma^M$ \emph{to the left} on a vacuum bra state $\bra{0}$. This may seem a bizarre choice, but it leads to the convenient identification of an $\g{OSp}(D,D|2s)$ spinor with the standard expansion of a superspace differential form, i.e.
\begin{align}
\bra{\cC} = \sum_p \frac{1}{p!} \bra{0} \psiGamma^{M_1} \cdots \psiGamma^{M_p} \, \cC_{M_p \cdots M_1}
\quad
\Leftrightarrow
\quad
\cC = \sum_p \frac{1}{p!} \rd z^{M_1} \cdots \rd z^{M_p} \, \cC_{M_p \cdots M_1}~.
\end{align}
Because superspace forms are typically written in this fashion with other related
changes (e.g. the de Rham differential acting from the right), it is natural to define
an orthosymplectic spinor as a bra rather than a ket. This will lead us to transpose a number of equations relative to the bosonic case. While this mirroring of bosonic formulae is inconvenient at first glance,  it has the practical effect of eliminating a number of minus signs that would otherwise occur.

A key feature of orthosymplectic spinors is that because the Clifford algebra is infinite dimensional, spinors will necessarily also be infinite dimensional. This corresponds to the notion that a superform can be of arbitrary rank as there is no upper bound on the number of fermionic legs $\rd \theta^\hmu$. Following what we did with the Clifford algebra, it will be convenient to decompose an orthosymplectic spinor depending on how many $\psiGamma^\hmu$ oscillators they involve. That is, we take
\begin{align}\label{E:superC.decompose.p}
\bra{\cC} = \sum_{p=0}^\infty \bra{\cC}_p~,
\end{align}
where $\bra{\cC}_p$ involves $\psiGamma^{\hmu_1} \cdots \psiGamma^{\hmu_p}$.

The natural action for an $\g{OSp}(D,D|2s)$ rotation is
$
\delta \bra{\cC} = \frac{1}{4} \bra{\cC} \Gamma^{\cM\cN} \Lambda_{\cN\cM}
$.
For a generalized diffeomorphism, the parameter $\Lambda$ is given by
$\Lambda_\cM{}^\cN = \pa_\cM \xi^\cN - \pa^\cN \xi_\cM (-)^{nm}$. This suggests that we
define the generalized diffeomorphism of $\cC$ in analogy to the bosonic case as
\begin{align}\label{E:RamondC.Diffeo}
\delta \bra\cC 
    &= \xi^\cN \pa_\cN \bra{\cC}
    + \frac{1}{2} \bra{\cC} \Gamma^{\cM\cN} \pa_\cN \xi_\cM
    + \frac{1}{2} \bra{\cC} \pa_\cM \xi^\cM (-)^m \eol
    &= \xi^\cN \pa_\cN \bra{\cC}
    + \frac{1}{2} \bra{\cC} \Gamma^{\cM} \Gamma^\cN \pa_\cN \xi_\cM~.
\end{align}
The relative normalization of the $\mathbb R^+$ term in the first line is chosen so they combine in the second line, similar to (but mirrored from) the bosonic case. This ensures that both $\bra{\cC}$ and $\bra{\cF} := \bra{\cC} \lDirac$ transform as spinors, where $\slashed{\pa} = \psiGamma^\cM \pa_\cM$. Further, upon solving the section condition as $\tilde\pa^M = 0$, we recover the expected transformation of the complex $\cC$ of super-$p$-forms, 
\begin{align}
\delta \bra{\cC} = \xi^N \pa_N \bra{\cC}
    + \bra{\cC} \psiGamma_M \psiGamma^N \pa_N \xi^M
    + \bra{\cC} \psiGamma^M \psiGamma^N \pa_N \tilde \xi_M \quad \implies \quad
\delta \cC = \Lie_\xi \cC + \cC \wedge \rd \tilde \xi~.
\end{align}

\subsection{Flat orthosymplectic bispinors}
\label{S:OSpSpinor.FlatOSPspinors}

Let us briefly recall how flat $\g{O}(D,D)$ bispinors arise in the bosonic case. We follow the same conventions as \cite{Butter:2022sfh}, but transposing the Fock space so that bras become kets, etc. Thus, we introduce the spinorial version of the double vielbein $\ket{\slashed{V}}$ which obeys
\begin{align}
\Gamma^\hm \ket{\slashed{V}} = \ket{\slashed{V}} \cdot \Gamma^\ha \,V_\ha{}^\hm~.
\end{align}
Here $\ket{\slashed{V}}$ is a bispinor-valued ket. For $D=10$, this ket decomposes as
\begin{align}
\ket{\slashed{V}} =
\begin{pmatrix}
\ket{V_\alpha{}^\balpha} & \ket{V_{\alpha \balpha}} \\
\ket{V^{\alpha\balpha}} & \ket{V^\alpha{}_{\balpha}}
\end{pmatrix}
\end{align}
where $\alpha$ and $\balpha$ are 16-component Weyl spinor indices of $\g{SO}(9,1)$ and $\g{SO}(1,9)$. The flat gamma matrix $\Gamma^\ha$ acts to the left on $\ket{\slashed{V}}$ (and similarly on any bispinor)
\begin{align}\label{E:flatGammaAction.bosonic}
\ket{\slashed{V}} \cdot \Gamma^\ra = \gamma^\ra \ket{\slashed{V}}~, \qquad
\ket{\slashed{V}} \cdot  \Gamma^\rba = \gamma_* \ket{\slashed{V}} \bar\gamma^\rba~,
\end{align}
where $\gamma^\ra$ and $\bar\gamma^\rba$ are gamma matrices of $\g{SO}(9,1)$ and $\g{SO}(1,9)$ respectively.\footnote{Because the action of $\Gamma^\ha$ is defined to the left on any bispinor, we have e.g. $\ket{\slashed{V}} \cdot \Gamma^\ra \Gamma^\rb = (\gamma^\ra \ket{\slashed{V}}) \cdot \Gamma^\rb = \gamma^\rb \gamma^\ra \ket{\slashed{V}}$, and so the order of the left-handed gamma matrices gets reversed. This is in contrast to the discussion in \cite{Butter:2022sfh}, where the Fock space structure was transposed and so the action of right-handed gamma matrices was reversed.}

The flattened bispinor $\slashed{\widehat C}$ is then built from $\bra{C}$ as
\begin{align}\label{E:FlatC.Bispinor}
\slashed{\widehat C} := e^{d} \, \braket{C}{\slashed{V}} = 
\begin{pmatrix}
\widehat C_\alpha{}^\balpha & \widehat C_{\alpha \balpha} \\
\widehat C^{\alpha \balpha} & \widehat C^\alpha{}_\balpha
\end{pmatrix}~.
\end{align}
The relations \eqref{E:flatGammaAction.bosonic} lead to
$\slashed{\widehat C} \cdot \Gamma^\ra = \gamma^\ra \slashed{\widehat C}$
and
$\slashed{\widehat C} \cdot  \Gamma^\rba = \gamma_* \slashed{\widehat C} \bar\gamma^\rba$.

While flat $\g{O}(D,D)$ spinors are finite dimensional bispinors, flat orthosymplectic spinors must be infinite dimensional. In the flat basis, we treat $\Gamma^\ha$ and $\Gamma^\halpha$ differently; the former lead to a finite dimensional Fock space and can be described by matrices, as before, while the latter are infinite dimensional and so we will retain a Fock space structure for the spinorial indices. A flat orthosymplectic spinor is then written, similar to \eqref{E:superC.decompose.p}, as 
\begin{align}
\bra{\widehat{\slashed{\cC}}} = \sum_p \bra{\fvac} b^{\halpha_1} \cdots b^{\halpha_p} 
    \widehat{\slashed{\cC}}_{\halpha_p \cdots \halpha_1}
\end{align}
where $b^{\halpha} = (b^\alpha, b^\balpha)$ are bosonic raising operators, acting to the left on a spinor vacuum state $\bra{\fvac}$, so that they obey
\begin{align}
[b^\halpha, b_\hbeta] = \delta_\hbeta{}^\halpha~, \qquad
\bra{\fvac} b_\halpha = 0~.
\end{align}
The quantity $\widehat{\slashed{\cC}}_{\halpha_p \cdots \halpha_1}$ is a bispinor carrying additional symmetric spinor indices; we can write it as
\begin{align}\label{E:SuperBispinorMatrix}
\widehat{\slashed{\cC}}_{\halpha_p \cdots \halpha_1} = 
\begin{pmatrix}
\widehat\cC_{\halpha_p \cdots \halpha_1,}{}_\alpha{}^\balpha 
    & \widehat\cC_{\halpha_p \cdots \halpha_1,}{}_{\alpha \balpha} \\
\widehat\cC_{\halpha_p \cdots \halpha_1,}{}^{\alpha \balpha} 
    & \widehat\cC_{\halpha_p \cdots \halpha_1,}{}^\alpha{}_\balpha
\end{pmatrix}
\end{align}
with $\halpha_i$ including both barred and unbarred indices.
The action of $\Gamma^\cA$ on $\bra{\widehat{\slashed{\cC}}}$ is defined as
\begin{alignat}{2}
\bra{\widehat{\slashed{\cC}}} \cdot \Gamma^\ra &= \gamma^\ra \bra{\widehat{\slashed{\cC}}}~, &\qquad
\bra{\widehat{\slashed{\cC}}} \cdot \Gamma^\rba &= \gamma_* \bra{\widehat{\slashed{\cC}}} \bar\gamma^\rba~, \eol
\bra{\widehat{\slashed{\cC}}} \cdot \Gamma^\halpha &= \sqrt{2}\, \gamma_* \bra{\widehat{\slashed{\cC}}} \bar\gamma_* 
\,b^\halpha~, &\qquad
\bra{\widehat{\slashed{\cC}}} \cdot \Gamma_\halpha &= \sqrt{2}\, \gamma_* \bra{\widehat{\slashed{\cC}}} \bar\gamma_* \,b_\halpha~.
\label{E:flatGammaAction}
\end{alignat}
The various factors of $\gamma_*$ and $\bar\gamma_*$ are necessary to reproduce the flat Clifford algebra. The $\cdot{}$ action above is to the left on any bispinor-valued Fock space and not just $\bra{\widehat{\slashed{\cC}}}$ itself, so for example,
$\bra{\widehat{\slashed{\cC}}} \cdot \Gamma^\ra \Gamma^\rb =
\Big( \gamma^\ra \bra{\widehat{\slashed{\cC}}} \Big) \cdot \Gamma^\rb =
\gamma^\rb \gamma^\ra \bra{\widehat{\slashed{\cC}}} 
$,
leading to a reversal of ordering on the left-handed spinor sector.

We introduce the bispinorial operator $\cswV$ that converts to the spinor Fock space, so that
\begin{align}
\bra{\widehat{\slashed{\cC}}} := \Phi^{-1/2}\,\bra{\cC} \cswV~.
\end{align}
The action of the $\Gamma^\cM$ matrices on $\cswV$ is defined as
\begin{align}
\Gamma^\cM \cswV = \cswV \cdot \Gamma^\cA\, \cV_\cA{}^\cM
\end{align}
where $\Gamma^\cA$ acts to the left on $\cswV$ just as on $\bra{\widehat{\slashed{\cC}}}$ in \eqref{E:flatGammaAction}.

The specific dictionary between the curved and flat Fock vacuum states follows when taking $\cswV$ to be that associated with the identity supervielbein; for shorthand, we denote this as $\widehat{\slashed{1}}$. Then we have
\begin{align}
\bra{0} \widehat{\slashed{1}} = \mathbf 1 \otimes \bra{\fvac} = \frac{1}{2^{5/2}}
\begin{pmatrix}
\delta_\alpha{}^\balpha \bra{\fvac} & 0 \\
0 & \delta^\alpha{}_\balpha \bra{\fvac}
\end{pmatrix} ~.
\end{align}

\subsection{The Ramond-Ramond sector and its curvature constraints}
Now suppose that $\bra{\cC}$ describes the complex of Ramond-Ramond $p$-forms. In addition to diffeomorphisms \eqref{E:RamondC.Diffeo}, it transforms under abelian gauge transformations as
\begin{align}
\delta \bra{\cC} 
    = \bra{\lambda} \lDirac
    = \frac{1}{\sqrt 2} \bra{\lambda} \Gamma^\cM \stackrel{\leftarrow}{\pa_\cM}
\end{align}
so that the field strength
$\bra{\cF} = \bra{\cC} \lDirac$
is invariant. As is typical in superspace formulations, we impose constraints on the \emph{covariant} field strengths -- that is, constraints on the flat components of the
field strength tensor flattened with the supervielbein, e.g. $\cF_{A_1 \cdots A_p}$.
For the Ramond-Ramond sector, the analogous constraints must then be imposed on the flattened bispinor
\begin{align}
\bra{\widehat{\slashed{\cF}}} := \Phi^{-1/2} \bra{\cF} \cswV~.
\end{align}
A remarkable simplification occurs here. In type II supergravities, there are only three nonzero components of $\cF_{A_1 \cdots A_p}$: these are $\cF_{\halpha \hbeta\, a_1 \cdots a_{p-2}}$ (which is constant and given by $\gamma$-matrices), $\cF_{\halpha\, a_1 \cdots a_{p-1}}$ (which is given by the dilatino), and $\cF_{a_1 \cdots a_p}$ (which is the covariant field strength). We have already claimed that the covariant field strengths and the dilatino will be encoded in the supervielbein; since these will have non-tensorial transformations under superdiffeomorphisms, we cannot use them to build $\bra{\widehat{\slashed{\cF}}}$. This suggests that the only natural choice for $\bra{\widehat{\slashed{\cF}}}$ is a constant! Structurally, there is only one way to do this.
Because the fermionic oscillators $b^\halpha$ carry the same type of index as those on the bispinor, it is possible to write down one constant bispinor which involves no dynamical information aside from the structure of the flat fermionic Fock space:
\begin{align}\label{E:RRF.constant}
\bra{\widehat{\slashed{\cF}}} =
-4k
\begin{pmatrix}
0 & 0 \\
\bra{\fvac} b^\alpha b^{\balpha} & 0
\end{pmatrix}~.
\end{align}
We choose the normalization to involve the constant $k$, which is the same that appears in the torsion tensor constraint \eqref{E:BasicTorsionConstraint}; this will lead to standard normalization conventions for the Ramond-Ramond sector in type II supergravity. This coincides with the expansion \eqref{E:SuperBispinorMatrix} if we identify the sole non-vanishing component as
\begin{align}\label{E:RRF.constant.v2}
\widehat\cF_{\alpha_2 \balpha_1,}{}^{\alpha \balpha} = 
-4k\,\delta_{\alpha_2}{}^{\alpha} \,\delta_{\balpha_1}{}^{\balpha}~.
\end{align}

This is a rather simple expression, and we should perform a few sanity checks. First, the field strength must be closed. In covariant notation, this condition reads
\begin{align}
0 =  \frac{1}{\sqrt 2} \bra{\widehat{\slashed{\cF}}} \Big(
    \Gamma^\cA \stackrel{\leftarrow}{\nabla}_\cA
    + \frac{1}{12} \Gamma^{\cA \cB \cC} \cT_{\cC \cB \cA}
    + \frac{1}{2} \Gamma^{\cA} \cT_{\cA}
    \Big)~.
\end{align}
The only non-vanishing torsion is $\cT_{\hgamma \hbeta \ha}$, so only the second term contributes:
\begin{align}
 0 &= \gamma^\rc \bra{\widehat{\slashed{\cF}}} b^{\alpha} b^{\beta} (\gamma_{\rc})_{\beta \alpha}
    + \gamma_* \bra{\widehat{\slashed{\cF}}} \bar\gamma^{\rbc} b^{\balpha} b^{\bbeta} (\bar\gamma_{\rbc})_{\ol{\beta \alpha}}~.
\end{align}
The first term is proportional to
\begin{align}
\begin{pmatrix}
0 & (\gamma^\rc)_{\alpha \beta} \\
(\gamma^\rc)^{\alpha \beta} & 0 
\end{pmatrix} \times
\begin{pmatrix}
0 & 0 \\
\bra{\fvac} b^\beta b^{\balpha} & 0
\end{pmatrix} \times
b^\gamma b^\delta (\gamma_\rc)_{\gamma \delta}
=
\begin{pmatrix}
\bra{\fvac} (\gamma^\rc)_{\alpha\beta} b^\beta b^\gamma b^\delta (\gamma_\rc)_{\gamma \delta}b^{\balpha} 
     & 0 \\
0 & 0 \\
\end{pmatrix}
\end{align}
but this vanishes using the fundamental 10D gamma matrix identity $(\gamma^\rc)_{\alpha (\beta} (\gamma_\rc)_{\gamma \delta)} = 0$. A similar cancellation occurs for the second term. This identifies $\bra{\widehat{\slashed{\cF}}}$ as a preferred covariantly closed and constant orthosymplectic spinor. 

As a second check, we can verify that $\bra{\widehat{\slashed\cF}}$ is Lorentz covariant. This is more or less obvious when we interpret it as \eqref{E:RRF.constant.v2}, but it is useful to understand how this works in ket language. An infinitesimal left-handed Lorentz transformation acts as 
\begin{align}
\delta_\lambda \bra{\widehat{\slashed\cF}}
    &= \frac{1}{2} \bra{\widehat{\slashed\cF}} \Big(
        \frac{1}{2} \Gamma^{\ra \rb} 
        - \frac{1}{4} \Gamma_\beta \Gamma^\gamma (\gamma^{\ra\rb})_\gamma{}^\beta \Big) 
        \lambda_{\rb\ra} 
    = \frac{1}{4} \lambda_{\ra\rb} \Big(\gamma^{\ra\rb} \bra{\widehat{\slashed\cF}}
        + \bra{\widehat{\slashed\cF}} b_\beta b^\gamma (\gamma^{\ra\rb})_\gamma{}^\beta \Big)= 0~.
\end{align}
The term in parentheses is the embedding of the $\g{SO}(9,1)$ generator $M_{\ra\rb}$ into the orthosymplectic group: it consists of the piece that rotates vectors and the piece rotating spinors. Together these cancel out when the explicit form of $\bra{\widehat{\slashed{\cF}}}$ is used. The same occurs for the right-handed sector. In fact, $\bra{\widehat{\slashed\cF}}$ can be seen to be invariant under the full $\g{H}_L \times \g{H}_R$ group. We leave this as an instructive exercise.

To confirm this result, we will verify in due course that it correctly reproduces the supersymmetry transformation of component Ramond-Ramond bispinor given in \cite{Jeon:2012hp}. We will also show that it leads to the correct Ramond-Ramond polyform in conventional type II superspace.

\section{Component fields and SUSY transformations of type II DFT}
\label{S:DFTcomps}
\nocollect{Gates:1997kr}

The component structure of type II DFT was given by Jeon et al. in \cite{Jeon:2012hp}, where the action and supersymmetry transformations were laid out in detail. Our goal in this section is to recover their results for the supersymmetry transformations from superspace. We will not address the construction of the action for two reasons. First, as with any on-shell component theory, this can be tricky because on-shell supersymmetry necessarily implies equations of motion. Second, we have not yet discovered a generic schema for constructing invariant actions in $\g{OSp}(D,D|2s)$ superspace; this is in contrast to $\g{GL}(D|s)$ superspace where invariant actions are associated with closed super $D$-forms
\cite{DAuria:1982mkx,RheonomicBook,Gates:1997kr,Gates:1997ag}. Nevertheless, one could still derive the equations of motion from superspace; for the sake of brevity, we will not exhaustively analyze these here, as this would mostly mirror the type I analysis \cite{Butter:2021dtu}, and would necessarily lead to the results in \cite{Jeon:2012hp} (since those equations of motion are implied by closure of the algebra).

\subsection{Decomposing the supervielbein}
In order to derive the physical component fields, we must first arrange the supervielbein in a specific way. The generators of $\g{OSp}(10,10|64)$ can be decomposed with respect to
$\g{O}(10,10) \times \g{Sp}(64, \mathbb R)$ and assigned levels as
\begin{align}
\underbrace{T^{\hmu \hnu}}_{\text{level -2}}, \quad 
\underbrace{T^{\hmu \hn}}_{\text{level -1}}, \quad 
\underbrace{T^{\hm \hn}~, T_{\hmu}{}^{\hnu}}_{\text{level 0}}~, \quad 
\underbrace{T_{\hmu \hn}}_{\text{level +1}}~, \quad 
\underbrace{T_{\hmu \hnu}}_{\text{level +2}}~.
\end{align}
The level corresponds to the difference between the number of lowered and
raised fermionic indices. Then the DFT supervielbein $\cV_\cM{}^\cA$ can be arranged in factors of increasing levels, i.e.
\begin{align}
\cV = \cV_{-2} \cV_{-1} \cV_0 \cV_{+1} \cV_{+2}~.
\end{align}
One consequence is that the fields and generators assigned to $\cV_{+1}$ and $\cV_{+2}$
will be more naturally written with tangent space indices.
We enumerate the fields in Table \ref{T:OSpV.Pieces}.
\begin{table}[t]
\centering
\renewcommand{\arraystretch}{1.3}
\begin{tabular}{ccc}
\toprule
field & generator & level \\ \midrule
$\cS^{\halpha \hbeta}$ & $T_{\halpha \hbeta}$ & $+2$ \\
$\Psi_{\ha}{}^{\hbeta}$ & $T_\hbeta{}^{\ha}$ & $+1$ \\
$V_{\hm}{}^{\ha}$ & $T_{\hm}{}^{\hn}$ & $0$ \\
$\phi_{\hmu}{}^{\halpha}$ & $T_{\hmu}{}^\hnu$ & $0$ \\
$\Xi_{\hmu}{}^\hm$ & $T_\hm{}^{\mu}$ & $-1$ \\
$\cB_{\hmu \hnu}$ & $T^{\hmu \hnu}$ & $-2$ \\
\bottomrule
\end{tabular}
\captionsetup{width=0.6\textwidth}
\caption{Constituent fields of the supervielbein.
Positive level fields are written with Lorentz indices.}
\label{T:OSpV.Pieces}
\end{table}
This arrangement is rather different from the superspace decomposition given in \eqref{E:SuperVielbein.BES}. That parametrization (and its generalization in section \ref{S:TypeIISS})
is more useful to recover type II superspace after solving the section condition, while the one here is useful for component DFT analysis. We will spell this connection out more clearly in section \ref{S:RelationSuperToComps}.

Let's briefly explain the field content in Table \ref{T:OSpV.Pieces}. The component fields lie in the first three lines. The bosonic double vielbein is $V_{\hm}{}^\ha$. The fermionic fields $\Psi_\ha{}^\hbeta$ consist of the gravitini $\Psi_\ra{}^{\bbeta}$ and $\Psi_\rba{}^\beta$ and the dilatini $\rho_\halpha$, which we define as
\begin{align}
\rho_\alpha := \Psi_\ra{}^\beta (\gamma^\ra)_{\beta \alpha}~, \qquad
\rho_\balpha := \Psi_\rba{}^\bbeta (\bar\gamma^\rba)_{\ol{\beta \alpha}}~.
\end{align}
The spin-3/2 pieces of $\Psi_\ra{}^{\beta}$ and $\Psi_\rba{}^{\bbeta}$ are pure gauge artifacts, as are $\cS^{\alpha\beta}$ and $\cS^{\ol{\alpha\beta}}$. $\cS^{\alpha \bbeta}$ encodes the covariantized Ramond-Ramond field strengths. 

The remaining fields turn out to be gauge artifacts of a different type: the lowest components of their $\theta$ expansions can be removed by $\theta$-dependent diffeomorphisms.
These are related (at leading order) to corresponding fields in conventional superspace.
For example, $\phi_\hmu{}^\halpha$ corresponds to the supervielbein component
$E_\hmu{}^\halpha$. The two pieces of $\Xi_\hmu{}^\hm = (\Xi_\hmu{}^m, \Xi_\hmu{}_m)$
correspond respectively to $E_\hmu{}^a$ and $B_{\hmu n}$ in conventional superspace.
Finally, $\cB_{\hmu \hnu}$ corresponds to the fermionic legs of the super 2-form $B_{MN}$.

Let's work out the explicit expressions for the supervielbein.
The fields at nonzero levels are normalized so that they fill out a graded symmetric element $\cA_{\cM \cN}$ of $\g{OSp}(10,10|64)$,
\begin{equation}\label{E:AGen}
\cA_{\cM \cN} =
\begin{pmatrix}
0  & \Xi_{\hm \hnu} & \Psi_\ha{}^\hbeta \\
-\Xi_{\hn \hmu} & \cB_{\hmu \hnu} & 0 \\
-\Psi_\hb{}^\halpha & 0 & \cS^{\halpha \hbeta} \\
\end{pmatrix}~.
\end{equation}
This notation is somewhat sloppy, as we are using flat indices for positive elements and curved indices for negative ones, but we trust this will not be confusing. Exponentiating the above generators using $\cV = \exp (\cA_{\cM}{}^\cN)$ for each level gives
\begin{align}
\cV_{+2} &= 
\begin{pmatrix}
1 & 0 & 0 \\
0 & 1 & 0 \\
0 & \cS & 1
\end{pmatrix}~, \qquad
\cV_{-2} = 
\begin{pmatrix}
1 & 0 & 0 \\
0 & 1 & -\cB \\
0 & 0 & 1
\end{pmatrix}~, \qquad
\cB = \cB_{\hmu \hnu}~, \qquad
\cS = \cS^{\halpha \hbeta}~,
\eol
\cV_{+1} &=
\begin{pmatrix}
1 & \Psi & 0 \\
0 & 1 & 0 \\
-\Psi^T & -\tfrac{1}{2} \Psi^T\Psi & 1
\end{pmatrix}~, \qquad
\begin{gathered}
\Psi = \Psi_\ha{}^\hbeta~,\qquad
\Psi^T = (\Psi^T)^{\halpha \hb} =  \Psi^{\hb \halpha}~, \quad
\end{gathered}~, \eol
\cV_{-1} &= 
\begin{pmatrix}
1 &  0 & -\Xi \\
-\Xi^T & 1 & \tfrac{1}{2} \Xi^T \Xi \\
0 & 0 & 1
\end{pmatrix}~, \qquad
\begin{gathered}
\Xi = \Xi_{\hm \hnu}~, \qquad
\Xi^T = (\Xi^T)_\hmu{}^{\hn} =  \Xi^{\hn}{}_{\hmu}~,
\end{gathered}
\end{align}
At level 0, there are two distinct commuting pieces, which we denote
\begin{align}
\cV_0  &=
\begin{pmatrix}
V_\hm{}^\ha & 0 & 0 \\
0 & 1 & 0 \\
0 & 0 & 1
\end{pmatrix}~, \qquad
\cV_\phi =
\begin{pmatrix}
1 & 0 & 0 \\
0 & \phi_\hmu{}^\halpha & 0 \\
0 & 0 & \phi_\halpha{}^\hmu
\end{pmatrix}~.
\end{align}
$V_\hm{}^\ha$ is the component DFT vielbein and we presume $\phi_\hmu{}^\halpha$ to be
invertible with inverse $\phi_\halpha{}^\hmu$.

Then a generic orthosymplectic element can be decomposed as
\begin{align}\label{E:SuperVDecompose}
\cV = \underbrace{\cV_{-2} \cV_{-1} \cV_\phi}_{\cV_\Xi} \times \cV_0 \times 
\underbrace{\cV_{+1} \cV_{+2}}_{\cV_\Psi}
    = \cV_\Xi \times \cV_0 \times\cV_\Psi
\end{align}
We emphasize that $\cV_0$ contains the bosonic double vielbein, $\cV_\Psi$
involves the gravitini, dilatini, and Ramond-Ramond bispinor,
and $\cV_\Xi$ involves fields with no component analogues
since their $\theta=0$ parts can be eliminated. Note that $\cV_0$ denotes
a generic element of $\g{O}(D,D) \subset \g{OSp}(D,D|2s)$ while
$\cV_{-2} \cV_\phi \cV_{+2}$ is a (nearly) generic element of
$\g{Sp}(2s,\mathbb R) \subset \g{OSp}(D,D|2s)$.

The organization of elements ensures that all flat indices above transform straightforwardly under the double Lorentz group. Meanwhile, the higher $\lambda$ transformations are entirely soaked up in $\cV_\Psi = \cV_{+1} \cV_{+2}$. The fields there transform as
\begin{alignat}{2}
\label{E:lambda.Trafo.1}
\delta \Psi_\ra{}^\beta &= -\lambda_\ra{}^\beta~, &\qquad
\delta \Psi_\rba{}^\bbeta &= -\lambda_\rba{}^\bbeta~, \\
\label{E:lambda.Trafo.2}
\delta \cS^{\alpha \beta} &= -\lambda^{\alpha\beta} + \Psi^{\rc (\alpha} \lambda_\rc{}^{\beta)}~, &\quad
\delta \cS^{\ol{\alpha \beta}} &= -\lambda^{\ol{\alpha\beta}} + \Psi^{\rbc (\balpha} \lambda_\rbc{}^{\bbeta)}~, \\
\label{E:lambda.Trafo.3}
\delta \cS^{\alpha \bbeta} &= 
    \frac{1}{2} \Psi^{\rc \bbeta} \lambda_\rc{}^{\alpha}
    + \frac{1}{2} \Psi^{\rbc \alpha} \lambda_\rbc{}^{\bbeta}
~, 
\end{alignat}
with all other fields invariant.
Note that we can define a shifted version of $\cS^{\alpha\bbeta}$,
\begin{align}\label{E:zcS.Def}
\zcS^{\alpha\bbeta}  := \cS^{\alpha \bbeta}
    - \frac{1}{2} \Psi^{\rc \alpha} \Psi_\rc{}^{\bbeta} 
    + \frac{1}{2} \Psi^{\rbc \alpha} \Psi_{\rbc}{}^{\bbeta}
\end{align}
to eliminate its higher $\lambda$ transformation entirely. This will be the covariantized Ramond-Ramond bispinor that appears in the gravitini transformation.

The transformations \eqref{E:lambda.Trafo.1} and \eqref{E:lambda.Trafo.2} admit the gauge (which we denote with asterisks)
\begin{align}\label{E:Psi.gauge}
\Psi_\ra{}^\alpha \stackrel{*}{=} \frac{1}{10} (\gamma_\ra)^{\alpha \beta} \rho_\beta~, \qquad
\Psi_\rba{}^\balpha \stackrel{*}{=} \frac{1}{10} (\gamma_\rba)^{\ol{\alpha\beta}} \rho_\bbeta~, \qquad
\cS^{\alpha\beta} \stackrel{*}{=} 0~, \qquad
\cS^{\ol{\alpha\beta}} \stackrel{*}{=} 0~.
\end{align}
We will actually avoid imposing this gauge, so that we can track how gauge invariance emerges.

In the decomposition \eqref{E:SuperVDecompose}, let us split off the $\cV_\Psi$ part and denote $\zcV = \cV_\Xi \cV_0$. Its inverse is explicitly
\begin{equation}
\begin{aligned}
(\zcV^{-1})_\cA{}^\cM &=
\begin{pmatrix}
V_\ha{}^\hm & 0 & \bullet \\
\bullet & \phi_\halpha{}^\hmu & \bullet \\
0 & 0 & \phi_\hmu{}^\halpha
\end{pmatrix}
\end{aligned}
~, \qquad
\begin{aligned}
\zcV_\halpha{}^{\hn} &= -\phi_{\halpha}{}^\hnu \Xi^{\hn}{}_\hnu~, \qquad
\zcV_{\ha}{}_\hnu = V_{\ha}{}^{\hm} \Xi_{\hm \nu}~, \\
\zcV_{\halpha}{}_\hnu &= \phi_\halpha{}^\hmu (\cB_{\hmu \hnu} 
    + \tfrac{1}{2} \Xi_{\hm \hmu} \Xi^{\hm}{}_\hnu)~.
\end{aligned}
\end{equation}
with the bulleted entries given explicitly above.
As we will solve the fermionic part of the section condition 
by taking $\tilde\pa^\hmu = 0$, it follows that
\begin{align}\label{E:zD.Def}
\zD_\cA := 
\zcV_\cA{}^\cM \pa_\cM &=
\begin{pmatrix}
V_{\ha}{}^{\hm} \pa_{\hm} \\ 
\zcV_\halpha{}^{\hm} \pa_{\hm} + \phi_\halpha{}^\hmu \pa_\hmu \\
0
\end{pmatrix}
\equiv
\begin{pmatrix}
\zD_{\ha} \\ 
\zD_{\halpha} \\
0
\end{pmatrix}~.
\end{align}
This defines $\zD_{\ha}$ and $\zD_\halpha$.
Note that we can build $D_\cA = \cV_\cA{}^\cM \pa_\cM = (\cV_\Psi^{-1})_\cA{}^\cB \zD_\cB$ as
\begin{align}\label{E:DtoComp}
D_\cA  = (\cV_\Psi^{-1})_\cA{}^\cB \zD_\cB =
\begin{pmatrix}
\zD_\ha - \Psi_\ha{}^\halpha \zD_\halpha \\
\zD_\halpha \\
\Psi^{\hb \halpha} \zD_{\hb}
- \Big(\cS^{\halpha \hbeta} + \frac{1}{2} \Psi^{\hc \halpha} \Psi_{\hc}{}^\hbeta \Big) \zD_\hbeta
\end{pmatrix}~.
\end{align}
It is quite convenient that $D_\halpha = \zD_\halpha$, as this lets us write
\begin{align}\label{E:CompDs}
D_\ha = \zD_\ha - \Psi_\ha{}^\hbeta D_\hbeta~, \qquad
D^\halpha = 
\Psi^{\hb \halpha} D_{\hb}
- \Big(\cS^{\halpha \hbeta} - \tfrac{1}{2} \Psi^{\ha \halpha} \Psi_{\ha}{}^\hbeta \Big) D_\hbeta~.
\end{align}
The bosonic derivative $D_\ha$ is now directly analogous to the conventional superspace
derivative, where it is given by the component flat derivative $\zD_\ha$ modified by
a gravitino (and dilatino) piece. The additional double fermionic derivative is $D^\halpha$: its explicit form  will be useful in subsequent computations.
The key property of \eqref{E:CompDs} is that the pieces of $\cV_\Xi$ appear only implicitly
via $D_\halpha$.
\emph{These simple expressions crucially follow only upon imposing $\pa^\hmu=0$, which we
will presume henceforth in this section}.

\subsection{Generalized diffeomorphisms}
There are three types of diffeomorphisms to consider: bosonic generalized diffeomorphisms of component double field theory and the fermionic and dual fermionic diffeomorphisms. We treat the fermionic diffeomorphisms as \emph{covariantized diffeomorphisms} and keep the bosonic and dual fermionic diffeomorphisms uncovariantized. That means we parametrize the latter as standard diffeomorphisms with
\begin{align}\label{E:StandardDiffeoParam}
\xi^\cM = (\xi^\hm, 0, -\tilde\xi_\hmu)
\end{align}
and encode the former as
\begin{align}\label{E:CovDiffeoParam}
\xi^\cA = (0, \eps^\halpha, 0)~.
\end{align}
The parameter $\xi^\hm$ describes double diffeomorphisms in the component theory, $\eps^\halpha$ describes supersymmetry, and $\tilde \xi_\hmu$ is a residual dual fermionic symmetry that plays no role for the physical fields.

Let's first address the transformations \eqref{E:StandardDiffeoParam}, as these are quite simple to understand. Denoting the diffeomorphism as
\begin{align}
\delta\cV_{\cM}{}^\cA = \xi^\cN \pa_\cN \cV_\cM{}^\cA
    + \cK_\cM{}^\cN \cV_\cN{}^\cA~, \qquad
\cK_\cM{}^\cN = \pa_\cM \xi^\cN - \pa^\cN \xi_\cM (-)^{nm}~,
\end{align}
we observe that only non-positive levels contribute to $\cK$ on account of $\pa^\hmu=0$, and it follows that
\begin{gather}
\delta \cV_{\Xi} = \xi^\hm \pa_\hm \cV_\Xi + (\cK_{-2} + \cK_{-1}) \cV_{\Xi}  + [\cK_0, \cV_{\Xi}]~, \quad
\delta \cV_0 = \xi^\hm \pa_\hm \cV_0 + \cK_0 \cV_0~, \quad
\delta \cV_{\Psi} = \xi^\hm \pa_\hm \cV_\Psi~.
\end{gather}
It is useful here that the level zero element $\cK_0$ only involves
$\pa_\hm \xi^\hn$ and not $\pa_\hmu \xi^\hnu$ or $\pa^\hmu \xi_\hnu$, as this
simplifies the transformation of $\cV_\Xi$.
Because the $\cK_0$ piece is just the bosonic $\g{O}(10,10)$ element, 
$\cV_0$ transforms as a DFT vielbein
should, while the gravitino, dilatino, and RR bispinor are scalar fields.
The negative level transformations are
completely soaked up by the fields in $\cV_\Xi$, which transform 
as
\begin{align}
\delta \Xi_{\hm \hnu} &= 
    \xi^\hn \pa_\hn \Xi_{\hm \hnu}
    + \pa_{\hm} \tilde\xi_\hnu -\pa_\nu \xi_{\hm}
    + (\pa_\hm \xi^\hn - \pa^\hn \xi_\hm) \Xi_{\hn \hnu}~, \eol
\delta \cB_{\hmu \hnu} &= 
    \xi^\hn \pa_\hn \cB_{\hmu \hnu}
    + \pa_\hmu \tilde\xi_\hnu + \pa_\hnu \tilde\xi_\hmu 
    + \Xi^{\hm}{}_{(\hmu} (\pa_\hm \tilde \xi_{\hnu)} - \pa_{\hnu)} \xi_\hm)~.
\end{align}
As in conventional superspace, these fields can be set to zero at lowest level in $\theta$.

Because the fields in $\cV_\Psi$ are inert under $\tilde\xi_\hmu$ and transform as scalars
under $\xi^\hm$, any gauge-fixing there is undisturbed by diffeomorphisms. For example,
if we have set $\cS^{\alpha \beta}$ to zero, it remains so.
This means component diffeomorphisms match superspace ones, with no compensating
transformations needed to preserve the gauges \eqref{E:Psi.gauge}. This will not be the case for supersymmetry transformations.

\subsection{The physical and composite fields of component DFT}
We have identified the vielbein, gravitini, and dilatini as components of the supervielbein. We still need to identify the dilaton and the Ramond-Ramond sector and verify that they transform sensibly under standard diffeomorphisms \eqref{E:StandardDiffeoParam}.

\paragraph{The dilaton.} Recall the superdilaton $\Phi$ transforms under diffeomorphisms as
\begin{align}
\delta \log \Phi = \xi^\hm \pa_\hm \log\Phi + \pa_\hm \xi^\hm - \pa_\hmu \xi^\hmu + \tilde\pa^\hmu \tilde \xi_\hmu
\end{align}
for a completely generic $\xi^\cM$.
At the component level, the last term can be dropped because we take $\tilde\pa^\hmu=0$, but 
the second term remains problematic as it will obstruct the construction of a
sensible supersymmetry transformation. This suggests that we define the component dilaton as
\begin{align}\label{E:CompDilaton}
e^{-2d} := \Phi \times \det \phi_\hmu{}^\halpha \quad \implies \quad
e^{2d} \delta e^{-2d}= \pa_\hm \xi^\hm + 2 \,\tilde\pa^\hmu \tilde\xi_\hmu~.
\end{align}
Now the second term drops out when $\tilde\pa^\hmu=0$
and the component dilaton transforms as a scalar field.\footnote{Another
way of arriving at this same conclusion is to recall that
the component DFT dilaton is related to the supergravity
dilaton by a factor of $e = \det e_\rmm{}^\ra$, that is,
$e^{-2d} = e\, e^{-2 \varphi}$.
The superdilaton in Siegel's superspace DFT is similarly related to
the conventional (non-density) superspace dilaton by a factor of $E = \sdet E_\tM{}^\tA$, i.e.
$\Phi = E\, e^{-2 \varphi}$. In conventional superspace, the component
and superspace dilatons coincide (hence both are $\varphi$ above).
This implies that $e^{-2d} = \Phi \times e / E = \Phi \times \det \phi_\mu{}^\alpha$.}

\paragraph{The Ramond-Ramond sector.} For the Ramond-Ramond sector, we proceed analogously to how superspace $p$-forms reduce to component $p$-forms. Recall for these, we have the notion of a double bar projection, taking both $\theta^\mu=0$ and $\rd \theta^\mu = 0$. Thus we have
\begin{align}
\cC = \sum_p \frac{1}{p!} \rd z^{M_1} \cdots \rd z^{M_p} \cC_{M_p \cdots M_1}(x,\theta) \quad \implies \quad
C = \cC \dbar = \sum_p \frac{1}{p!} \rd x^{m_1} \cdots \rd x^{m_p} C_{m_p \cdots m_1}(x)~,
\end{align}
where $C_{m_p \cdots m_1}(x) = \cC_{m_p \cdots m_1}(x,0)$. For the case of an orthosymplectic spinor, the analogous operation is to project to $\psiGamma^\hmu = 0$. That is, we define
\begin{align}
\bra{C} := \bra{\cC} \Dbar = \bra{\cC} \Big\vert_{\theta^\hmu = \psiGamma^\hmu=0}
\end{align}
The spinor $\bra{C(x)}$ transforms precisely as a component $\g{O}(D,D)$ spinor
when $\xi^\cM$ is given by \eqref{E:StandardDiffeoParam} and we set $\tilde\pa^\hmu=0$:
\begin{align}
\delta \bra{\cC} &= \xi^\hn \pa_\hn \bra{\cC}
    + \frac{1}{2} \bra{\cC} \Gamma^{\hm} \Gamma^\cN \pa_\cN \xi_\hm 
    + \frac{1}{2} \bra{\cC} \Gamma^{\hmu} \Gamma^\cN \pa_\cN \tilde\xi_\hmu
    \quad \stackrel{\rm proj.}{\implies} \quad \eol
\delta \bra{C} &= \xi^\hn \pa_\hn \bra{C}
    + \frac{1}{2} \bra{C} \Gamma^{\hm} \Gamma^\hn \pa_\hn \xi_\hm~.
\end{align}

\paragraph{Spin connection.}
In analogy to \eqref{E:zD.Def}, we define the component spin connection
\begin{align}\label{E:omega.Def}
\omega_{\ha \, \hb \hc} := 
    (\cV_\Psi)_\ha{}^\cD \Omega_{\cD\, \hb \hc}
    = \Omega_{\ha \, \hb \hc} + \Psi_\ha{}^\halpha \Omega_{\halpha\, \hb \hc}
\end{align}
The motivation for this is two-fold: first, it lets us define the component covariant derivative as
\begin{align}
\zcD_\ha := (\cV_\Psi)_\ha{}^\cB \cD_\cB = \cD_\ha + \Psi_\ha{}^\hbeta \cD_\hbeta
    = \zD_\ha - \frac{1}{2} \omega_{\ha}{}^{\hb \hc} M_{\hc \hb}~.
\end{align}
Second, it gives a very simple prescription for translating torsion constraints in superspace to torsion constraints in components. As shown in \cite{Butter:2021dtu}, the above definition of the spin connection implies that the component torsions are given by
\begin{align}
T_{\hc \hb \ha } &= (\cV_\Psi)_\ha{}^\cA (\cV_\Psi)_\hb{}^\cB (\cV_\Psi)_\hc{}^\cC \cT_{\cC \cB \cA}~.
\end{align}
A similar calculation shows that the component dilaton torsion is
\begin{align}
T_\ha = \cT_\ha + \Psi_\ha{}^\halpha (\cT_\halpha + \cT_{\halpha \hbeta}{}^\hbeta)
    + \Psi^{\hb \hbeta} \cT_{\hbeta \hb \ha}
    - \Psi_\ha{}^{\halpha} \Psi^{\hb \hbeta} \cT_{\hbeta \halpha \hb}
    + \cS^{\hbeta \hgamma} \Big(\cT_{\hgamma \hbeta \ha} + \Psi_\ha{}^\alpha \cT_{\hgamma\hbeta\ha}\Big)~.
\end{align}
Using the superspace constraints on the torsion tensors, we conclude that
\begin{subequations}
\begin{align}
T_{\ra \rb \rc} &= 3 k \,\Psi_{[\ra}{}^\alpha (\gamma_\rb)_{\alpha\beta}
    \Psi_{\rc]}{}^\beta~, \\
T_{\rba \rb \rc} &= 2 k \,\Psi_\rba{}^\alpha 
    (\gamma_{[\rb})_{\alpha\beta} \Psi_{\rc]}{}^\beta
    - k\,\Psi_\rb{}^\balpha (\bar\gamma_\rba)_{\ol{\alpha\beta}} \Psi_\rc{}^{\bbeta} ~, \\
T_\ra &= -k \,\Psi_\ra{}^\alpha \rho_\alpha
    - k \, \Psi_\ra{}^\balpha \rho_\balpha
    + k (\gamma_{\ra})_{\beta \gamma} \cS^{\beta \gamma}~,
\end{align}
\end{subequations}
and similarly for their barred versions. We could further impose the gauge conditions
\eqref{E:Psi.gauge}, but it will be enlightening to avoid that here. The above conditions imply that $\omega_{\ra \rb \rc}$ and $\omega_{\rba \rb \rc}$ receive additional contributions $\Delta\omega$ relative to the expressions $\omega(V)$ that follow from the bosonic theory:
\begin{subequations}
\label{E:Deltaomega}
\begin{align}
\label{E:Deltaomega.a}
\Delta \omega_{[\ra \rb \rc]} &= k \, \bar\Psi_{[\ra} \gamma_\rb \Psi_{\rc]}~, \\
\label{E:Deltaomega.b}
\Delta \omega^\rb{}_{\rb \ra} &= -k \,\bar\Psi_\ra \rho
    - k \, \bar\Psi'_\ra \rho'
    + k (\gamma_{\ra})_{\beta \gamma} \cS^{\beta \gamma}~, \\
\label{E:Deltaomega.c}
\Delta \omega_{\rba \rb \rc} &= 2k \,\bar\Psi_{\rba} \gamma_{[\rb} \Psi_{\rc]}
    - k \bar\Psi'_\rb \bar\gamma_\rba \Psi'_{\rc}~.
\end{align}
\end{subequations}
This means that this spin connection is \emph{not} invariant under the higher $\lambda$ transformations. This had to be the case from its definition \eqref{E:omega.Def} and the transformations \eqref{E:Omega.lambda.a} of $\Omega_{\cM \rb \rc}$, which imply that
\begin{align}
\delta \omega_{\ra \rb \rc} 
    &= -2k \,\bar\Psi_{\ra} \gamma_{[\rb} \lambda_{\rc]}
    - \frac{2k}{9} \,\eta_{\ra [\rb} (\gamma_{\rc]} )_{\alpha \beta} \lambda^{\alpha \beta}
    + \Lambda_{\ra|\rb\rc}~,
\end{align}
consistent with the contributions $\Delta\omega$.

\paragraph{The covariant Ramond-Ramond field strength.}
To uncover the relation between the component Ramond-Ramond bispinor field strength $\slashed{\widehat F}$ and the bispinor component $\zcS^{\alpha\bbeta}$ of the supervielbein, we follow again the conventional supergravity dictionary by projecting to $\psiGamma^\hmu = 0$. That is, we identify
\begin{align}
\bra{F} := \bra{\cF} \Dbar = \bra{\widehat{\slashed{\cF}}} \cswV{}^{-1} \Phi^{1/2} \,\Dbar~.
\end{align}
Now we decompose the spinorial vielbein operator, following \eqref{E:SuperVDecompose}, as
\begin{align}\label{E:SuperVDecompose.Spinorial}
\cswV = \cswV{}_\Xi \times \cswV{}_0 \times \cswV{}_\Psi
\end{align}
The projection to $\psiGamma^\hmu = 0$ effectively dispenses with the $\cV_\Xi$ factor except for the piece that generates $\phi_\hmu{}^\halpha = \exp(a_\hmu{}^\hnu) \delta_\hnu{}^\halpha$. This factor leads to
\begin{align}
\cswV_\phi^{-1} \vert_{\psiGamma^\hmu=0}
    =  \exp \Big(
        \tfrac{1}{2} (\psiGamma_\hmu \psiGamma^\hnu + \psiGamma^\hnu \psiGamma_\hmu) a_\hnu{}^\hmu
    \Big) \vert_{\psiGamma^\hmu=0}
    = \exp \Big(
        \tfrac{1}{2} a_\hmu{}^\hmu
    \Big) 
    = (\det \phi_\hmu{}^\halpha)^{1/2}~.
\end{align}
We identify the flat bispinor
\begin{align}\label{E:FlatBispinorF}
\slashed{\widehat F} = e^{-d} \braket{F}{\slashed{V}} = \bra{\widehat{\slashed{\cF}}} \cswV_\Psi^{-1} \Big\vert_{b^\halpha = 0}
\end{align}
Expanding this out using
$\cswV{}_\Psi^{-1} = \exp\left( \tfrac{1}{2} \Gamma_\hbeta \Gamma^\ha \Psi_\ha{}^\hbeta +\tfrac{1}{4} \Gamma_{\halpha} \Gamma_\hbeta \cS^{\hbeta \halpha} \right)
$,
we find
\begin{align}
\slashed{\widehat F} &= 
-4k
\begin{pmatrix}
0 &
    - \tfrac{1}{2} (\gamma^\rb \Psi_\rba)_\alpha (\Psi_{\rb} \bar\gamma^\rba)_{\balpha}
    + \tfrac{1}{2} \rho_\alpha \bar \rho_\balpha \\[1ex]
\cS^{\alpha \balpha} 
    + \tfrac{1}{2} (\gamma^{\ra\rb} \Psi_\rb)^\alpha \Psi_\ra{}^\balpha 
    - \tfrac{1}{2} \Psi_\rba{}^\alpha (\Psi_\rbb \bar\gamma^{\ol{\rb \ra}})^{\balpha}
    & 
    0
\end{pmatrix} \eol[2ex]
&= 
-4k\Big(
    \slashed{\cS}
    + \tfrac{1}{2} \gamma^{\ra\rb} \Psi_\rb \times \bar\Psi'_\ra
    - \tfrac{1}{2} \Psi_\rba \times \bar\Psi'_\rbb \bar\gamma^{\ol{\rb \ra}}
    - \tfrac{1}{2} \gamma^\rb \Psi_\rba \times \bar \Psi'_{\rb} \bar\gamma^\rba
    + \tfrac{1}{2} \rho \times \bar\rho'
    \Big)
\end{align}
where we have written the result first in 16-component Weyl notation and then in 32-component Dirac notation. Now employing the redefinition \eqref{E:zcS.Def} of the Ramond-Ramond bispinor $\cS$ to the $\lambda$-invariant $\zcS$, we find
\begin{align}\label{E:Fslash.DFT}
\slashed{\widehat F} &= 
-4k
\begin{pmatrix}
0 &
    - \tfrac{1}{2} (\gamma^\rb \Psi_\rba)_\alpha (\Psi_{\rb} \bar\gamma^\rba)_{\balpha}
    + \tfrac{1}{2} \rho_\alpha \bar \rho_\balpha \\[1ex]
\zcS^{\alpha \balpha} 
    + \tfrac{1}{2} (\gamma^{\ra} \rho)^\alpha \Psi_\ra{}^\balpha 
    - \tfrac{1}{2} \Psi_\rba{}^\alpha (\rho \bar\gamma^\rba)^{\balpha}
    & 
    0
\end{pmatrix} \eol[2ex]
&=
-4k\Big(
    \mathring{\slashed{\cS}}
    + \tfrac{1}{2} \gamma^{\ra} \rho \times \bar\Psi'_\ra
    - \tfrac{1}{2} \Psi_\rba \times \bar \rho' \bar\gamma^{\rba}
    - \tfrac{1}{2} \gamma^\rb \Psi_\rba \times \bar \Psi'_{\rb} \bar\gamma^\rba
    + \tfrac{1}{2} \rho \times \bar\rho'
    \Big)~.
\end{align}
This expression is clearly invariant under the higher $\lambda$ transformations.
We can use it to identify
\begin{align}
\zcS^{\alpha \balpha}
    = -\frac{1}{4k} \widehat F^{\alpha \balpha}
    - \frac{1}{2} (\gamma^\ra \rho)^\alpha \Psi_\ra{}^{\balpha}
    + \frac{1}{2} \Psi_\rba{}^{\alpha}\, (\bar\gamma^\rba\rho)^{\balpha}
\end{align}

We can identify two constraints on $\slashed{\widehat F}$ from its Weyl decomposition. The first is
$\gamma_* \slashed{\widehat F} \bar\gamma_* = - \slashed{\widehat F}$, which constrains the diagonal to vanish.
The second constrains the upper right block:
\begin{align}
0 &= 
(1+\gamma_*) \Big(
    \slashed{\widehat F} 
    - \tfrac{4k}{2} \gamma^\rb \Psi_\rba \times \bar \Psi'_{\rb} \bar\gamma^\rba
    + \tfrac{4k}{2} \rho \times \bar \rho'
    \Big)
(1-\bar\gamma_*) ~.
\end{align}

\subsection{Supersymmetry transformations}
\label{S:SUSYtrafo}
Now we can compute the supersymmetry transformations arising from a covariant diffeomorphism with parameter \eqref{E:CovDiffeoParam}. Including a compensating tangent space transformation $\lambda_\cA{}^\cB$ (which will be necessary in this case),
the supervielbein transforms as $\delta \cV = \cV \,(\cK - \lambda)$
where
\begin{align}
\cK_{\cA}{}^{\cB} = \nabla_\cA \xi^\cB - \nabla^\cB \xi_\cA (-1)^{ab}
    + \xi^\cC \cT_{\cC \cA}{}^{\cB}~.
\end{align}
As we are restricting our choice of $\xi^\cA$,
we should be careful about expressions like 
$\nabla_\cA \xi^\cB = D_{\cA} \xi^{\cB} - \Omega_{\cA}{}^{\cB \cC} \xi_\cC$,
which might lead to an unexpected contribution when the $\Omega$ connection
is not purely Lorentz. Luckily, for $\xi^\cA = (0, \eps^\halpha, 0)$, the only
contribution is the Lorentz piece 
$\Omega_{\cA}{}^{\beta}{}_\gamma \eps^\gamma
    = \frac{1}{4} (\gamma^{\rb\rc})_\gamma{}^\beta \Omega_{\cA \,\rb\rc}\, \eps^\gamma$.
To emphasize this, we replace $\nabla$ with $\cD$ where $\cD$ carries 
only the double Lorentz connection.
For the torsion term, only $\cT_{\hgamma \hbeta \ha}$ is non-vanishing. 
Thus the only nonzero elements of $\cK_\cA{}^\cB$ are 
\begin{alignat}{5}
\cK^{\halpha \hbeta} &= 2\, \cD^{(\halpha} \eps^{\hbeta)}~, &\quad
\cK_\ha{}^\hbeta &= \cD_\ha \eps^\hbeta~, &\quad
\cK_\halpha{}^\hbeta &= \cD_\halpha \eps^\hbeta~, &\quad
\cK_{\halpha}{}^\hb &= k (\gamma^{\hb})_{\halpha \hgamma} \eps^\hgamma~, \eol
&& 
\cK^\halpha{}^\hb &= -\cD^\hb \eps^\halpha~, &\quad
\cK^\halpha{}_\hbeta &= -\cD_\hbeta \eps^\halpha~, &\quad
\cK_{\ha \hbeta} &= k (\gamma_{\ha})_{\hbeta \hgamma} \eps^\hgamma~,
\end{alignat}
corresponding to levels $+2$, $+1$, $0$, and $-1$. The possible
compensating $\lambda$ transformations lie at levels $+2$, $+1$ and $0$.
We parametrize an arbitrary variation as $\cJ = \cV^{-1} \delta \cV$, so that
\begin{align}
\delta \cV = \cV \times \cJ = \cV_{\Xi} \cV_0 \times \zJ \times \cV_\Psi \quad \implies \quad
\zJ := \cV_\Psi \cJ \cV_\Psi^{-1}~.
\end{align}
For the case of a supersymmetry transformation, $\cJ = \cK - \lambda$, and we can
read off $\zJ$ level-by-level, using the fact that 
$\cV_\Psi \equiv \exp \Psi \exp \cS = \exp (\Psi + \cS)$
for fields $\Psi$ at level $+1$ and $\cS$ at level $+2$:
\begin{align}
\zJ_{+2} &= 
    \cK_{+2} - \lambda_{+2} 
    + [\Psi, (\cK_{+1} - \lambda_{+1})] 
    + \tfrac{1}{2!} [\Psi, [\Psi, (\cK_{0} - \lambda_{0})]] 
    + \tfrac{1}{3!} [\Psi, [\Psi, [\Psi, \cK_{-1}]]] 
    \eol & \quad
    + [\cS, (\cK_{0} - \lambda_{0})] ~, \eol
\zJ_{+1} &= \cK_{+1} - \lambda_{+1}
    + [\Psi, (\cK_0-\lambda_0)]
    + \tfrac{1}{2} [\Psi, [\Psi, \cK_{-1}]] 
    + [\cS, \cK_{-1}]
    ~, \eol
\zJ_0 &= \cK_0 - \lambda_0 + [\Psi, \cK_{-1}] ~, \eol
\zJ_{-1} &= \cK_{-1}, \eol
\zJ_{-2} &= 0~.
\end{align}
From the explicit expressions for $\cV_\ell$, we find for the non-negative levels,
\begin{alignat}{2}
(\zJ_2)^{\halpha \hbeta} &= \delta \cS^{\halpha \hbeta} 
    -\Psi^\ha{}^{(\halpha} \delta \Psi_\ha{}^{\hbeta)}~, &\qquad
(\zJ_1)_\ha{}^\hbeta &= \delta \Psi_\ha{}^{\hbeta} ~, \eol
(\zJ_0)_\ha{}^\hb &= V_\ha{}^\hm \delta V_\hm{}^\hb \equiv J_\ha{}^\hb~, &\qquad
(\zJ_0)_\halpha{}^\hbeta &= \phi_\halpha{}^\hmu \delta \phi_\hmu{}^\hbeta~.
\end{alignat}

These are rather complicated expressions, but only some of them are relevant. For example,
the full expressions for $(\zJ_2)^{\alpha \beta}$ and $(\zJ_2)^{\ol{\alpha \beta}}$
tell us about $\delta \cS^{\halpha \hbeta}$ but $\cS^{\alpha\beta}$ and $\cS^{\ol{\alpha\beta}}$ are pure gauge degrees of freedom, so are not really relevant; in effect, these transformations would just identify what $\lambda^{\alpha\beta}$ and $\lambda^{\ol{\alpha\beta}}$ would need to be in order to maintain a specific gauge choice. And while $\cS^{\alpha\bbeta}$ contains the Ramond-Ramond field strength, we will actually derive the transformation of the potential directly. The transformation of $\phi_\mu{}^\alpha$ isn't really relevant either; it involves a leading term $\pa_\hmu \eps^\halpha$ implying that at lowest order in $\theta$ one is free to fix $\phi_\hmu{}^\halpha = \delta_\hmu{}^\halpha$.

That leaves the transformations of the bosonic vielbein and the gravitini and dilatini. We discuss these below.

\paragraph{DFT vielbein.}
For the bosonic DFT vielbein, we find (in Dirac notation)
\begin{align}\label{E:SUSY.vielbein.J}
J_{\ra \rbb} = k\, (\bar\eps \gamma_\ra \Psi_\rbb) 
    - k\, (\bar\eps' \bar\gamma_\rbb \Psi'_\ra) ~, \qquad
J_{\ra \rb}
    = 2 k\,(\bar\eps \gamma_{\ra} \Psi_\rb) - \lambda_{\ra \rb}~, \qquad
J_{\overline{\ra\rb}} = 2 k\,(\bar\eps' \gamma_{[\rba} \Psi'_{\rbb]}) - \lambda_{\overline{\ra\rb}}~.
\end{align}
The expression for $J_{\ra \rbb}$ is exactly as expected from \cite{Jeon:2012hp}.
The nonzero expressions for $J_{\ra\rb}$ and $J_{\ol{\ra\rb}}$ indicate that in order to make 
contact with supersymmetric type II DFT \cite{Jeon:2012hp} (where they are taken to vanish) 
one should choose $\lambda$ appropriately. This is rather natural to do since these expressions involve $\Psi_\rb{}^\alpha$ and $\Psi_\rbb{}^\balpha$, which transform under the higher $\lambda$-transformations. Imposing $\lambda_{\ra\rb}$ and $\lambda_{\ol{\ra\rb}}$ to kill these, we find
\begin{subequations}
\label{E:SUSY.vielbein}
\begin{align}
\delta V_\hm{}^\ra &= 
    k \, V_\hm{}^\rbb \Big(
        \eps^\balpha (\bar\gamma_\rbb)_{\ol{\alpha \beta}} \Psi^\ra{}^{\bbeta}
        - \eps^\alpha (\gamma^\ra)_{\alpha \beta} \Psi_\rbb{}^{\beta} \Big)~, \\
\delta V_\hm{}^\rba &= 
    k \,V_\hm{}^\rb \, \Big(\eps^\alpha (\gamma_\rb)_{\alpha \beta} \Psi^\rba{}^{\beta}
    - \eps^\balpha (\bar\gamma^\rba)_{\ol{\alpha \beta}} \Psi_\rb{}^{\bbeta} \Big)~.
\end{align}
\end{subequations}

\paragraph{Gravitini and dilatini.}
The gravitini and dilatini transformations are a good bit more complicated. 
The general expression for both is
\begin{align}\label{E:MasterPsiSUSY}
\delta \Psi_\ha{}^\hbeta
    &= \zcD_\ha \eps^\hbeta
    - k \, \eps^\hdelta (\gamma_\ha)_{\hdelta \hgamma} \,\cS^{\hgamma \hbeta}
    + k \, (\bar\eps \gamma^\hb \Psi_\ha)\, \Psi_\hb{}^\hbeta
    - \frac{1}{2} k (\bar\eps \gamma_\ha \Psi^\hb)\, \Psi_\hb{}^\hbeta
    - \lambda_\ha{}^\hbeta 
    + \lambda_\ha{}^\hb \Psi_\hb{}^\hbeta
    - \Psi_\ha{}^\hgamma \lambda_\hgamma{}^\halpha
~.
\end{align}
For the gravitini, we find
\begin{align}
\delta \Psi_\rba{}^\beta
    &= \zcD_\rba \eps^\beta 
    - k (\bar\eps' \bar\gamma_\rba)_\bgamma \,\cS^{\bgamma \beta}
    + k \, (\bar\eps \gamma^\rb \Psi_\rba)\, \Psi_\rb{}^\beta
    + k \, (\bar\eps' \bar\gamma^\rbb \Psi'_\rba)\, \Psi_\rbb{}^\beta
    \eol & \quad
    - \tfrac{1}{2} k \,(\bar\eps' \bar\gamma_\rba \bar\Psi'_\rb)\, \Psi^\rb{}^\beta
    - \tfrac{1}{2} k \,(\bar\eps' \bar\gamma_\rba \Psi'_\rbb)\, \Psi^\rbb{}^\beta
    + \lambda_\rba{}^\rbb \Psi_{\rbb}{}^\beta
    - \frac{1}{4} \Psi_\rba{}^\gamma (\gamma^{\rb\rc})_\gamma{}^\alpha\, \lambda_{\rb\rc}
    ~.
\end{align}
If we apply the redefinition \eqref{E:zcS.Def} for the Ramond-Ramond bispinor and the expressions for $\lambda_{\ra\rb}$ implied by \eqref{E:SUSY.vielbein.J}, the expression becomes 
\begin{align}
\delta \Psi_\rba{}^\beta
    &= \zcD_\rba \eps^\beta 
    -\frac{k}{2} (\bar\eps \gamma^{\rb \rc})^\beta (\bar\Psi_\rba \gamma_\rb \Psi_\rc)
    + \frac{k}{2} (\bar\eps \rho) \Psi_\rba{}^\beta
    - \frac{k}{2} (\bar\Psi_\rba \rho) \eps^\beta
    + \frac{1}{2} (\bar\eps \gamma_\rb \Psi_\rba ) (\bar\rho\gamma^\rb)^{\beta}
    - k (\bar\eps \bar\gamma_\rba)_\bgamma \,\zcS^{\bgamma \beta}
\end{align}
or, after a Fierz rearrangement,
\begin{align}
\delta \Psi_\rba{}^\beta
    &= \zcD_\rba \eps^\beta 
    + \frac{1}{4} (\bar\eps \gamma^{\rb \rc})^\beta \Big(
        \frac{k}{2} \bar\Psi_\rba \gamma_{\rb \rc} \rho 
        - 2k \bar\Psi_\rba \gamma_\rb \Psi_\rc
    \Big)
    - \frac{k}{2} (\bar\eps \rho) \Psi_\rba{}^\beta
    - \frac{k}{4} (\bar\Psi_\rba \rho) \eps^\beta
    - k (\bar\eps' \gamma_\rba)_\bgamma \,\zcS^{\bgamma \beta}
\end{align}
We have arranged the non-gauge invariant $\Psi_\ra{}^\beta$ terms so that they overlap with the contribution of $\omega_{\rba \rb \rc}$ from $\zcD_\rba \eps^\beta$. It is easy to see that the non-gauge invariant pieces cancel against those in \eqref{E:Deltaomega.c}.
Replacing $\omega_{\rba \rb \rc}$ with $\omega(V)_{\rba \rb \rc}$ that depends only on the double vielbein, we find
\begin{align}
\label{E:SUSY.gravitino}
\delta \Psi_\rba{}^\beta
    &= \zcD_\rba \eps^\beta\vert_{\omega(V)}
    + \frac{1}{4} (\bar\eps \gamma^{\rb \rc})^\beta \Big(
        \frac{k}{2} \bar\Psi_\rba \gamma_{\rb \rc} \rho 
        - k \bar\Psi'_\rb \bar\gamma_\rba \Psi'_{\rc}
    \Big)
    - \frac{k}{2} (\bar\eps \rho) \Psi_\rba{}^\beta
    - \frac{k}{4} (\bar\Psi_\rba \rho) \eps^\beta
    - k (\bar\eps' \gamma_\rba)_\bgamma \,\zcS^{\bgamma \beta}~.
\end{align}

For the dilatini, we find (after a Fierz rearrangement)
\begin{align}
\delta \rho_\alpha
    &= (\gamma^\ra)_{\alpha \beta} \zcD_\ra \eps^\beta 
    + \frac{1}{4} (\bar\eps \gamma^{\ra\rb\rc})_\alpha \Big(
        \frac{k}{48} \bar\rho \gamma_{\ra\rb\rc} \rho
        - k \bar\Psi_\ra \gamma_\rb \Psi_\rc
        + \frac{k}{12} \bar\Psi_{\rbd} \gamma_{\ra\rb\rc} \Psi^\rbd
    \Big)
    \eol & \quad
    - \frac{1}{2} (\bar\eps \gamma^\ra)_\alpha \Big(
        k \,\bar\Psi_\ra \rho
        - k \,(\gamma_\ra)_{\beta \gamma}\,S^{\beta \gamma} 
    \Big)
    + k (\bar \eps' \bar\gamma^\rba \Psi'_\rb)
    \, (\gamma^\rb \Psi_\rba)_\alpha
\end{align}
The second and third terms overlap with the contribution of $\omega_{[\ra\rb\rc]}$ and
$\omega^\rb{}_{\rb \ra}$; again, the non-gauge-invariant pieces cancel when we trade
$\omega$ for $\omega(V)$, giving
\begin{align}
\label{E:SUSY.dilatino}
\delta \rho_\alpha
    &= (\gamma^\ra)_{\alpha \beta} \zcD_\ra \eps^\beta\vert_{\omega(V)}
    + \frac{1}{4} (\bar\eps \gamma^{\ra\rb\rc})_\alpha \Big(
        \frac{k}{48} \bar\rho \gamma_{\ra\rb\rc} \rho
        + \frac{k}{12} \bar\Psi_{\rbd} \gamma_{\ra\rb\rc} \Psi^\rbd
    \Big)
    \eol & \quad
    - \frac{1}{2} (\bar\eps \gamma^\ra)_\alpha \Big(
    - k \, \bar\Psi'_\ra \rho'
    \Big)
    + k (\bar \eps' \bar\gamma^\rba \Psi'_\rb)
    \, (\gamma^\rb \Psi_\rba)_\alpha~.
\end{align}

Similar equations to $\Psi_\ra{}^\bbeta$ and $\rho_\balpha$ by adding/removing bars over the indices.

\paragraph{Dilaton.}
To derive the supersymmetry transformation of the component dilation $e^{-2d}$, we first compute
\begin{align}
\delta \log \det \phi_\hmu{}^\halpha = 
\phi_\halpha{}^\hmu \delta \phi_\hmu{}^\halpha & = (\zJ_0)_\halpha{}^\halpha 
    = \cD_\halpha \eps^\halpha - k\, \eps^\halpha \rho_\halpha~.
\end{align}
Combining with the transformation of the superdilaton,
\begin{align}
\delta \log \Phi = \xi^\cA \cT_\cA + \nabla_\cA \xi^\cA \,(-1)^a
    = - \cD_\halpha \eps^\halpha
\end{align}
we recover the expected component transformation,
\begin{align}
\label{E:SUSY.dilaton}
e^{2d} \delta e^{-2d} = -k\, \eps^\halpha \rho_\halpha~.
\end{align}

\paragraph{Ramond-Ramond sector.}
In order to derive a supersymmetry transformation of a $p$-form in conventional superspace, 
one first converts a superdiffeomorphism to a covariant superdiffeomorphism by subtracting 
off a local gauge transformation. Explicitly, this reads
\begin{align}
\delta_\xi \cC = \mathbb L_\xi \cC = \xi \lrcorner \rd \cC + \rd (\xi\lrcorner \cC) \quad \implies \quad
\delta^{\rm cov}_\xi \cC = \xi \lrcorner \rd \cC = \xi \lrcorner \cF ~.
\end{align}
This generalizes easily for an orthosymplectic spinor. The transformation 
\eqref{E:RamondC.Diffeo} can be rewritten
\begin{align}
\delta_\xi \bra{\cC}
    = \frac{1}{2} \Big(\bra{\cC} \Gamma^\cN \stackrel{\leftarrow}{\pa}_\cN \Big)\Gamma^\cM \xi_\cM
    + \frac{1}{2} \Big(\bra{\cC} \Gamma^\cM \xi_\cM\Big) \Gamma^\cN \stackrel{\leftarrow}{\pa}_\cN 
    = \bra{\cF} \slashed{\xi}
    + \bra{\lambda} \lDirac
\end{align}
where 
$\bra{\lambda} = \bra{\cC}\slashed{\xi} = \frac{1}{\sqrt 2} \bra{\cC}\Gamma^\cM \xi_\cM$ is a special parameter for an abelian transformation. This leads to the definition of a covariant diffeomorphism,
\begin{align}
\delta^{\rm cov}_\xi \bra{\cC} = 
     \bra{\cF} \slashed{\xi} =  \frac{1}{\sqrt 2} \bra{\cF} \Gamma^\cM \xi_\cM~.
\end{align}
To proceed to components, it helps to again recall what we do in conventional superspace. Starting with a 1-form, for example, we write
\begin{align}\label{E:dCm1}
\delta^{\rm cov}_\xi \cC_M = \xi^N \cF_{NM}
    = E_M{}^B \xi^A \cF_{A B}
\end{align}
keeping $\cC_M$ with a curved index but rewriting $\xi$ and $\cF$ with flat indices.
Subsequently projecting to $\rd \theta=0$ gives
\begin{align}\label{E:dCm2}
\delta^{\rm cov}_\xi C_m 
    = E_m{}^B \eps^\alpha \cF_{\alpha B} \vert_{\theta=0}
\end{align}
The steps for an orthosymplectic spinor are similar.
The analogous procedure of \eqref{E:dCm1} is to flatten $\bra{\cF}$, writing 
$\bra{\cF} = \bra{\widehat{\slashed{\cF}}} \cswV^{-1} \,\Phi^{1/2}$. This leads to
\begin{align}\label{E:RR.Vary1}
\delta^{\rm cov}_\xi \bra{\cC}
    = \frac{1}{\sqrt 2} \bra{\widehat{\slashed{\cF}}} \cswV{}^{-1} \Gamma^\cM \xi_\cM \Phi^{1/2} 
    = \frac{1}{\sqrt 2} \bra{\widehat{\slashed{\cF}}} \Gamma_\halpha \eps^\halpha\, 
        \cswV{}^{-1} \Phi^{1/2}
\end{align}
Using a very similar computation as that leading to \eqref{E:FlatBispinorF}, we find
\begin{align}
e^d \times \braket{\delta^{\rm cov}_\xi C}{\slashed{V}} 
&= \frac{1}{\sqrt 2} 
    \bra{\widehat{\slashed{\cF}}} \Gamma_\halpha \eps^\halpha\, \cswV{}_\Psi^{-1} \Dbar~.
\end{align}
The left-hand side can be rewritten as the variation of 
$\slashed{\widehat C} = e^d \braket{C}{\slashed{V}}$:
\begin{align}
e^d \times \braket{\delta^{\rm cov}_\xi C}{\slashed{V}} 
    &= \delta \slashed{\widehat C}
    - \slashed{\widehat C} \delta d
    + \frac{1}{4} \slashed{\widehat C} \cdot \Gamma^{\ha \hb} J_{\hb \ha}
\end{align}
where $J_{\hb \ha} = V_\hb{}^\hm \delta V_{\hm \ha}$.
Let's evaluate the right-hand side in two steps. First,
\begin{align}\label{E:deltaCStep1}
\frac{1}{\sqrt 2} \bra{\widehat{\slashed{\cF}}} \Gamma_\halpha \eps^\halpha = 
-4k
\begin{pmatrix}
0 & 0 \\[1ex]
\bra{\fvac} (\eps^\alpha b^{\balpha} - b^\alpha \eps^{\balpha}) & 0
\end{pmatrix}~.
\end{align}
Next, we evaluate $\cV_\Psi^{-1}$. Only the level one terms contribute because \eqref{E:deltaCStep1} involves only a single raising operator. This means effectively we have
$\cV_\Psi^{-1} = 1 + \frac{1}{2} \Gamma_\hbeta \Gamma^\ha \Psi_\ha{}^\hbeta + \cdots$.
Evaluating this leads to
\begin{align}
e^d \times \braket{\delta^{\rm cov}_\xi C}{\slashed{V}} 
= -4k\sqrt{2}
\begin{pmatrix}
\rho_\alpha \eps^\balpha - (\gamma^\rb \eps)_\alpha\, \Psi_\rb{}^\balpha & 0 \\
0 & \eps^\alpha \rho_\balpha - \Psi_\rbb{}^\alpha (\eps \gamma^\rbb)_{\balpha}
\end{pmatrix}
\end{align}
Employing Dirac notation, we arrive at
\begin{align}
\label{E:SUSY.RRsector}
\delta \slashed{\widehat C}
    &= 
    -4k \sqrt{2}
    \Big(\rho \times \bar \eps' - \gamma^\rb \eps \times \bar \Psi_\rb'
    + \eps \times \bar \rho' - \Psi_\rbb \times \bar\eps' \gamma^\rbb\Big)
    +\slashed{\widehat C} \delta d
    - \frac{1}{2} \gamma_* \gamma^\ra \slashed{\widehat C} \bar\gamma^{\rbb} \, 
    V_\rbb{}^\hm \delta V_{\hm \ra}
\end{align}

\subsection{Comparison to component results}
The component supersymmetry transformations for type II DFT were given in different conventions in \cite{Jeon:2012hp}. To match those results, we trade their indices $p$ and $\bar p$ for $\ra$ and $\rba$ here, while $A$ there corresponds to $\hm$ here. The $\gamma$-matrices are related as
\begin{gather}
\gamma^p = \gamma^\ra~, \qquad
\bar\gamma^{\bar p} = \bar\gamma^{\rba}~, \qquad
\gamma^{11} = - \gamma_*~, \qquad
\bar\gamma^{11} = - \bar\gamma_*~.
\end{gather}
We fix the constant $k = i$ in the constant torsion tensor \eqref{E:BasicTorsionConstraint}, flip the sign for the dilatini,
\begin{align}
\rho_{\rm JLPS} = -\rho_\alpha~, \qquad
\rho'_{\rm JLPS} = -\rho_\balpha
\end{align}
and rescale the Ramond-Ramond sector fields as
\begin{align}\label{E:ParkRescaleCF}
\cC_{\rm JLPS} = \frac{1}{4 \sqrt 2} \slashed{\widehat C} \bar\gamma_*~, \qquad
\cF_{\rm JLPS} = \frac{1}{4} \slashed{\widehat F} \bar\gamma_*~.
\end{align}
The factors of $\gamma_*$ and $\sqrt{2}$ are necessary to recover our conventions for the relation between $\slashed{\widehat F}$ and $\slashed{\widehat C}$,
\begin{align}
\slashed{\widehat F}
    = \frac{1}{\sqrt 2} \Big(\gamma^\ra \zcD_\ra \slashed{\widehat C} + \gamma_{*} \zcD_\rba \slashed{\widehat C} \bgamma^{\rba} \Big)~,
\end{align}
and the overall factor of $1/4$ is for normalization of the kinetic terms.
Taking into account these changes as well as the differences in our spin connection versus that given in \cite{Jeon:2012hp}, one can show that the supersymmetry transformations of the component fields 
\eqref{E:SUSY.vielbein},
\eqref{E:SUSY.gravitino},
\eqref{E:SUSY.dilatino},
\eqref{E:SUSY.dilaton}, and
\eqref{E:SUSY.RRsector}
all match precisely.

We should make a final comment about the normalization of the action of Jeon et al.~\cite{Jeon:2012hp} in comparison to the bosonic action of Hohm et al.~\cite{Hohm:2011zr, Hohm:2011dv}. The latter action is normalized so that
\begin{align}
S_{\rm{HKZ}} &= \int \rd^{10} x\, \rd^{10} \tilde x\, \Big(
    e^{-2d} \cR(\cH, d) + \frac{1}{4} \bra{F} \mathbb S \ket{F}
\Big) \eol
    &= \int \rd^{10} x\, \sqrt{-g}\Big[
    e^{-2\varphi} \Big(
        R + 4 (\pa\phi)^2 - \frac{1}{2} |H^{(3)}|^2
    \Big)
    - \frac{1}{4} \sum_p |\widehat F^{(p)}|^2
    \Big]
\end{align}
where the norm on $p$-forms includes a factor of $1/p!$ and we work in the democratic formulation for the Ramond-Ramond sector.\footnote{In these formulae alone, we use the conventions of Hohm et al. for $R$ and $\cR$, which differ from ours by a sign. }
The corresponding action of Jeon et al. is normalized so that
\begin{align}
S_{\rm JLPS} &= \int \rd^{10} x\, \rd^{10} \tilde x\, e^{-2d} \Big(
    \frac{1}{8} \cR(\cH, d)
    + \frac{1}{2} \tr (\cF_{\rm JLPS} \bar\cF_{\rm JLPS})
    + \text{fermions}
\Big) \eol
&= \int \rd^{10} x\, \rd^{10} \tilde x\, e^{-2d} \frac{1}{8}  \Big(
    \cR(\cH, d)
    - \frac{1}{4} \tr (\slashed{\widehat F} \overline{\slashed{\widehat F}})
    + \text{fermions}
\Big) \eol
    &= \frac{1}{8} S_{\rm HKZ} + \text{fermions}
\end{align}
The rescaling in \eqref{E:ParkRescaleCF} is crucial to recover the same normalizations.
Note that $|H^{(3)}|^2$ and $\frac{1}{2} |\widehat F^{(3)}|^2 + \frac{1}{2} |\widehat F^{(7)}|^2$ are typically normalized the same in the IIB duality frame so that S-duality takes a simple form.

\section{Democratic type II superspace}
\label{S:TypeIISS}
Our last major task is to recover type II superspace directly from super-DFT.
Unlike the type I situation, type II supergravity is not unique: not only do we have IIA and IIB supergravities, characterized by even rank or odd rank $p$-form field strengths in the Ramond-Ramond sector, but also their timelike T-duals, denoted IIB$^*$ and IIA$^*$, whose Ramond-Ramond sector is characterized by the wrong sign kinetic terms.\footnote{Type IIA supergravity was formulated in \cite{Chamseddine:1980cp,Bergshoeff:1981um} by dimensional reduction of 11D supergravity \cite{Cremmer:1978km}. Its massive deformation was introduced in \cite{Romans:1985tz}. Type IIB supergravity was formulated in \cite{Schwarz:1983wa, Howe:1983sra}. The starred cases were proposed by Hull \cite{Hull:1998vg}. Type IIA supergravity  was discussed in superspace in \cite{Nilsson:1981bn, Carr:1986tk} and the IIB superspace was already employed in \cite{Howe:1983sra}.} 
Superspace formulations of any supergravity theory are in 1:1 correspondence with their component formulations; that is, given the component supersymmetry transformations, one can always rebuild the superspace and vice-versa. While there are a number of references on type II superspace, we will focus on the appendices of Wulff \cite{Wulff:2013kga}, which are useful for two reasons: they are formulated in the string frame, and they treat IIA and IIB very similarly. Both of these features are natural when we descend from double field theory.

The descent from super-DFT to conventional superspace is a somewhat involved procedure, with the primary technical hurdle being the parametrization of the supervielbein in a convenient way.  The main result in this section will be to recover a type II superspace that is \emph{fully democratic}, meaning not only that it treats Ramond-Ramond potentials and their duals simultaneously (i.e. democratic in the sense of \cite{Bergshoeff:2001pv}), but also IIA and IIB (as well as IIA$^*$ and IIB$^*$) rather in parallel. This will match (after a simple rewriting of some formulae) Wulff's formulation of type II \cite{Wulff:2013kga}.

\subsection{Double vielbein decomposition and the Ramond-Ramond sector in bosonic DFT}
\label{S:TypeIISS.Bosonic}
As a first step to understanding how type II superspace emerges, we will review how conventional gravity emerges from the bosonic double vielbein. In the chiral tangent frame basis, the double vielbein may always be decomposed as (see e.g. \cite{Jeon:2012kd,Jeon:2012hp})
\begin{subequations}\label{E:genericVchiral.bosonic}
\begin{align}
V_\hm{}^\ha &=
\frac{1}{\sqrt 2}
\begin{pmatrix}
\delta_m{}^n & b_{mn} \\
0 & \delta^m{}_n
\end{pmatrix}
\times
\begin{pmatrix}
e_n{}^\ra & \bar e_n{}^{\rba} \\
\eta^{\ra \rb} e_\rb{}^n & \,\eta^{\ol{\ra\rb}} \bar e_{\rbb}{}^n
\end{pmatrix}~, \\
V_\ha{}^\hm &=
\frac{1}{\sqrt 2}
\begin{pmatrix}
e_\ra{}^n & \,\eta_{\ra \rb} e_{n}{}^\rb \\
\bar e_\rba{}^n & \,\eta_{\ol{\ra\rb}} \bar e_{n}{}^{\rbb}
\end{pmatrix} \times
\begin{pmatrix}
\delta_n{}^m & -b_{nm} \\
0 & \delta^n{}_m
\end{pmatrix}~.
\end{align}
\end{subequations}
The two vielbeins $e_m{}^\ra$ and $\bar e_m{}^\rba$ rotate separately under the two Lorentz groups, while the Kalb-Ramond two-form $b_{mn}$ is invariant. This is a rather generic decomposition, and it follows simply by assuming that 
$V_\ra{}^m = \frac{1}{\sqrt2} e_\ra{}^m$ and $V_\rba{}^m = \frac{1}{\sqrt 2} \bar e_\rba{}^m$ are both invertible matrices. (This is essentially equivalent to assuming that the component $\cH^{\hm\hn}$ of the generalized metric is invertible.) The requirement that this be an $\g{O}(10,10)$ element amounts to demanding that $e_m{}^\ra$ and $\bar e_m{}^\rba$ both give the same metric,
\begin{align}
g_{mn} = e_m{}^\ra e_n{}^\rb \eta_{\ra\rb} = -\bar e_m{}^\rba \bar e_n{}^\rbb \eta_{\ol{\ra\rb}}~.
\end{align}
Equivalently, $(e^{-1} \bar e)_\ra{}^\rbb$ is an element of $\g{O}(1,9)$.

Using this observation, we may further separate $V_\hm{}^\ha$ into three factors, schematically, $V = V_b \times V_e \times V_\Lambda$:
\begin{align}\label{E:VbeLambda.Bosonic}
V_\hm{}^\ha &=
\frac{1}{\sqrt 2}
\begin{pmatrix}
\delta_m{}^n & b_{mn} \\
0 & \delta^m{}_n
\end{pmatrix}
\times
\begin{pmatrix}
e_n{}^\rb & e_n{}^{\rb} \\
\eta^{\ra \rb} e_\rb{}^n & \,-\eta^{\ra\rb} e_{\rb}{}^n
\end{pmatrix} 
\times
\begin{pmatrix}
\delta_\rb{}^\ra & 0 \\
0 & (e^{-1} \bar e)_\rb{}^{\rba}
\end{pmatrix} ~.
\end{align}
The last factor is an element of $\g{O}(1,9) \subset \g{O}(10,10)$. If the full double Lorentz  group is gauged, then we may always discard this as a gauge choice. However, if only $\g{SO}(1,9)$ or $\g{SO}^+(1,9)$ are gauged, we must more carefully account for it. As argued in \cite{Jeon:2012kd, Jeon:2012hp, Butter:2022sfh}, this factor is crucial for distinguishing between the various duality frames for the Ramond-Ramond sector.

The spinorial form of the vielbein may also be written as a product of three factors,
\begin{align}\label{E:SpinorialVielbein.Decompose}
\ket{\slashed{V}} = \mathbb S_b \times \ket{\slashed{V}_e} \times \slashed \Lambda
\end{align}
where $\mathbb S_b$ is a Fock-space operator (i.e. it carries curved spinor indices on both sides), $\ket{\slashed{V}_e}$ is a bispinor-valued ket (i.e. it carries a single curved spinor index on the left), and $\slashed \Lambda$ is spinor Lorentz transformation. We are treating $\ket{\slashed{V}_e}$ here as a bispinor-valued ket rather than a ket with an additional flat spinor index on the right. Thus it is crucial here that $V_\Lambda$ is purely a right-handed Lorentz transformation, and so we may write $\slashed \Lambda$ simply by right multiplication. In treating the spinorial vielbein asymmetrically in this way, we have to give prescriptions for how each of these objects behaves. The Kalb-Ramond factor is
\begin{align}
\mathbb S_b &= \exp\Big(-\tfrac{1}{4} \Gamma^{mn} b_{nm}\Big)=
\exp\Big(-\tfrac{1}{2} \psiGamma^m \psiGamma^n b_{nm}\Big)
\end{align}
and acts as
$\mathbb S_b \psiGamma^m \mathbb S_b^{-1} = \psiGamma^m$ and
$\mathbb S_b \psiGamma_m \mathbb S_b^{-1} = \psiGamma_m + \psiGamma^n b_{nm}$.
The bispinor valued ket acts as
\begin{align}\label{E:BosonicVe.Action}
\Gamma^m \ket{\slashed{V}_e} = \frac{1}{\sqrt 2} \Big(
    \gamma^\ra \ket{\slashed{V}_e}  
    + \gamma_* \ket{\slashed{V}_e} \bar\gamma^{\rba} 
    \Big) e_\ra{}^m ~, \qquad
\Gamma_m \ket{\slashed{V}_e} = \frac{1}{\sqrt 2} \Big(
    \gamma^\ra \ket{\slashed{V}_e}  
    - \gamma_* \ket{\slashed{V}_e} \bar\gamma^{\rba} 
    \Big) \eta_{ab} \,e_m{}^\rb~.
\end{align}
The Lorentz transformation acts as
\begin{align}
\slashed \Lambda \bar \gamma^\rba\slashed \Lambda^{-1}
    = \bar\gamma^\rbb \Lambda_\rb{}^\rba
    = \bar\gamma^\rbb (e^{-1} \bar e)_\rb{}^\rba
\end{align}
As a sanity check the Lorentz transformation acting on \eqref{E:BosonicVe.Action} from the right gives, for $\ket{\slashed{V}_{e\Lambda}} = \ket{\slashed{V}_e} \slashed{S}_\Lambda$,
\begin{align}
\Gamma^m \ket{\slashed{V}_{e\Lambda}}= \frac{1}{\sqrt 2} \Big(
    \gamma^\ra e_\ra{}^m \ket{\slashed{V}_{e\Lambda}}  
    + \gamma_* \ket{\slashed{V}_{e\Lambda}} \bar\gamma^{\rba} 
    \bar e_\rba{}^m
    \Big)~, \eol
\Gamma_m \ket{\slashed{V}_{e\Lambda}} = \frac{1}{\sqrt 2} \Big(
    e_m{}^\ra \gamma_\ra  \ket{\slashed{V}_{e\Lambda}} 
    + \gamma_* \ket{\slashed{V}_{e\Lambda}} e_m{}^\rba \bar\gamma_{\rba} 
    \Big)~,
\end{align}
as we would expect.

The Ramond-Ramond field strength $\bra{F}$ decomposes as
\begin{align}
\bra{F} = \sum_p \frac{1}{p!} \bra{0} \psiGamma^{m_1} \cdots \psiGamma^{m_p} F_{m_p \cdots m_1}
\end{align}
where we employ the democratic formulation with every $p$-form field strength appearing (with $p$ even or odd depending on the duality frame).
The flattened field strength is generated by contracting with the ket $\ket{\slashed{V}}$ and multiplying by a factor of the dilaton, $e^d$. The effect of the Kalb-Ramond factor $\mathbb S_b$ is to replace $\bra{F}$ with $\bra{\widehat F}$ where $\widehat F = F e^{-b}$. The effect of the ket $\ket{\slashed{V}_e}$ leads to
\begin{align}
\bra{0} \psiGamma^{m_p} \cdots \psiGamma^{m_1} \ket{\slashed{V}_e}
    = (\det e)^{1/2} \, \gamma^{\ra_1 \cdots \ra_p} e_{a_1}{}^{m_1} \cdots e_{a_p}{}^{m_p} 
\end{align}
which follows from $\bra{0} \Gamma_m = 0$ and $\braket{0}{\slashed{V}_e} = \frac{1}{2^{5/2}}(\det e)^{1/2} \times \mathbf 1$. The final Lorentz transformation gives
\begin{align}
\slashed{\widehat F} &= e^\varphi  \sum_p \frac{1}{p!}
    \gamma^{\ra_1 \cdots \ra_p} \widehat F_{\ra_1 \cdots \ra_p} \slashed{Z}
\end{align}
where $\widehat F_{\ra_1 \cdots \ra_p} := e_{\ra_1}{}^{m_1} \cdots e_{\ra_p}{}^{m_p} 
    \widehat F_{m_1\cdots m_p}$ and 
    $\slashed{Z} = \frac{1}{2^{5/2}} \slashed \Lambda$ is the flat vacuum.

\subsection{Decomposing the double supervielbein}
\label{S:TypeIISS.Decomposing}

We want to repeat the above steps for the supervielbein. The details are given in appendix \ref{A:SuperV} and we just give the results here.
The superspace analogue to the parametrization \eqref{E:genericVchiral.bosonic} is a product of three factors:
\begin{align}\label{E:V.BELambdaS}
\cV
    = \cV_B \times \cV_{E\Lambda} \times \cV_S
\end{align}
The first is built out of the Kalb-Ramond super two-form,
\begin{align}
(\cV_B)_\cM{}^\cN =
\begin{pmatrix}
\delta_M{}^N & B_{MN} (-)^n \\
0 & \delta^N{}_M
\end{pmatrix}~, \qquad
\end{align}
The second factor $\cV_{E\Lambda}$ is written, in a chiral decomposition of the indices, as
\begin{align}
(\cV_{E\Lambda})_\cM{}^\cA =
\renewcommand{\arraystretch}{1.5}
\left(\begin{array}{ccc|ccc}
\frac{1}{\sqrt 2} E_M{}^\ra & E_M{}^\alpha & 0 
    & \frac{1}{\sqrt 2} E_M{}^\rba & E_M{}^\balpha & 0 \\ 
\frac{1}{\sqrt 2} E^\ra{}^M & 0 & - E_\alpha{}^M (-)^m \!\!\!
    & \frac{1}{\sqrt 2} E^\rba{}^M & 0 & - E_\balpha{}^M (-)^m
\end{array}\right)~.
\end{align}
The two superfields $E_M{}^\ra$ and $E_M{}^\rba$ are related by a Lorentz transformation,
\begin{align}
E_M{}^\rba = E_M{}^\rb \Lambda_\rb{}^\rba~,
\end{align}
a clear generalization of the bosonic condition,
and the inverse vielbeins are defined by
\begin{align}
E_\halpha{}^M E_M{}^\hbeta = \delta_\halpha{}^\hbeta~, \qquad
E_\ra{}^M E_M{}^\rb = \delta_\ra{}^\rb~, \qquad
E_\rba{}^M E_M{}^\rbb = \delta_\rba{}^\rbb~, \qquad
E_\halpha{}^M E_M{}^\hb = E_\ha{}^M E_M{}^\hbeta = 0~.
\end{align}
The $\cV_S$ factor is given, also in a chiral decomposition, as
\begin{align}
(\cV_S)_\cA{}^\cB &=
\renewcommand{\arraystretch}{1.5}
\left(\begin{array}{ccc|ccc}
\delta_\ra{}^\rb & \sqrt{2} S_\ra{}^\beta & 0 
    & 0 & 0 & 0 \\
0 & \delta_\alpha{}^\beta & 0
    & 0 & 0 & 0 \\
-\sqrt{2} S^{\rb \alpha} & S^{\alpha\beta} - S^{\rc\alpha} S_\rc{}^\beta &  \delta^\alpha{}_\beta
    & 0 & S^{\alpha \bbeta} & 0 \\ \hline
0 & 0 & 0
    &\delta_\rba{}^\rbb & \sqrt{2} S_\rba{}^\bbeta & 0  \\
0 & 0 & 0
    & 0 & \delta_\balpha{}^\bbeta & 0 \\
0 & S^{\balpha \beta}  & 0 
    & -\sqrt{2} S^{\ol{\rb \alpha}} & S^{\ol{\alpha\beta}} - S^{\rbc\balpha} S_\rbc{}^\bbeta &  \delta^\balpha{}_\bbeta
\end{array} \right).
\end{align}
It consists of fermionic superfields $S_\ra{}^\alpha$ and $S_\rba{}^{\balpha}$ as well as the symmetric bosonic superfields $S^{\alpha\beta}$, $S^{\ol{\alpha\beta}}$, and $S^{\alpha \bbeta}$. All these constituents transform as their indices imply under double Lorentz transformations, while under the additional $\g{H}_L \times \g{H}_R$ transformations,
\begin{subequations}
\begin{alignat}{2}
\delta S_\ra{}^\alpha &= -\frac{1}{\sqrt2} \lambda_\ra{}^\alpha~, 
&\qquad
\delta S_\rba{}^\balpha &= -\frac{1}{\sqrt2} \lambda_\rba{}^\balpha~,
\\
\delta S^{\alpha \beta} &= -\lambda^{\alpha\beta}
    + \sqrt{2}\, S^{c (\alpha} \lambda_\rc{}^{\beta)}~, 
&\qquad
\delta S^{\ol{\alpha \beta}} &= -\lambda^{\ol{\alpha\beta}}
    + \sqrt{2}\, S^{\rbc (\balpha} \lambda_\rbc{}^{\bbeta)}~.
\end{alignat}
\end{subequations}

The Lorentz transformation $\Lambda_\ra{}^\rbb$ belongs to $\g{O}(1,9)$, and so is characterized by two signs, corresponding to the presence of timelike and/or spacelike orientation reversals. In the event that it lies in the connected part (with no orientation reversals), it can be gauged to the identity and then $\cV_{E\Lambda}$ becomes, in the toroidal decomposition,
\begin{align}\label{E:Ve.Toroidal}
(\cV_E)_\cM{}^\cA =
\begin{pmatrix}
E_M{}^A & 0 \\
0 & E_A{}^M (-)^{am+a}
\end{pmatrix}~.
\end{align}
This leads to the conventional decomposition \eqref{E:SuperVielbein.BES} modulo some redefinitions,
\begin{align}
E'_M{}^\alpha &= E_M{}^\alpha + E_M{}^b \eta_{bc} \,S^{\rc\alpha}~, \eol
E'_M{}^\balpha &= E_M{}^\balpha - E_M{}^b \eta_{bc} \,S^{\ol{\rc\alpha}}~, \eol
S'^{\alpha \bbeta} &= S^{\alpha \bbeta} + S^{\ra\alpha} \eta_{ab} \,S^{\ol{\rb\beta}}~,
\end{align}
with the primed fields belonging to \eqref{E:SuperVielbein.BES}.

When the Lorentz transformation is more general, we encounter a bit of a puzzle. Normally, we would like to factor out all right-handed Lorentz transformations to define a physical supervielbein that transforms only under left Lorentz transformations.
This would suggest introducing a spinorial Lorentz transformation $\Lambda_\talpha{}^\bbeta$ and defining a new gravitino $E_M{}^\talpha$ transforming under the left Lorentz group by
\begin{align}
E_M{}^{\balpha} = E_M{}^{\tbeta} \Lambda_{\tbeta}{}^{\balpha}~.
\end{align}
We would expect this spinorial $\Lambda$ to obey
\begin{align}\label{E:putativeLambdaTrafo}
\Lambda_{\talpha}{}^{\bgamma} \,(\bar\gamma^\rba)_{\ol{\gamma \delta}} \,
        \Lambda_{\tbeta}{}^{\bdelta} 
        \stackrel{?}{=} (\bar\gamma^{\rbb})_{\ol{\alpha \beta}} 
    \Lambda_{\rb}{}^{\rba}~.
\end{align}
This fails! One obvious reason is that we are employing a chiral basis for the 10D $\gamma$-matrices, which cannot account for Lorentz transformations that flip chirality (i.e. with an odd number of orientation reversals). However, this fails to be possible even for a combined time and space orientation reversal, where one would find a minus sign in the above equation. The reason is that the correct relation for $\gamma$ matrices would read (in Dirac notation)
$\slashed{\Lambda} \bar\gamma^\rba  \slashed{\Lambda}^{-1} = \bar\gamma^\rbb \Lambda_\rb{}^{\rba}$.
For \eqref{E:putativeLambdaTrafo} to be satisfied, we need 
$(\slashed\Lambda^{-1})^{\bdelta}{}_{\tbeta} = (\slashed{\Lambda})_\tbeta{}^\bdelta$
and this holds only for the connected part, $\g{SO}^+(1,9) = \g{O}^{(+,+)}(1,9)$.

Luckily, all $\g{O}(1,9)$ transformations may be understood as a fixed element $\zLambda$ in one of the four connected sectors times an $\g{SO}^+(1,9)$ transformation $\dot\Lambda$. We write this as
\begin{align}
\Lambda_\ra{}^{\rbb} = \zLambda_\ra{}^{\rbc} \,\dot\Lambda_\rbc{}^\rbb
\end{align}
Then we choose $\dot\Lambda_\talpha{}^{\bbeta}$ to obey
\begin{align}
\dot\Lambda_{\talpha}{}^{\bgamma} \,(\bar\gamma^\rba)_{\ol{\gamma \delta}} \,
        \dot \Lambda_{\tbeta}{}^{\bdelta} 
        = (\bar\gamma^{\rbb})_{\ol{\alpha \beta}} 
    \dot \Lambda_{\rbb}{}^{\rba}~.
\end{align}
This leads us to define the following constant quantities:
\begin{align}\label{E:tgamma.def}
(\gamma^a)_{\talpha \tbeta} := (\gamma^\rbb)_{\ol{\alpha\beta}} \,(\zLambda^{-1})_{\rbb}{}^{\rba}~, \qquad
(\gamma^a)^{\talpha \tbeta} := -(\gamma^\rbb)^{\ol{\alpha\beta}} \, (\zLambda^{-1})_{\rbb}{}^{\rba}~.
\end{align}
This is equivalent using the full vector and spinorial $\Lambda$:
\begin{align}
(\gamma^a)_{\talpha \tbeta} = 
    \dot\Lambda_{\talpha}{}^{\bgamma} \dot\Lambda_{\tbeta}{}^{\bdelta}\,
    (\gamma^\rbb)_{\ol{\gamma\delta}} \,(\Lambda^{-1})_{\rbb}{}^{\rba}~, \qquad
(\gamma^a)^{\talpha \tbeta} = -(\gamma^\rbb)^{\ol{\gamma\delta}} \, (\Lambda^{-1})_{\rbb}{}^{\rba}\, (\dot\Lambda^{-1})_{\bgamma}{}^{\talpha} (\dot\Lambda^{-1})_{\bdelta}{}^{\tbeta}~.
\end{align}
The reason we have included an additional sign for $(\gamma^a)^{\talpha\tbeta}$ is to recover the same Clifford algebra as the standard $\gamma$-matrices
\begin{align}
(\gamma^a)_{\talpha \tbeta} (\gamma^b)^{\tbeta \tgamma} + 
(\gamma^b)_{\talpha \tbeta} (\gamma^a)^{\tbeta \tgamma}
= 2 \,\eta^{a b} \,\delta_\talpha{}^\tgamma~.
\end{align}

The four different possibilities correspond to each of the four duality frames IIB, IIA, IIB$^*$, and IIA$^*$. We are free to make whatever choice we wish for $\zLambda$. The most convenient choices are as follows:
\begin{align}
\zLambda =
\begin{cases}
1 & \text{IIB} \\
R_{0 1 \cdots 9} & \text{IIB${}^*$} \\
R_{1 \cdots 9} & \text{IIA} \\
R_{0} & \text{IIA${}^*$}
\end{cases}
\end{align}
where $R_a$ denotes a sign flip of the $a$ direction. Note that $R_{1\cdots 9}$ is equivalent to $R_9$ up to a constant $\g{SO}^+(1,9)$ transformation, but the former is more convenient for our choice of $\gamma$-matrices. Keeping in mind that
$(\bar\gamma^\rba)_{\ol{\alpha \beta}} = (\gamma^a)_{\alpha\beta}$ and
$(\bar\gamma^\rba)^{\ol{\alpha \beta}} = -(\gamma^a)^{\alpha\beta}$,
these lead to
\begin{align}\label{E:tildegamma}
(\gamma^c)_{\talpha \tbeta} = 
\begin{cases}
\phantom{+} (\gamma^c)_{\alpha \beta} & \text{IIB} \\
- (\gamma^c)_{\alpha \beta} & \text{IIB${}^*$} \\
- (\gamma^c)^{\alpha \beta} & \text{IIA} \\
\phantom{+} (\gamma^c)^{\alpha \beta} & \text{IIA${}^*$}
\end{cases}~, \qquad
(\gamma^c)^{\talpha \tbeta} = 
\begin{cases}
\phantom{+} (\gamma^c)^{\alpha \beta} & \text{IIB} \\
-(\gamma^c)^{\alpha \beta} & \text{IIB${}^*$} \\
- (\gamma^c)_{\alpha \beta} & \text{IIA} \\
\phantom{+} (\gamma^c)_{\alpha \beta} & \text{IIA${}^*$}
\end{cases}~.
\end{align}
The $\talpha$ index for IIA/IIA$^*$ must be understood as the opposite chirality as $\alpha$. Thus we denote for the gravitino, for example,
\begin{align}
E_M{}^\talpha =
\begin{cases}
E_M{}^{\alpha'} & \text{IIB/IIB${}^*$} \\
E_M{}_\alpha & \text{IIA/IIA${}^*$}
\end{cases}
\end{align}
where prime for IIB/IIB$^*$ denotes this being the second gravitino.

One way of understanding these expressions for $(\gamma^a)_{\talpha \tbeta}$ is via the supersymmetry algebra,
\begin{align}\label{E:TypeIISusyAlgebra}
\{Q_\alpha, Q_\beta\} = -\frac{k}{\sqrt 2} (\gamma^c)_{\alpha \beta} P_c~, \qquad
\{Q_\talpha, Q_\tbeta\} = -\frac{k}{\sqrt 2} (\gamma^c)_{\talpha \tbeta} P_c~.
\end{align}
For IIB, the second SUSY is just a copy of the first. For IIA, the extra sign factor in \eqref{E:tildegamma} is needed so that both chiralities may be combined into a single 32-component supercharge with
\begin{align}
\{Q, Q\} 
    = \frac{k}{\sqrt 2} \gamma^c \gamma_* C^{-1} \, P_c
    = \frac{k}{\sqrt 2} \gamma^c C_{\rm 11D}^{-1} \,P_c~,
\end{align}
which is the natural truncation from the $D=11$ supersymmetry algebra. But we are also allowing here for the possibility of the starred supergravities. In the formulation we are using, these have the wrong sign in the supersymmetry algebra from their unstarred analogues. Alternatively, one can multiply all upper/lower tilded spinors by $\pm i$ (i.e. make an imaginary similarity transformation) to restore the conventional sign for supersymmetry, but at the cost of changing the reality condition for tilded spinors (and the Ramond-Ramond bispinor).

The upshot is that we now can, without any gauge-fixing, split $\cV_{E\Lambda}$ into $\cV_E\times \cV_\Lambda$ where $\cV_E$ is given in the toroidal basis by \eqref{E:Ve.Toroidal} with
\begin{align}
E_M{}^A = (E_M{}^\ra, E_M{}^\alpha, E_M{}^\talpha)
\end{align}
and $\cV_\Lambda$ is given in the chiral basis as
\begin{align}
(\cV_\Lambda)_\cA{}^\cB &=
\renewcommand{\arraystretch}{1.5}
\left(\begin{array}{ccc|ccc}
\delta_\ra{}^\rb & 0 & 0 
    & 0 & 0 & 0 \\
0 & \delta_\alpha{}^\beta & 0
    & 0 & 0 & 0 \\
0 & 0 & \delta^\alpha{}_\beta
    & 0 & 0 & 0 \\ \hline
0 & 0 & 0
    &\Lambda_\ra{}^\rbb & 0 & 0  \\
0 & 0 & 0
    & 0 & \dot\Lambda_\talpha{}^\bbeta & 0 \\
0 & 0  & 0 
    & 0 & 0 & (\dot{\Lambda}^{-T})^\talpha{}_\bbeta
\end{array} \right), \qquad
\Lambda_\ra{}^\rbb = \zLambda_\ra{}^\rbc \dot\Lambda_\rbc{}^\rbb
\end{align}
Part of this is an $\g{SO}^+(1,9)$ transformation, which can be eliminated by a gauge transformation; the remainder can be thought of as a constant similarity transformation on barred vector indices, converting barred gamma matrices $(\bar\gamma^\rba)_{\ol{\alpha \beta}}$ to $(\gamma^a)_{\talpha\tbeta}$ in \eqref{E:tildegamma}.

\subsection{Gauge-fixing to democratic type II superspace}
\label{S:TypeIISS.Gaugefixing}
Let us now analyze the structure of type II superspace that emerges from double field theory. This will be a democratic formulation with the constant Lorentz transformation $\zLambda$ defining which duality frame we are part of. Although it is possible to do this analysis without fixing any gauge, it will be significantly simpler if we impose
\begin{align}\label{E:Sgauge}
S^{\alpha\beta} = S^{\ol{\alpha\beta}} = 0~, \qquad
S^{\ra \alpha} = \frac{i}{10} (\gamma^\ra)^{\alpha\beta}  \chi_\beta~, \qquad
S^{\rba \balpha} = \frac{i}{10} (\bar\gamma^\rba)^{\ol{\alpha\beta}} \chi_\bbeta~,
\end{align}
where $\chi_\alpha$ and  $\chi_\talpha = \Lambda_\talpha{}^\bbeta \chi_\bbeta$ will be
the component dilatini. Note that these can be given gauge-invariant definitions as
\begin{align}
\chi_\alpha := -i S^{\ra \beta} (\gamma_{\ra})_{\beta\alpha}~, \qquad
\chi_\balpha := -i S^{\ol{\ra \beta}} (\gamma_{\rba})_{\ol{\beta\alpha}}~.
\end{align}

It will also be computationally simpler if we gauge fix $\cV_\Lambda$, fixing $\dot\Lambda=1$, leaving $\cV_\Lambda$ to just be the constant transformation $\zLambda$ on barred vector indices. This means that tilde spinor indices are identical to barred spinor indices (because $\dot\Lambda_\talpha{}^{\bbeta} = \delta_\talpha{}^\bbeta$), but it will be useful to keep the tilde notation anyway.

The simplest way of handling the constant $\zLambda$ transformation is to simply declare that we will treat it as a constant similarity transformation on all barred vector indices. Pushing that through $\cV_S$, we find
\begin{align}
\cV_\cM{}^\cA =
\begin{pmatrix}
1 & B_{M N} (-)^n \\
0 & 1
\end{pmatrix} 
\begin{pmatrix}
E_N{}^B & 0 \\
0 & E_B{}^N (-)^{bn+b}
\end{pmatrix}
\times
(\cV_S)_\cB{}^\cA
\end{align}
where $(\cV_S)_\cB{}^\cA$ has the non-vanishing components (aside from the identity)
\emph{written in the chiral basis} as
\begin{alignat}{2}
(\cV_S)_\rb{}^\alpha &= \sqrt{2} \,\eta_{bc} S^{c \alpha}~, &\qquad
(\cV_S)_\rbb{}^\balpha &= -\sqrt{2} \,\eta_{bc} S^{c \balpha}~, \eol
(\cV_S)^{\alpha \rb} &= - \sqrt{2} S^{b\alpha}~, &\qquad
(\cV_S)^{\balpha \rbb} &= -\sqrt{2} S^{b \balpha}~, \eol
(\cV_S)^{\alpha \beta} &= - S^{b \alpha} \eta_{b c} S^{c\beta}~, &\quad
(\cV_S)^{\balpha \bbeta} &= S^{b \balpha} \eta_{bc} S^{c \bbeta}~, \eol
(\cV_S)^{\alpha \bbeta} &= (\cV_S)^{\bbeta \alpha} 
    = S^{\alpha \bbeta} ~.
\end{alignat}
Note that because of the sign choice made in \eqref{E:tgamma.def}, the last term in \eqref{E:Sgauge} picks up an extra sign,
\begin{align}
S^{\ra \talpha} := 
    S^{\rbb \bbeta} (\Lambda^{-1})_{\rbb}{}^{\ra} (\Lambda^{-1})_\bbeta{}^\talpha 
    = -\frac{i}{10} (\gamma^\ra)^{\talpha \tbeta} \chi_\tbeta~.
\end{align}

Now imposing the section condition $\tilde \pa^M=0$, one can show that
\begin{align}
\mathring D_\cA := (\cV_S)_\cA{}^\cB D_\cB = 
\begin{pmatrix}
E_A{}^M \pa_M \\
0
\end{pmatrix}~.
\end{align}
Using $\cV_S$ to define shifted fluxes,
\begin{align}\label{E:ShiftedFlux}
\mathring \cF_{\cC \cB \cA} = (\cV_S)_\cC{}^{\cC'} (\cV_S)_\cB{}^{\cB'} (\cV_S)_\cA{}^{\cA'}
    \cF_{\cC' \cB' \cA'}~,
\end{align}
these turn out to be given in the toroidal basis as
\begin{subequations}
\begin{align}
\mathring \cF_{C B A}  &= H_{C B A}~, \\
\mathring \cF_{C B}{}^A &= C_{C B}{}^A + 2 \mathring D_{[C} (\cV_S)_{B]}{}^{A'} (\cV_S^{-1})_{A'}{}^A~, \\
\mathring \cF_{C}{}^{B A} &= \mathring D_C (\cV_S)^{[B| D} (\cV_S^{-1})_{D}{}^{|A]}
    + \mathring D_C (\cV_S)^{[B|}{}_D (\cV_S^{-1})^{D|A]} (-)^d ~, \\
\mathring \cF^{CBA} &= 0~.
\end{align}
\end{subequations}
The $H$-flux $H_{CBA}$ is given as usual by $H = \rd B = \frac{1}{3!} E^A E^B E^C H_{CBA}$, and $C_{CB}{}^A$ are the components of $C^A = \rd E^A = \frac{1}{2} E^B E^C C_{C B}{}^A$.
For the dilaton flux we find
\begin{align}\label{E:zFluxDil}
\mathring \cF_A 
    = -2 \mathring D_A \varphi + \mathring \cF_{A B}{}^B (-)^b~, \qquad
\mathring \cF^A    &= 
    \mathring \cF^A{}_{B}{}^{B} (-)^b
\end{align}
where we have related the DFT superdilaton $\Phi$ to the supergravity dilaton $\varphi$ via $\Phi = e^{-2 \varphi} \sdet(E_M{}^A)$. 

Henceforth we drop the $\mathring{}$ notation and denote simply $D_A = E_A{}^M \pa_M$.

\subsection{Torsion and $H$-flux constraints}
\label{S:TorsionHFlux}
The analysis of the torsion and $H$ curvatures is straightforward. We proceed by dimension.
It will be useful to write some expressions with DFT fluxes and connections that carry barred vector indices as intermediate formula. These should be dressed with $\zLambda$. Rather than clutter formulae with numerous $\zLambda$ we will simply write such barred vector indices as tilded vector indices, i.e. $V_{\tra} = \zLambda_{\tra}{}^\rbb V_\rbb$.

\paragraph{Dimension $\leq 0$.}
At dimension $-\tfrac{1}{2}$ and $0$, the $H$-flux is given by
\begin{subequations}
\begin{align}
H_{\hgamma \hbeta \halpha} &= H_{\gamma \tbeta a} = 0~, \\
H_{\gamma \beta a} &= \frac{1}{\sqrt{2}} \cF_{\gamma \beta \ra}
    = \frac{k}{\sqrt 2} (\gamma_a)_{\gamma \beta}, \\
H_{\tgamma \tbeta a} 
    &= \frac{1}{\sqrt{2}} \cF_{\ol{\gamma\beta} \tra}
    = -\frac{k}{\sqrt{2}} (\gamma_{a})_{\tgamma \tbeta}~.
\end{align}
\end{subequations}
The dimension 0 torsion components are
\begin{subequations}
\begin{align}
T_{\gamma \beta}{}^a &= C_{\gamma \beta}{}^a = \frac{1}{\sqrt 2} \cF_{\gamma \beta}{}^\ra = \frac{k}{\sqrt 2} (\gamma^a)_{\gamma \beta}~,\\
T_{\tgamma \tbeta}{}^a &= C_{\tgamma \tbeta}{}^a 
    = \frac{1}{\sqrt 2} \cF_{\ol{\gamma\beta}}{}^{\tra}
    = \frac{k}{\sqrt 2} (\gamma^a)_{\tgamma \tbeta}~, \\
T_{\gamma \tbeta}{}^a &= C_{\gamma \tbeta}{}^a = 0
\end{align}
\end{subequations}

\paragraph{Dimension 1/2.}
At dimension 1/2, we find
\begin{align}
H_{\hgamma b a} &= \mathring \cF_{\hgamma b a}
    = \frac{1}{2} \Big(
        \mathring \cF_{\hgamma \rb \ra}
        + \mathring \cF_{\hgamma \rb \tra}
        + \mathring \cF_{\hgamma \trb \ra}
        + \mathring \cF_{\hgamma \trb \tra}
    \Big)~, \eol
C_{\hgamma b}{}^a &= 
    \mathring \cF_{\hgamma b}{}^a
    = \frac{1}{2} \Big(
        \mathring \cF_{\hgamma \rb}{}^{\ra}
        + \mathring \cF_{\hgamma \rb}{}^{\tra}
        + \mathring \cF_{\hgamma \trb}{}^{\ra}
        + \mathring \cF_{\hgamma \trb}{}^{\tra}
    \Big)~, \eol
0 &= \mathring \cF_{\hgamma}{}^{b a} = 
    \frac{1}{2} \Big(
        \mathring \cF_{\hgamma}{}^{\rb \ra}
        + \mathring \cF_{\hgamma}{}^{\rb \tra}
        + \mathring \cF_{\hgamma}{}^{\trb \ra}
        + \mathring \cF_{\hgamma}{}^{\trb \tra}
    \Big)~. \label{E:SugraRed.Constraint1}
\end{align}
Making use of the third relation, one can show that
\begin{align}
H_{\hgamma b a} &= \mathring \cF_{\hgamma \rb \tra} - \mathring \cF_{\hgamma \trb \ra}
    = \cF_{\hgamma \rb \tra} - \cF_{\hgamma \trb \ra}
    = 0~, \eol
C_{\gamma b}{}^a
    &= \mathring \cF_{\gamma \trb}{}^{\ra}  + \mathring \cF_{\gamma \trb}{}^{\tra}
    = - \Omega_{\gamma \trb}{}^{\tra} \quad \implies \quad
\Omega_{\gamma b}{}^a := \Omega_{\gamma \trb}{}^{\tra} 
\quad \implies \quad T_{\gamma b}{}^a = 0 \eol
C_{\tgamma b}{}^a 
    &= \mathring \cF_{\bgamma \rb}{}^{\ra} + \mathring \cF_{\bgamma \rb}{}^{\tra}  
    = - \Omega_{\bgamma \rb}{}^\ra \quad \implies \quad
\Omega_{\tgamma b}{}^a := \Omega_{\bgamma \rb}{}^{\ra}
\quad \implies \quad T_{\tgamma b}{}^a = 0 
\end{align}
In the second and third lines, 
we have fixed the supergravity spin connection $\Omega_{\hgamma b}{}^a$
in terms of the DFT spin connection so that the corresponding torsion vanishes.
It is also useful to use \eqref{E:SugraRed.Constraint1} to show that
\begin{align}
\Omega_{\gamma \rb \ra} + \Omega_{\gamma \trb \tra} 
    = \frac{i k \sqrt 2}{5} (\gamma_{ba} \chi)_{\gamma}~, \qquad
\Omega_{\tgamma \rb \ra} + \Omega_{\tgamma \trb \tra} 
    = -\frac{i k \sqrt 2}{5} (\gamma_{ba} \chi)_{\tgamma}~.
\end{align}
The remaining torsion components $T_{\hgamma \hbeta}{}^\halpha$ can now be computed:
\begin{subequations}
\begin{align}
T_{\tgamma \beta}{}^\alpha &= T_{\tgamma \tbeta}{}^\alpha = 
T_{\gamma \tbeta}{}^{\talpha} = T_{\gamma \beta}{}^{\talpha} = 0~, \\
T_{\gamma \beta}{}^\alpha &= \frac{ik}{\sqrt 2} \Big(
    2 \,\chi_{(\gamma} \delta_{\beta)}{}^\alpha
    - (\gamma_a)_{\gamma \beta} (\gamma^a \chi)^{\alpha}\Big)~, \\
T_{\tgamma \tbeta}{}^\talpha &= \frac{ik}{\sqrt 2} \Big(
    2\,\chi_{(\tgamma} \delta_{\tbeta)}{}^\talpha
    - (\gamma_a)_{\tgamma \tbeta} (\gamma^a \chi)^{\talpha} \Big)~.
\end{align}
\end{subequations}
The spinor derivative of the dilaton follows from \eqref{E:zFluxDil} and is identified as the dilatini,
\begin{align}
D_\alpha \varphi = \frac{i k}{\sqrt 2} \chi_\alpha~, \qquad
D_\talpha \varphi = \frac{i k}{\sqrt 2} \chi_\talpha~.
\end{align}

\paragraph{Dimension 1.}
At dimension 1, we have
\begin{align}
\mathring \cF_{c b a} = H_{cba}~, \qquad
\mathring \cF_{c b}{}^a = C_{c b}{}^a~, \qquad
\mathring \cF_c{}^{ba} = 0~, \qquad
\mathring \cF^{c b a} = 0~.
\end{align}
This implies, using $T_{cba} = 0$ to compute $\Omega_{cba}$,
\begin{align}
\mathring \cF_{\trc \rb \ra} &= 
    \frac{1}{2 \sqrt 2} \Big(
    H_{c b a} + C_{c b a} - C_{c a b} - C_{b a c}
    \Big) 
    = \frac{1}{2 \sqrt 2} \Big(
    H_{c b a} 
    - 2\, \Omega_{c b a} 
    \Big)~,\eol
\mathring \cF_{\rc \trb \tra} 
    &= \frac{1}{2 \sqrt 2} \Big(
    H_{c b a} - C_{c b a} + C_{c a b} + C_{b a c}
    \Big) 
    = \frac{1}{2 \sqrt 2} \Big(
    H_{c b a} 
    + 2\, \Omega_{c b a} 
    \Big)~.
\end{align}
This gives the supergravity $\Omega$ in terms of the DFT one:
\begin{align}
\Omega_{c b a} &= \frac{1}{\sqrt 2} (\mathring \cF_{\rc \trb \tra} - \mathring \cF_{\trc \rb \ra})     = \frac{1}{\sqrt 2} (\Omega_{\trc \rb \ra} - \Omega_{\rc \trb \tra})
    - \eta_{c d} S^{d\gamma} \Omega_{\gamma \trb \tra}
    - \eta_{c d} S^{d\tgamma} \Omega_{\tgamma \rb \ra}
\end{align}
The other dimension 1 torsion components involve $T_{\hgamma b}{}^{\halpha}$.
First, we compute
\begin{align}
T_{\gamma b}{}^\alpha
    &= \sqrt{2} \mathring \cF_{\gamma \trb}{}^\alpha
    - \frac{1}{4} \Omega_{b c d} (\gamma^{c d})_{\gamma}{}^\alpha 
    = 
    - \frac{1}{4 \sqrt 2} (\mathring \cF_{\rb \trc\trd} + \mathring \cF_{\trb \rc \rd}) (\gamma^{c d})_{\gamma}{}^\alpha 
    = 
    - \frac{1}{8} H_{bcd} (\gamma^{c d})_{\gamma}{}^\alpha~,
\end{align}
and similarly
\begin{align}     
T_{\tgamma b}{}^\talpha
    &= + \frac{1}{8} H_{bcd} (\gamma^{c d})_{\tgamma}{}^\talpha~.
\end{align}
For the remaining torsion components, we have
\begin{align}
T_{\tgamma b}{}^\alpha = C_{\tgamma b}{}^\alpha 
    &= \sqrt{2} \mathring \cF_{\bgamma \trb}{}^\alpha 
    = 
    - \sqrt{2} (\cV_S)^{\alpha \bbeta} \cF_{\bbeta\bgamma \trb}
    = k \sqrt{2} \,S^{\alpha \tbeta} (\gamma_b)_{\tbeta\tgamma}~,
\end{align}
and similarly,
\begin{align}
T_{\gamma b}{}^\talpha &= -k \sqrt{2} S^{\talpha \beta} (\gamma_b)_{\beta\gamma}~.
\end{align}

We emphasize that the results given here for the torsion and $H$-flux tensors are actually gauge-invariant under the higher $\lambda$ symmetries and the right Lorentz group (as they must be, since they are built out of invariant potentials) and could have been derived without imposing the gauge \eqref{E:Sgauge}.

\subsection{The spinorial supervielbein and Ramond-Ramond field strengths}
To completely characterize type II supergravity we must also derive the Ramond-Ramond field strengths. This will require some knowledge of the spinorial supervielbein.
It turns out to be easy enough to work in a generic gauge here, so we will not impose
\eqref{E:Sgauge}.

Recall that we had introduced the spinorial supervielbein in section \ref{S:OSpSpinor.FlatOSPspinors} as an operator $\cswV$ converting the super-Fock space to the spinor Fock space. Just as we decomposed the bosonic spinorial vielbein in \eqref{E:SpinorialVielbein.Decompose}, we will split up the spinorial supervielbein in a somewhat asymmetric way, writing
\begin{align}
\cswV = \mathbb S_B \times \cswV_{E\Lambda} \times \cswV_{S}~.
\end{align}
The first factor $\mathbb S_B$ is a pure Fock space operator:
\begin{align}
\mathbb S_B &= \exp\Big(-\tfrac{1}{4} \Gamma^{MN} B_{NM}\Big)=
\exp\Big(-\tfrac{1}{2} \psiGamma^M \psiGamma^N B_{NM}\Big)
\end{align}
Next, $\cswV_{E\Lambda}$ is an operator converting the Fock space to the spinor Fock space. It acts as
\begin{align}
\Gamma^\cM \cswV_{E\Lambda}  &= \cswV_{E\Lambda} \cdot \Gamma^\cA (\cV_{E\Lambda})_\cA{}^\cM~. 
\end{align}
Decomposing this explicitly gives
\begin{align}
\Gamma^M \cswV_{E\Lambda}  &= 
    \frac{1}{\sqrt 2} \gamma^\ra \cswV_{E\Lambda} E_\ra{}^M
    + \frac{1}{\sqrt 2} \gamma_* \cswV_{E\Lambda} \bar\gamma^{\rba} E_\rba{}^M
    + \sqrt{2}\, \gamma_* \cswV_{E\Lambda} \bar\gamma_* b^\alpha E_\alpha{}^M
    + \sqrt{2}\, \gamma_* \cswV_{E\Lambda} \bar\gamma_* b^\balpha E_\balpha{}^M
    ~, \eol
(-)^m \Gamma_M \cswV_{E\Lambda}  &= 
    \frac{1}{\sqrt 2} E_{M}{}^{\ra} \gamma_\ra \cswV_{E\Lambda}
    + \frac{1}{\sqrt 2} E_{M}{}^{\rba} \gamma_* \cswV_{E\Lambda} \bar\gamma_{\rba}
    - \sqrt{2}\, E_M{}^{\alpha} \,\gamma_* \cswV_{E\Lambda} \bar\gamma_* b_\alpha 
    - \sqrt{2}\, E_M{}^{\balpha}\, \gamma_* \cswV_{E\Lambda} \bar\gamma_* b_\balpha ~.
\end{align}
As a sanity check, one can confirm that the Clifford algebra is satisfied.
Finally the factor $\cswV_S$ is given in terms of flat $\Gamma^\cA$ matrices as
\begin{align}
\cswV_S =
\exp\Big(
- \frac{1}{2} \Gamma_\alpha \Gamma_\bbeta S^{\bbeta \alpha}
- \frac{1}{4} \Gamma_\alpha \Gamma_\beta S^{\beta \alpha}
- \frac{1}{4} \Gamma_\balpha \Gamma_\bbeta S^{\ol{\beta \alpha}}
- \frac{1}{2} \sqrt{2} \Gamma_\alpha \Gamma_\rb S^{\rb \alpha}
- \frac{1}{2} \sqrt{2} \Gamma_\balpha \Gamma_\rbb S^{\ol{\rb \alpha}}
\Big)~.
\end{align}
This must be evaluated acting to the left on a bispinor-valued Fock space.

Now we can analyze the connection between the orthosymplectic Ramond-Ramond spinor field strength $\bra{\cF}$ and its flattened version. First, let us recall that $\bra{\cF}$ is given by the conventional expansion
\begin{align}\label{E:braF.Expansion}
\bra{\cF} = \sum_p \frac{1}{p!} \bra{0} \psiGamma^{M_1} \cdots \psiGamma^{M_p} \cF_{M_p \cdots M_1}~.
\end{align}
We have asserted its flattened version to be given by
\begin{align}\label{E:RRFS.Start}
\bra{\widehat{\slashed{\cF}}} = \bra{\cF} \cswV \,\Phi^{-1/2}
    = -4k
\begin{pmatrix}
0 & 0 \\
\bra{\fvac} b^\alpha b^{\balpha} & 0
\end{pmatrix}~.
\end{align}
Using the decomposed spinorial vielbein, this can be rewritten as
\begin{align}\label{E:F.TwoVersions}
\bra{\widehat{\slashed{\cF}}} \cswV_S^{-1} = \bra{\widehat\cF} \cswV_{E\Lambda} \,\Phi^{-1/2}~,
\end{align}
where $\bra{\widehat \cF} = \bra{\cF} \cswV_B$ is given by \eqref{E:braF.Expansion} with $\cF$ replaced by $\widehat\cF = \cF e^{-B}$. 

What has happened here is identical to what occurs in the bosonic analysis: the polyform $\cF = \sum_p \cF_p$ that naturally appears in double field theory is a complex of closed $p$-forms that transform into each other under the $B$-field gauge transformations. They are related to invariant field strengths $\widehat \cF_p$ by
\begin{align}
\widehat \cF = \sum_p \widehat \cF_p = \sum_p \cF_p \,e^{-B} = \cF \, e^{-B}~.
\end{align}
The polyform $\widehat \cF$ is not closed but obeys $\rd \widehat \cF = - \widehat\cF \wedge H$. 

We are now going to separately analyze the two sides of \eqref{E:F.TwoVersions}. Let's start with the left:
\begin{align}\label{E:F.TwoVersions.LHS}
\bra{\widehat{\slashed{\cF}}} \slashed{\cV}_S^{-1} 
&= -4k\,
\bra{\fvac}
\begin{pmatrix}
-i b^\balpha \chi_\alpha 
    & - \chi_\alpha \chi_\balpha \\
b^\alpha b^\balpha  + S^{\alpha \balpha}
    & i b^\alpha \chi_\balpha
\end{pmatrix} \eol
&= -4k\,
\bra{\fvac} \Big(
b \bar b'
+ i b \bar\chi'
+ i \chi \bar b'
+ \slashed{S}
- \chi \bar\chi'
\Big)~.
\end{align}
In the first line we have written the expression in Weyl notation and in the second line we have repeated it in Dirac notation. Recall here that in our convention for Dirac notation, a Majorana fermion $\psi$ decomposes as
$\psi= (\psi_\alpha, \psi^\alpha)$ and
$\bar\psi= \psi^T C = (\psi^\alpha, \psi_\alpha)$,
and similarly for barred indices, with these spinors denoted by primes. We emphasize that no gauge fixing was required here; this is a consequence of \eqref{E:RRFS.Start} being invariant under $\g{H}_L\times \g{H}_R$.

For the right-hand side of \eqref{E:F.TwoVersions}, we will need to do some work. 
First, we define, here in this section alone, flattened versions of the Fock space raising operators,
\begin{align}
\psiGamma^\ra := \psiGamma^M E_M{}^\ra~, \qquad
\psiGamma^\alpha := \psiGamma^M E_M{}^\alpha~, \qquad
\psiGamma^\balpha := \psiGamma^M E_M{}^\balpha~.
\end{align}
Note that we are using $E_M{}^\balpha$ rather than $E_M{}^\talpha$ here. This will be temporary but very convenient to avoid introducing just yet the Lorentz transformation taking us from tilde to barred spinor indices. Now we require the following lemmas:
\begin{subequations}
\label{E:FlattenPsiA}
\begin{align}
\label{E:FlattenPsiA.a}
\bra{0} \psiGamma^{\ra_1} \cdots \psiGamma^{\ra_p} \cswV_{E\Lambda} 
    &= \gamma^{\ra_p \cdots \ra_1} \bra{\slashed{Z}} \, \times E^{1/2} ~, \\[2ex]
\label{E:FlattenPsiA.b}
\bra{0} \psiGamma^{\ra_1} \cdots \psiGamma^{\ra_{p}} \psiGamma^\halpha \cswV_{E\Lambda} 
    &= \gamma_* \gamma^{\ra_{p} \cdots \ra_1} \bra{\slashed{Z}} \bar\gamma_* b^\halpha \times E^{1/2} ~, \\[2ex]
\label{E:FlattenPsiA.c}
\bra{0} \psiGamma^{\ra_1} \cdots \psiGamma^{\ra_{p}} \psiGamma^\halpha \psiGamma^\hbeta\cswV_{E\Lambda} 
    &= \gamma^{\ra_{p} \cdots \ra_1} \bra{\slashed{Z}} 
    b^\halpha b^\hbeta \times E^{1/2} ~.
\end{align}
\end{subequations}
These expressions involve the flat bispinor vacuum
\begin{align}\label{E:slashedZ.vac}
\bra{\slashed{Z}}  := \bra{0} \cswV_{E\Lambda}
    = \frac{1}{2^{5/2}} \bra{\fvac} \slashed{\Lambda}
\end{align}
where $\slashed{\Lambda}$ is the \emph{32-component Dirac bispinor} corresponding to the Lorentz transformation $\Lambda_\ra{}^{\rbb}$. Note that the vielbein superdeterminant $E$ is given by
\begin{align}
E = \sdet \Big( E_M{}^\ra, E_M{}^\alpha, E_M{}^\talpha\Big) = \sdet \Big( E_M{}^\ra, E_M{}^\alpha, E_M{}^\balpha\Big)~,
\end{align}
and can be equally written with $E_M{}^{\talpha}$ in place of $E_M{}^\balpha$, since $\Lambda_\talpha{}^\bbeta$ lies in the connected part of the Lorentz group.
The proofs of \eqref{E:FlattenPsiA.a}--\eqref{E:FlattenPsiA.c} are fairly straightforward.
One can easily show that \eqref{E:FlattenPsiA.c} follows from 
\eqref{E:FlattenPsiA.b}, which follows from \eqref{E:FlattenPsiA.a}.
The proof of \eqref{E:FlattenPsiA.a} is inductive.
The initial step is 
\begin{align}
\bra{0} \cswV_{E\Lambda}
    = \frac{1}{2^{5/2}} \bra{\fvac} \slashed{\Lambda} \times E^{1/2}~.
\end{align}
To show this, we first decompose $\cswV_{E\Lambda}$ into the product of $\cswV_{E}$ and $\cswV_\Lambda$ where $E_M{}^A = (E_M{}^a, E_M{}^\alpha, E_M{}^\talpha)$ lies in the connected part $\g{GL}^+(10|32)$. Then 
$\bra{0}\cswV_E=  E^{1/2} \times \frac{1}{2^{5/2}} \bra{\fvac} \mathbf 1$ follows exactly as in the component calculation. Then the action of $\cswV_\Lambda$ decomposes into two parts: one is the vectorial $\Lambda_\ra{}^{\rbb}$ transformation, which generates the Dirac bilinear $\slashed{\Lambda}$; the second is the connected spinorial transformation $\dot\Lambda_\talpha{}^{\bbeta}$, which generates its determinant (which is unity) upon acting on the spinorial Fock vacuum. The inductive step follows by observing
\begin{align}
\bra{0} \psiGamma^{[\ra_1} \cdots \psiGamma^{\ra_{p}]} \cswV_{E\Lambda} 
    &= \frac{1}{2} \gamma^{[\ra_p} \Big(\bra{0} \psiGamma^{\ra_1} \cdots \psiGamma^{\ra_{p-1}]} \cswV_{E\Lambda} 
    \Big)
    + \frac{1}{2} \gamma_* \Big(\bra{0} \psiGamma^{[\ra_1} \cdots \psiGamma^{\ra_{p-1}} \cswV_{E\Lambda} 
    \Big) \bar\gamma^{\rbb} \Lambda_{\rbb}{}^{\ra_p]} \eol
    &= E^{1/2} \Big(\frac{1}{2} \gamma^{\ra_p \cdots \ra_{1}} \bra{\slashed{Z}}
    + \frac{1}{2} \gamma_* \gamma^{[\ra_{p-1} \cdots \ra_1} \bra{\slashed{Z}} \bar\gamma^{\rbb} \Lambda_{\rbb}{}^{\ra_p]} \Big)\eol
    &= E^{1/2} \Big(\frac{1}{2} \gamma^{\ra_p \cdots \ra_{1}} \bra{\slashed{Z}}
    + \frac{1}{2} \gamma_* \gamma^{[\ra_{p-1} \cdots \ra_1} \gamma_* \gamma^{\ra_p]} \bra{\slashed{Z}} \Big)\eol
    &= E^{1/2} \gamma^{\ra_p \cdots \ra_{1}} \bra{\slashed{Z}}~.
\end{align}

Now we can apply the lemma \eqref{E:FlattenPsiA}. Let the $p$-form complex be given in flat indices by
\begin{align}
\widehat \cF = \sum_p \frac{1}{p!} \bra{0} \psiGamma^{A_1} \cdots \psiGamma^{A_P} \widehat \cF_{A_P \cdots A_1}
\end{align}
where for now we continue to use $E_M{}^A = (E_M{}^a, E_M{}^\alpha, E_M{}^\balpha)$ to flatten indices. Comparing this expression to \eqref{E:F.TwoVersions.LHS}, it is clear that the only non-vanishing components are
$\widehat\cF_{\ra_P \cdots \ra_1}$, 
$\widehat\cF_{\alpha \,\ra_{p-1} \cdots \ra_1}$, 
$\widehat\cF_{\balpha \,\ra_{p-1} \cdots \ra_1}$, and
$\widehat\cF_{\alpha \bbeta \,\ra_{p-2} \cdots \ra_1}$.
This means that we can directly compute
\begin{align}
\bra{\widehat\cF} \cswV_{E\Lambda} \,\Phi^{-1/2}
&= \frac{1}{2^{5/2}} e^{\varphi} \bra{\fvac} \sum_p \Big(
\frac{1}{(p-2)!} \gamma^{\ra_{p-2} \cdots \ra_1} \slashed{\Lambda} \,\,b^{\alpha} b^{\bbeta}
    \widehat\cF_{\bbeta \alpha\, \ra_{p-2} \cdots \ra_1}
+ \frac{1}{(p-1)!} \gamma^{\ra_{p-1} \cdots \ra_1} \slashed{\Lambda} \,\,b^{\alpha}
    \widehat\cF_{\alpha\, \ra_{p-1} \cdots \ra_1}
    \eol & \qquad
+ \frac{1}{(p-1)!} \gamma^{\ra_{p-1} \cdots \ra_1} \slashed{\Lambda} \,\,b^{\balpha}
    \widehat\cF_{\balpha\, \ra_{p-1} \cdots \ra_1}
+ \frac{1}{p!} \gamma^{\ra_{p} \cdots \ra_1} \slashed{\Lambda}\,\,
    \widehat\cF_{\ra_{p} \cdots \ra_1}
    \Big)~.
\end{align}
The factors of $\gamma_*$ in \eqref{E:FlattenPsiA} vanish when we account for the chirality of the bispinor vacuum and the even/odd degree of the $p$-forms $\cF_p$.
To fully match this to \eqref{E:F.TwoVersions.LHS}, it helps significantly to first rewrite this expression to use 32-component Dirac indices. We arrange them as follows:
\begin{align}
\bra{\widehat\cF} \cswV_{E\Lambda} \,\Phi^{-1/2}
&= 
\frac{1}{2^{5/2}} e^{\varphi} \bra{\fvac} \sum_p \Big(
\frac{1}{(p-2)!} \gamma^{\ra_{p-2} \cdots \ra_1} \slashed{\Lambda} \,\,
    (\bar b' \widehat{\slashed{\cF}}_{\ra_{p-2} \cdots \ra_1} b)
- \frac{1}{(p-1)!} \gamma^{\ra_{p-1} \cdots \ra_1} \slashed{\Lambda} \,\,
    (\bar\Psi^{\widehat \cF}_{\ra_{p-1} \cdots \ra_1} b)
    \eol & \qquad
+ \frac{1}{(p-1)!} \gamma^{\ra_{p-1} \cdots \ra_1} \slashed{\Lambda} \,\,
    (\bar b' \Psi'{}^{\widehat \cF}_{\ra_{p-1} \cdots \ra_1}) 
+ \frac{1}{p!} \gamma^{\ra_{p} \cdots \ra_1} \slashed{\Lambda}\,\,
    \widehat\cF_{\ra_{p} \cdots \ra_1}
    \Big)
\end{align}
where we introduce notation for the Dirac spinors and bispinors here:
\begin{align}
\Psi'{}^{\widehat \cF}_{\ra_{p-1} \cdots \ra_1} &=
\begin{pmatrix}
\widehat\cF_{\balpha\, \ra_{p-1} \cdots \ra_1} \\[1ex]
0
\end{pmatrix}~, \qquad
\bar\Psi^{\widehat \cF}_{\ra_{p-1} \cdots \ra_1}  =
\begin{pmatrix}
0 & \widehat\cF_{\alpha\, \ra_{p-1} \cdots \ra_1}
\end{pmatrix}~, \eol
\widehat{\slashed{\cF}}_{\ra_{p-2} \cdots \ra_1}
&=
\begin{pmatrix}
0 & \widehat \cF_{\balpha \alpha\, \ra_{p-2} \cdots \ra_1} \\[1ex]
0 & 0
\end{pmatrix}~,
\end{align}
Next we rewrite \eqref{E:F.TwoVersions.LHS} as a sum over $\gamma$-matrices, using the completeness relation
\begin{align}\label{E:CompletenessRelation}
\slashed \cO
    = \frac{1}{32} \sum_p \frac{1}{p!} 
        \gamma^{\ra_{p} \cdots \ra_{1}} \slashed{\Lambda} \times
        \Tr \Big(\slashed{\cO}  \slashed\Lambda^{-1} \gamma_{\ra_{1} \cdots \ra_{p}}\Big)
\end{align}
where $\slashed{\cO}$ is a bispinor with a left spinor index on the left and a right spinor index on the right. Applying this to  \eqref{E:F.TwoVersions.LHS} gives
\begin{align}
\bra{\widehat{\slashed{\cF}}} \slashed{\cV}_S^{-1} 
&= -\frac{k}{8} \sum_p \frac{1}{p!}
\gamma^{\ra_{p} \cdots \ra_{1}} \slashed{\Lambda}
\times
\Tr\Big(\bra{\fvac} (
b \bar b'
+ i b \bar\chi'
+ i \chi \bar b'
+ \slashed{S}
- \chi \bar \chi'
\Big) \slashed\Lambda^{-1} \gamma_{\ra_{1} \cdots \ra_{p}} \Big)
\end{align}
Now comparing terms is straightforward:
\begin{subequations}
\begin{align}
\widehat{\slashed{\cF}}_{\ra_{p-2} \cdots \ra_1}
    &= -\frac{k}{2\sqrt 2} e^{-\varphi} \,\slashed{\Lambda}^{-1} \gamma_{\ra_1 \cdots \ra_{p-2}}(1-\gamma_*) \\
\bar\Psi^{\widehat \cF}_{\ra_{p-1} \cdots \ra_1}
    &= -\frac{i k}{\sqrt 2} e^{-\varphi} \,
    \bar \chi' \slashed{\Lambda}^{-1}
    \gamma_{\ra_1 \cdots \ra_{p-1}} \\
\Psi'{}^{\widehat \cF}_{\ra_{p-1} \cdots \ra_1}
    &= +\frac{i k}{\sqrt 2} e^{-\varphi} \,
    \slashed{\Lambda}^{-1} \gamma_{\ra_1 \cdots \ra_{p-1}} \chi \\
\widehat\cF_{\ra_p \cdots \ra_1}
    &= -\frac{k}{\sqrt 2} e^{-\varphi}\,\Tr\Big(
    (\slashed{S} - \chi \bar\chi')
    \slashed{\Lambda}^{-1} \gamma_{\ra_1 \cdots \ra_p}\Big) 
\end{align}
\end{subequations}
An explicit Weyl projector is needed in the first field strength to emphasize that only certain chiralities are present; for the other field strengths, the chirality restriction follows from the chirality of $\chi$ and $S$.

These field strengths are not quite the ones we want, because the barred spinor indices transform under the right-handed Lorentz transformations. If we apply the Lorentz transformation $\dot\Lambda$ to the field strengths (and also to $\chi_\balpha$), we find that $\Lambda$ is replaced by the constant $\zLambda$ above.
Now, what kind of object is $\slashed\zLambda$? It is a constant spinorial Lorentz transformation corresponding to the vectorial $\zLambda_\ra{}^{\rbb}$.
For the four duality frames, it is given by
\begin{subequations}
\begin{alignat}{2}
&\text{IIB} &\qquad 
\slashed\zLambda^{-1} &=
\begin{pmatrix}
\delta_\talpha{}^\beta & 0 \\
0 & \delta^\talpha{}_\beta 
\end{pmatrix}~, \\
&\text{IIB$^*$} &\qquad 
\slashed\zLambda^{-1}  &=
\begin{pmatrix}
\delta_\talpha{}^\beta & 0 \\
0 & -\delta^\talpha{}_\beta 
\end{pmatrix}~, \\
&\text{IIA} &\qquad 
\slashed\zLambda^{-1} &=
\begin{pmatrix}
0 & \delta_{\talpha \beta} \\
\delta^{\talpha \beta} & 0 
\end{pmatrix}~, \\
&\text{IIA$^*$} &\qquad 
\slashed\zLambda^{-1}  &=
\begin{pmatrix}
0 & \delta_{\talpha \beta} \\
-\delta^{\talpha \beta} & 0 
\end{pmatrix}~.
\end{alignat}
\end{subequations}
The overall sign choice for each of these is ambiguous, and corresponds to the $\mathbb Z_2$ ambiguity in the Ramond-Ramond sector. The choice we have made above is rather simple and lets us easily make contact with the results of Wulff \cite{Wulff:2013kga} .
In Weyl notation, we find
\begin{align}\label{E:RRF.dim0}
\widehat \cF_{\tbeta \alpha\, \ra_{p-2} \cdots \ra_1}
&= -\frac{k}{\sqrt 2} e^{-\varphi} \times
\begin{cases}
\delta_\tbeta{}^\beta (\gamma_{\ra_1 \cdots \ra_{p-2}})_{\beta \alpha} & \text{IIB/IIB$^*$  ($p$ odd)} \\
\delta_{\tbeta \beta} (\gamma_{\ra_1 \cdots \ra_{p-2}})^\beta{}_{\alpha} & \text{IIA/IIA$^*$ ($p$ even)} 
\end{cases}
\end{align}
with the relative sign conventions for starred and unstarred duality frames chosen to match here. The reversed ordering of the vector indices follows from the construction, and writing it this way eliminates additional sign factors in \cite{Wulff:2013kga}.
For the dimension 1/2 components, we find
\begin{subequations}\label{E:RRF.dim1/2}
\begin{align}
\widehat \cF_{\alpha \,a_{p-1} \cdots a_1} &= -\frac{ik}{\sqrt 2} e^{- \varphi} \times
\begin{cases}
+\chi_\tbeta \,\delta^\tbeta{}_\beta \,
    (\gamma_{a_{1} \cdots a_{p-1}} )^{\beta}{}_{\alpha}
    & \text{IIB ($p$ odd)} \\[1ex]
-\chi_\tbeta \,\delta^\tbeta{}_\beta \,
    (\gamma_{a_{1} \cdots a_{p-1}} )^{\beta}{}_{\alpha}
    & \text{IIB$^*$ ($p$ odd)} \\[1ex]
+\chi_\tbeta \,\delta^{\tbeta \beta} \,
    (\gamma_{a_{1} \cdots a_{p-1}} )_{\beta \alpha} 
    & \text{IIA ($p$ even)} \\[1ex]
-\chi_\tbeta \,\delta^{\tbeta \beta} \,
    (\gamma_{a_{1} \cdots a_{p-1}} )_{\beta \alpha} 
    & \text{IIA$^*$ ($p$ even)}
\end{cases} ~, \\[2ex]
\widehat \cF_{\talpha \,a_{p-1} \cdots a_1} &= 
\frac{ik}{\sqrt 2} e^{- \varphi} \times
\begin{cases}
\delta_\talpha{}^\alpha \,(\gamma_{a_{1} \cdots a_{p-1}} )_{\alpha}{}^{\beta}\chi_\beta
    & \text{IIB/IIB$^*$ ($p$ odd)} \\[1ex]
\delta_{\talpha \alpha} \,(\gamma_{a_{1} \cdots a_{p-1}} )^{\alpha \beta} \,\chi_\beta
    & \text{IIA/IIA$^*$ ($p$ even)}
\end{cases} ~.
\end{align}
\end{subequations}
Finally, the component with all vector indices, which is often called the supercovariant field strength, is given by
\begin{align}\label{E:RRF.dim1}
\widehat\cF_{\ra_p \cdots \ra_1}
&= -\frac{k}{\sqrt 2} e^{-\varphi} \times
\begin{cases}
    S^{\alpha \tbeta} \,\delta_{\tbeta}{}^{\beta} 
        (\gamma_{\ra_1 \cdots \ra_p})_{\beta \alpha}
        - \chi_\alpha \chi_\tbeta\, \delta^\tbeta{}_\beta 
        (\gamma_{\ra_1 \cdots \ra_p})^{\beta \alpha}    
& \text{IIB ($p$ odd)} \\[1ex]
S^{\alpha \tbeta} \,\delta_{\tbeta}{}^{\beta} 
        (\gamma_{\ra_1 \cdots \ra_p})_{\beta \alpha}
        + \chi_\alpha \chi_\tbeta\, \delta^\tbeta{}_\beta 
        (\gamma_{\ra_1 \cdots \ra_p})^{\beta \alpha}
& \text{IIB$^*$ ($p$ odd)} \\[1ex]
S^{\alpha \tbeta} \,\delta_{\tbeta\beta}
        (\gamma_{\ra_1 \cdots \ra_p})^\beta{}_{\alpha}
        - \chi_\alpha \chi_\tbeta \,\delta^{\tbeta \beta} 
            (\gamma_{\ra_1 \cdots \ra_p})_\beta{}^{\alpha}
& \text{IIA ($p$ even)} \\[1ex]
S^{\alpha \tbeta} \,\delta_{\tbeta\beta}
        (\gamma_{\ra_1 \cdots \ra_p})^\beta{}_{\alpha}
        + \chi_\alpha \chi_\tbeta \,\delta^{\tbeta \beta} 
            (\gamma_{\ra_1 \cdots \ra_p})_\beta{}^{\alpha}
& \text{IIA$^*$ ($p$ even)}
\end{cases}
\end{align}
The results for IIA and IIB match the expressions given by Wulff \cite{Wulff:2013kga} for $k = -i\sqrt{2}$ (see also \cite{Wulff:2016tju}), up to a redefinition
\begin{align}
S_{\rm Wulff}^{\alpha \tbeta} = \frac{16k}{\sqrt 2} \,S^{\alpha \tbeta}~, \qquad
S_{\rm Wulff}^{\tbeta \alpha} = -\frac{16k}{\sqrt 2} \,S^{\tbeta \alpha }~.
\end{align}
Note that the bispinor of \cite{Wulff:2013kga, Wulff:2016tju} is chosen to be antisymmetric as opposed to symmetric.

Although the starred supergravities were not explicitly given in \cite{Wulff:2016tju},
they can be easily derived by analytic continuation from the unstarred cases. Starting from an unstarred supergravity, one relaxes the reality condition on all fields, then makes an imaginary similarity transformation in superspace,
\begin{align}
D_\talpha \rightarrow -i D_\talpha~, \qquad E_M{}^\talpha \rightarrow i E_M{}^\talpha~, \qquad
\chi_\talpha \rightarrow -i \chi_\talpha~, \qquad S^{\alpha \tbeta} \rightarrow i S^{\alpha\tbeta}~.
\end{align}
and then reimposes the original reality condition.\footnote{Alternatively, one can \emph{not} make the similarity transformation but simply Wick rotate $E_M{}^\talpha$ to now be imaginary (and all consequences of this). This keeps the the type II formulae, including the supersymmetry algebra, unchanged.}
The transformation of the dilatini follows because they are the spinor derivative of the dilaton, and the redefinition of $S$ follows from its embedding in the DFT supervielbein. Alternatively, both redefinitions arise by keeping the higher dimension torsion conditions in section \ref{S:TorsionHFlux} unchanged. (The dimension 0 torsion has flipped sign as we have mentioned in \eqref{E:TypeIISusyAlgebra}, so that the supersymmetry algebra in the tilde sector has the opposite sign.)
But we also must alter the Ramond-Ramond sector with an imaginary factor, because otherwise the dimension 0 constraint \eqref{E:RRF.dim0} would imply that the field strengths are imaginary. We choose to flip
\begin{align}
\cF \rightarrow i \cF
\end{align}
so that the dimension 0 constraint \eqref{E:RRF.dim0} is unchanged; this is then responsible for the well-known sign flip of the Ramond-Ramond Lagrangian.
One can then easily check that the conditions for the starred supergravities in \eqref{E:RRF.dim1/2} and \eqref{E:RRF.dim1}  follow from their unstarred analogues.

\subsection{Summary of democratic type II superspace}
\label{S:TypeIISS.Demo}
Let us now summarize the results for the democratic type II superspace that emerges from double field theory. It consists of a supervielbein $E_M{}^A$, a Kalb-Ramond super two-form $B_{MN}$, a scalar dilaton $e^{-2 \varphi}$, and a set of Ramond-Ramond super $(p-1)$-forms $\widehat \cC_{M_1 \cdots M_{p-1}}$ with $p$ even for IIA/IIA$^*$ and $p$ odd for IIB/IIB$^*$.

The supervielbein decomposes into a graviton 1-form $E_M{}^a$ and two gravitini $E_M{}^\alpha$ and $E_M{}^\talpha$ (both Majorana), where the $\talpha$ index is either the same chirality as $\alpha$ or opposite, depending on the duality frame:
\begin{align}
E_M{}^\talpha =
\begin{cases}
E'_M{}^{\alpha} & \text{IIB/IIB$^*$} \\[1ex]
E_{M \alpha} & \text{IIA/IIA$^*$}
\end{cases}~.
\end{align}
We employ tilde $\gamma$-matrices given by
\begin{align}
(\gamma^c)_{\talpha \tbeta} = 
\begin{cases}
\phantom{+} (\gamma^c)_{\alpha \beta} & \text{IIB} \\[1ex]
- (\gamma^c)_{\alpha \beta} & \text{IIB${}^*$} \\[1ex]
- (\gamma^c)^{\alpha \beta} & \text{IIA} \\[1ex]
\phantom{+} (\gamma^c)^{\alpha \beta} & \text{IIA${}^*$}
\end{cases}~, \qquad
(\gamma^c)^{\talpha \tbeta} = 
\begin{cases}
\phantom{+} (\gamma^c)^{\alpha \beta} & \text{IIB} \\[1ex]
- (\gamma^c)^{\alpha \beta} & \text{IIB${}^*$} \\[1ex]
- (\gamma^c)_{\alpha \beta} & \text{IIA} \\[1ex]
\phantom{+} (\gamma^c)_{\alpha \beta} & \text{IIA${}^*$}
\end{cases}~.
\end{align}
The supervielbein is subject to local Lorentz transformations
\begin{align}
\delta E_M{}^a = -E_M{}^b \lambda_b{}^a~, \qquad
\delta E_M{}^\alpha = -\frac{1}{4} E_M{}^\beta (\gamma^{ab})_{\beta}{}^\alpha \lambda_{ab}~, \qquad
\delta E_M{}^\talpha = -\frac{1}{4} E_M{}^\tbeta (\gamma^{ab})_{\tbeta}{}^\talpha \lambda_{ab}~.
\end{align}
The Lorentz group is restricted to $\g{SO}^+(9,1)$ and gauged by a composite spin connection $\Omega_{M A}{}^B$. The Kalb-Ramond two-form and Ramond-Ramond $p$-forms transform as
\begin{align}
\delta B &= \rd \tilde \xi~, \\
\delta \widehat \cC_{p-1} &= \rd \widehat \lambda_{p-2} + \widehat \lambda_{p-4} \wedge H~.
\end{align}
The torsion tensors $T^A$ and field strengths $H$ and $\widehat\cF_p$ are given by
\begin{alignat}{2}
T^A &= \rd E^A + E^B \wedge \Omega_B{}^A 
    &\,
    &= \frac{1}{2} E^B E^C T_{CB}{}^A~, \\
H &= \rd B 
    &\,
    &= \frac{1}{3!} E^A E^B E^C H_{CBA}~, \\
\widehat \cF_p &= \rd \widehat \cC_{p-1} + \widehat \cC_{p-3} \wedge H
    &\,
    &=\frac{1}{p!} E^{A_1} \cdots E^{A_p} \widehat \cF_{A_p \cdots A_1}~.
\end{alignat}

The non-vanishing torsion tensors are given through dimension 1 by
\begin{subequations}
\begin{alignat}{3}
T_{\alpha \beta}{}^c &= \frac{k}{\sqrt 2}\, (\gamma^c)_{\alpha \beta}~, &\qquad
T_{\talpha \tbeta}{}^c &= \frac{k}{\sqrt 2}\, (\gamma^c)_{\talpha \tbeta}~, \\
T_{\gamma \beta}{}^\alpha &= \frac{ik}{\sqrt 2} \Big(
    2 \,\chi_{(\gamma} \delta_{\beta)}{}^\alpha
    - (\gamma_a)_{\gamma \beta} (\gamma^a \chi)^{\alpha}\Big)~, &\quad
T_{\tgamma \tbeta}{}^\talpha &=  \frac{ik}{\sqrt 2} \Big(
    2\,\chi_{(\tgamma} \delta_{\tbeta)}{}^\talpha
    - (\gamma_a)_{\tgamma \tbeta} (\gamma^a \chi)^{\talpha} \Big)~, \\
T_{\gamma b}{}^\alpha
    &= - \frac{1}{8} H_{bcd} \,(\gamma^{c d})_{\gamma}{}^\alpha~, &\quad
T_{\tgamma b}{}^\talpha
    &= \frac{1}{8} H_{bcd} \,(\gamma^{c d})_{\tgamma}{}^\talpha~, \\
T_{\tgamma b}{}^\alpha 
    &= k \sqrt{2} \,S^{\alpha \tbeta} \,(\gamma_b)_{\tbeta\tgamma}~, &\qquad
T_{\gamma b}{}^\talpha &= -k \sqrt{2} \,S^{\talpha \beta} \,(\gamma_b)_{\beta\gamma}~.
\end{alignat}
\end{subequations}
The dilatini $\chi_\alpha$ and $\chi_\talpha$ are given by the spinor derivatives of the dilaton
\begin{align}
D_\alpha \varphi = \frac{i k}{\sqrt 2} \chi_\alpha~, \qquad
D_\talpha \varphi = \frac{i k}{\sqrt 2} \chi_\talpha~.
\end{align}
The non-vanishing components of the Kalb-Ramond field strength are
\begin{align}
H_{\gamma \beta a} 
    = \frac{k}{\sqrt 2} (\gamma_a)_{\gamma \beta}, \qquad
H_{\tgamma \tbeta a} 
    = -\frac{k}{\sqrt{2}} (\gamma_{a})_{\tgamma \tbeta}~, \qquad H_{abc}~.
\end{align}
The non-vanishing components of $\widehat \cF_{A_1\cdots A_p}$ are given in
\eqref{E:RRF.dim0}, \eqref{E:RRF.dim1/2}, and \eqref{E:RRF.dim1}, which we do not repeat here.
The supercovariant Ramond-Ramond bispinor is defined by
\begin{align}
S^{\alpha \tbeta} 
    &= -\frac{e^\varphi}{16 k \sqrt 2} \times
    \begin{cases}
    \sum_p
        \frac{1}{p!} \widehat \cF_{a_1 \cdots a_p} (\gamma^{a_1 \cdots a_p})^{\alpha \beta}\, \delta_\beta{}^\tbeta
        & \text{IIB/IIB$^*$ ($p$ odd)} \\[2ex]
    \sum_p
        \frac{1}{p!} \widehat \cF_{a_1 \cdots a_p} (\gamma^{a_1 \cdots a_p})^\alpha{}_\beta \, \delta^{\beta \tbeta}
        & \text{IIA/IIA$^*$ ($p$ even)}
    \end{cases}
\end{align}
To match conventions with \cite{Wulff:2013kga}, one should take $k=-i\sqrt{2}$.

A subtle issue here is that the $p$-forms in the Ramond-Ramond sector are not only democratic in the sense of including the dual forms \cite{Bergshoeff:2001pv}, but they also include so-called ``over the top'' forms \cite{Howe:2015hpa}, superforms of rank greater than the spacetime dimension.
This can be traced back to our use of the completeness relation for the $\gamma$-matrices \eqref{E:CompletenessRelation} in the expressions for the field strengths.
For IIB/IIB$^*$, the odd rank field strengths run from $\widehat \cF_1$ to $\widehat \cF_{11}$, where $\widehat \cF_{11}$ has one or two fermionic form indices. The latter is built from a 10-form $\cC_{10}$ \cite{Bergshoeff:2005ac,Bergshoeff:2007ma}.\footnote{Note that there are two 10-forms discussed in these references: a doublet and a quadruplet, which are implied by $\g{SU}(1,1)$ covariance, that is taking the full set of S- and T-duality transformations into account. Similar results are implied by $E_{11}$ \cite{Kleinschmidt:2003mf}. Here we have only a single (and singlet) 10-form implied by T-duality.} Similarly, IIA/IIA$^*$ runs from $\widehat \cF_0$ to $\widehat \cF_{12}$. The superform $\widehat \cC_{11}$ has no bosonic part. The 0-form field strength $\widehat \cF_0$ (which has no potential in a conventional sense) is the Romans mass \cite{Romans:1985tz}.

\subsection{Relation between superspace and DFT component parametrizations}
\label{S:RelationSuperToComps}
As a final step in this section, we will give the dictionary between the type II superspace and DFT component parametrizations of the DFT supervielbein. Incidentally, this will also yield as the dictionary between the component fields of type II supergravity and their
DFT analogues.

First, let us give a special decomposition of the square supervielbein 
$E_M{}^A = (E_M{}^\ra, E_M{}^\halpha)$ used in \eqref{E:V.BELambdaS}:
\begin{align}
E_M{}^A =
\begin{pmatrix}
\delta_m{}^n & 0 \\[1ex]
\Xi_\hmu{}^n & \delta_\hmu{}^\hnu
\end{pmatrix} \times
\begin{pmatrix}
e_n{}^\rb & \psi_n{}^\hbeta \\[1ex]
0 & \phi_\hnu{}^\hbeta
\end{pmatrix} 
= 
\begin{pmatrix}
e_m{}^\ra & \psi_m{}^\halpha \\[1ex]
\Xi_\hmu{}^n e_n{}^a & \,\,\phi_\hmu{}^\halpha + \Xi_\hmu{}^n \psi_n{}^\halpha
\end{pmatrix}~.
\end{align}
We presume $\phi$ and $e$ are invertible, so the inverse $E_A{}^M$ is given by
\begin{align} 
E_A{}^M =
\begin{pmatrix}
e_\ra{}^m + \psi_\ra{}^\hbeta \phi_\hbeta{}^\hnu \Xi_\hnu{}^m 
    & - \psi_\ra{}^\hbeta \phi_\hbeta{}^\hmu \\[1ex]
-\phi_\halpha{}^\hnu \Xi_\hnu{}^n  & \phi_\halpha{}^\hmu
\end{pmatrix}~, \qquad
\psi_\ra{}^\hbeta := e_\ra{}^m \psi_m{}^\hbeta~.
\end{align}
The field $e_m{}^\ra$ is the left-handed vielbein. Analogous formulae can be written down with $e_m{}^\rba = e_m{}^\rb \Lambda_\rb{}^\rba$ and $\psi_\rba{}^{\halpha} = \Lambda_\rba{}^\rb \psi_\rb{}^{\halpha}$.

The field $\phi_\hmu{}^\halpha$ appearing here can be identified with the same field in \eqref{E:SuperVDecompose}.
For the bosonic double vielbein in $\cV_0$ of \eqref{E:SuperVDecompose}, we use the decomposition \eqref{E:genericVchiral.bosonic}. Then $e_m{}^\ra$ defined in both formulations coincide, and $b_{mn}$ in \eqref{E:genericVchiral.bosonic} coincides with $B_{mn}$ (as we would wish). The remaining dictionary of fields in \eqref{E:SuperVDecompose} is as follows.
The fields that live purely in superspace (i.e. the components of $\cV_{-2}$ and $\cV_{-1}$ ) are given by
\begin{align}
\cB_{\hmu \hnu} &= B_{\hmu \hnu} - \Xi_{(\hmu}{}^m B_{m \hnu)}
    ~, \eol
\Xi^m{}_\hnu&= \phi_\hnu{}^\halpha E_\halpha{}^m = -\Xi_\hnu{}^m~, \eol
\Xi_{m \hnu} &= \phi_\hnu{}^\halpha E_{\halpha}{}^N B_{m N} = B_{m \hnu} -\Xi_\hnu{}^n B_{mn}~.
\end{align}
The constituents of $\cV_{+1}$ are given by
\begin{alignat}{2}
\Psi_\ra{}^\beta &= \frac{1}{\sqrt 2} e_\ra{}^m \psi_m{}^\beta + \sqrt{2} S_\ra{}^{\beta}~, 
&\qquad
\Psi_\rba{}^\beta &= \frac{1}{\sqrt 2} e_\rba{}^m \psi_m{}^\beta~, 
\eol
\Psi_\rba{}^\bbeta &= \frac{1}{\sqrt 2} e_\rba{}^m \psi_m{}^\bbeta + \sqrt{2} S_\rba{}^{\bbeta}~, 
&\qquad
\Psi_\ra{}^\bbeta &= \frac{1}{\sqrt 2} e_\ra{}^m \psi_m{}^\bbeta~
\end{alignat}
The component DFT gravitini lie in the second column. Taking the $\gamma$-trace of the first column gives the dilatini relations
\begin{align}
\rho_\alpha = \frac{1}{\sqrt 2} \psi_m{}^\beta (\gamma^m)_{\beta \alpha} + i \sqrt{2} \chi_\alpha~, \qquad
\rho_\balpha = \frac{1}{\sqrt 2} \psi_m{}^\bbeta (\bar\gamma^m)_{\bbeta \balpha} + i \sqrt{2} \chi_\balpha~,
\end{align}
where $\gamma^m := \gamma^\ra e_\ra{}^m$ and $\bar\gamma^m = \bar\gamma^\rba e_\rba{}^m$.
Finally the components of $\cV_{+2}$ are
\begin{align}
\cS^{\alpha \beta} &= S^{\alpha \beta} + S^{\rc (\alpha} \psi_\rc{}^{\beta)}~, \qquad
\cS^{\ol{\alpha \beta}} = S^{\ol{\alpha \beta}} 
    + S^{\rbc (\balpha} \psi_\rbc{}^{\bbeta)}~, \\
\cS^{\alpha \bbeta} &= S^{\alpha \bbeta} 
    + \frac{1}{2} S^{\rb \alpha} \psi_\rb{}^{\bbeta}
    + \frac{1}{2} S^{\rbb \bbeta} \psi_\rbb{}^{\alpha}~.
\end{align}
The last relation can be rewritten in terms of $\zcS$ as
\begin{align}
\zcS^{\alpha\bbeta} &= S^{\alpha \bbeta} 
    - \frac{1}{2} g^{mn} \,\psi_m{}^{\alpha} \psi_n{}^{\bbeta}~.
\end{align}
This relates the two component expressions for the Ramond-Ramond bispinor.

It is an interesting exercise to check this relation explicitly using the two different expressions we have for $\slashed{\widehat F}$. The expression in component DFT is
\eqref{E:Fslash.DFT}. To derive the analogous expression in type II supergravity requires
a bit of work. First, recall that the polyform $\widehat F$ is given as an expansion in
tangent space components, i.e.
\begin{align}
\widehat F &= \sum_p \frac{1}{p!} \,\rd x^{m_1} \cdots \rd x^{m_p} \widehat F_{m_p \cdots m_1} \eol
    &= \sum_p \frac{1}{p!} \,\rd x^{m_1} \cdots \rd x^{m_p} 
        \,e_{m_1}{}^{\ra_1} \cdots e_{m_{p-2}}{}^{\ra_{p-2}}
        \Big(
    e_{m_{p-1}}{}^{\ra_{p-1}} e_{m_p}{}^{\ra_p} \widehat F_{\ra_p \cdots \ra_1}
    + p \, 
        e_{m_{p-1}}{}^{\ra_{p-1}} \psi_{m_p}{}^{\alpha} 
        \widehat F_{\alpha \,\ra_{p-1} \cdots \ra_1}
    \eol & \qquad\qquad
    + p \, 
        e_{m_{p-1}}{}^{\ra_{p-1}} \psi_{m_p}{}^{\balpha} 
        \widehat F_{\balpha \,\ra_{p-1} \cdots \ra_1}
    + p (p-1)\, \psi_{m_{p-1}}{}^{\alpha} \psi_{m_p}{}^{\bbeta} 
        \widehat F_{\bbeta \alpha \,\ra_{p-1} \cdots \ra_1}
    \Big)~.
\end{align}
Using the explicit expressions \eqref{E:RRF.dim0}, \eqref{E:RRF.dim1/2}, and \eqref{E:RRF.dim1}, and then rewriting as a bispinor using
\begin{align}
\slashed{\widehat F}
    = e^{\varphi} \sum_p \frac{1}{p!} \widehat F_{\ra_1 \cdots \ra_p} \gamma^{\ra_1 \cdots \ra_p} \slashed{Z}
    = \frac{1}{2^{5/2}} e^{\varphi} \sum_p \frac{1}{p!} \widehat F_{\ra_1 \cdots \ra_p} \gamma^{\ra_1 \cdots \ra_p} \slashed{\Lambda}
\end{align}
we find
\begin{align}
\slashed{\widehat F}
    &= -4k \Big(
    \slashed{\cS}
    - \chi \,\bar\chi'
    + \frac{i}{2} \gamma^m \psi_m \, \bar \chi'
    + \frac{i}{2} \chi \, \bar\psi'_m \bar\gamma^m 
    - \frac{i}{2}  \psi_m \, \bar \chi'\bar\gamma^m
    + \frac{i}{2} \gamma^m \chi \, \bar\psi'_m
    \eol & \quad
    -\frac{1}{4} \gamma^{mn} \psi_m \, \bar\psi'_n
    + \frac{1}{4} \psi_m \, \bar\psi'_n \bar\gamma^{mn} 
    + \frac{1}{2} \gamma^{[m} \psi_m \, \bar\psi'_n \bar\gamma^{n]} \Big)
\end{align}
This indeed matches the component DFT result \eqref{E:Fslash.DFT} upon applying the dictionary above.

\section{Conclusion and open problems}
\label{S:Conclusion}

The goal in this paper was to define the supergeometry of type II double field theory in superspace and to provide the tools to derive component DFT as well as conventional type II superspace, building on the progress made in type I \cite{Butter:2021dtu}. As in type I, we showed that one can take all torsion and curvature tensors to vanish through dimension two, except for the basic dimension zero constant torsion tensor associated with supersymmetry.  It was crucial here that the tangent space group be extended beyond the double Lorentz group.

We further built on the discussion of orthosymplectic spinors in \cite{Cederwall:2016ukd}, giving a prescription for their constant field strength, as well as a complete description for how to transform between ``curved'' and ``flat'' orthosymplectic spinors. An interesting result that came for free was a unified description of democratic type II superspace in section \ref{S:TypeIISS.Demo}.

There are several additional avenues one could pursue. We highlight a few below.

\paragraph{Generalized type II DFT and supergravity.}
In analyzing the Bianchi identities in section \ref{S:TypeIISS}, we took pains to separate the Bianchi identities for $\cT_{\cA \cB \cC}$ from those of the dilaton torsion $\cT_\cA$. We showed that no data from $\cT_\cA$ was necessary to constrain the dilaton-independent torsion and curvature tensors through dimension two. Moreover, the constraints imposed on the dilaton torsion and dilaton curvatures (through dimension two at least) could be deduced purely from their Bianchi identities without supposing the existence of a superdilaton field.

The reason we organized the analysis in this way is that Tseytlin and Wulff have shown that if one starts with conventional type II superspace and supposes only the constraints of $\kappa$-symmetry, one can show that one arrives at generalized type II supergravity \cite{Wulff:2016tju} (see also \cite{Arutyunov:2015mqj}). This is a formulation of supergravity where the dilatini are not presumed to arise from the spinor derivative of a dilaton. In addition to the usual supergravity fields, one finds two vectors at dimension 1, denoted $X_a$ and $K^a$. The latter is a Killing vector of the entire supergravity multiplet. In standard supergravity, $K^a$ vanishes and $X_a = D_a \varphi$.

The constraints we employed for double field theory in superspace are the exact analogues of the $\kappa$-symmetry constraints, and so if we were to \emph{not} suppose the existence of the superdilaton, it is more or less obvious that we should recover the generalized type II supergeometry of Tseytlin and Wulff by generalizing the bosonic discussion of \cite{Sakatani:2016fvh} to superspace. The idea is one would replace $\pa_\cM \log\Phi$ in the definition of the dilaton flux $\cF_A$ with a more general vector that does not obey the section condition. The dilatini, the vector field $X_a$, and the Killing vector $K^a$ should then turn out to be various components of this quantity. It would be interesting to work this out in detail.

\paragraph{Green-Schwarz action in double superspace.}
The work of Tseytlin and Wulff was inspired by the question of whether $\kappa$-symmetry of the Green-Schwarz superstring uniquely selected out the constraints of 10D supergravity. In light of that, it would be natural to try to formulate a GS-type action using the doubled supergeometry introduced here. Such a doubled worldsheet was discussed by Park already in a flat space background \cite{Park:2016sbw} (see also the work by Bandos \cite{Bandos:2015cha}). This was generalized to include fermions to second order (along with the Ramond-Ramond fields) for more general backgrounds by Sakamoto and Sakatani \cite{Sakamoto:2018krs}. A complete formulation should be possible, and one would expect the $\kappa$-symmetry of this action to lead to the constraints we have imposed on $\cT_{\cA\cB\cC}$. A very similar idea (in Hamiltonian language) was the motivation for \cite{Hatsuda:2014qqa}, where constraints on torsion were motivated also by $\kappa$-symmetry. This is a topic we are currently exploring.

\paragraph{Non-geometric backgrounds.}
Recent work has emphasized that conventional supergravity backgrounds (i.e. with an invertible metric and two-form) are not the only allowed generalized metrics in double field theory \cite{Morand:2017fnv,Cho:2019ofr,Park:2020ixf} (see also \cite{Berman:2019izh} in exceptional field theory). One such example is the Gomis-Ooguri non-relativistic string background \cite{Gomis:2000bd}, which in the classification scheme of \cite{Morand:2017fnv} is a $(1,1)$ non-Riemannian background. This was already discussed in the context of the doubled Green-Schwarz superstring some time ago \cite{Park:2016sbw}. This should be able to be addressed in type II supersymmetric double field theory, both at the component level and in superspace.\footnote{We thank Jeong-Hyuck Park for discussions on this point.}

\paragraph{Exceptional superspace.}
A final fascinating topic is the generalization to exceptional field theory. It is well-known that the exceptional groups $\g{E}_{D(D)}$ possess $\g{O}(D,D)$ subgroups, and so one might consider the embedding of $\g{O}(10,10)$ into $\g{E}_{11}$. In fact, an early discussion of the Ramond-Ramond sector of double field theory (at least in the IIA duality frame) was found in a level decomposition \cite{Rocen:2010bk} of the $\g{E}_{11}$ formulation of West \cite{West:2003fc,West:2010ev} (see also the recent work on $\g{E}_{11}$ \cite{Bossard:2017wxl, Bossard:2021ebg} and its supersymmetrization \cite{Bossard:2019ksx}). A natural avenue would be to explore the lower levels of the supersymmetric version of $\g{E}_{11}$ by attempting to encode $\g{OSp}(10,10|64)$ within it. This would involve geometrizing the abelian $\bra{\lambda}$ transformation of the Ramond-Ramond sector and encoding the Ramond-Ramond \emph{potentials} $\bra{\cC}$ into the supervielbein itself. Such an approach has already been explored implicitly by Hatsuda, Kamimura, and Siegel \cite{Hatsuda:2014aza}, who attempted to geometrize the Ramond-Ramond charges that appear in the supersymmetry algebra. This could be a fascinating springboard to formulating $\g{E}_{11}$ in superspace.

\section*{Acknowledgements}
It is a pleasure to thank Falk Hassler and Jeong-Hyuck Park for discussions related to this work, and especially Yuho Sakatani for alerting me to \cite{Sakamoto:2018krs}.
This work was partially supported by the NSF under grant NSF-2112859 and the Mitchell Institute for Fundamental Physics and Astronomy at Texas A\&M University.

\appendix

\section{Conventions for spinors and $\gamma$-matrices of $\g{SO}(9,1)$}
\label{A:Conventions}
We summarize below our conventions for $\g{SO}(9,1)$. The metric $\eta^{ab}$ has a mostly positive signature. The 32-component gamma matrices $\gamma^a$, charge conjugation matrix $C$, and chirality matrix $\gamma_*$ are given by
\begin{align}
\gamma^a =
\begin{pmatrix}
0 & (\gamma^a)_{\alpha \beta} \\
(\gamma^a)^{\alpha \beta} & 0
\end{pmatrix}~, \qquad
\gamma_* =
\begin{pmatrix}
\delta_\alpha{}^\beta & 0 \\
0 & -\delta^\alpha{}_\beta
\end{pmatrix}~, \qquad
C &=
\begin{pmatrix}
0 & \delta^\alpha{}_\beta \\
\delta_\alpha{}^\beta & 0
\end{pmatrix}
\end{align}
and obey
\begin{align}
\{\gamma^a,\gamma^b\} = 2 \,\eta^{ab}~, \qquad
(\gamma^a)^T = C \gamma^a C^{-1}~, \qquad
(\gamma_*)^T = -C \gamma_* C^{-1}~, \qquad
\gamma^{[a_1} \cdots \gamma^{a_{10}]} = \veps^{a_1 \cdots a_{10}} \gamma_*
\end{align}
The matrices $(\gamma^a)_{\alpha\beta}$ and $(\gamma^a)^{\alpha\beta}$ are 16-component Weyl sigma matrices in 10D, but we denote them $\gamma^a$ for convenience.
A 32-component Dirac spinor $\psi$ is written in terms of 16-component Weyl spinors as
\begin{align}
\psi =
\begin{pmatrix}
\psi_\alpha \\
\psi^\alpha
\end{pmatrix}
\end{align}
For a Majorana spinor, its Dirac conjugate is the same as its Majorana conjugate, with
\begin{align}
\bar \psi = \psi^T C =
\begin{pmatrix}
\psi^\alpha & \psi_\alpha
\end{pmatrix}~.
\end{align}

Our $\gamma$-matrices are chosen to be Majorana (i.e. real), so that the $B$-matrix is the identity. Then the Majorana condition is simply $\psi^* = \psi$. An explicit realization of 10D Majorana $\gamma$-matrices follows from the Majorana representation of $\g{SO}(8)$. There one employs 8-component Weyl matrices $(\sigma^i)_{\ul{\alpha} \ul{\dot\beta}}$, where here (and here alone) $\ul{\alpha}$ and $\ul{\dalpha}$ denote the $\rep{8}_s$ and $\rep{8}_c$ of $\g{SO}(8)$. All $\g{SO}(8)$ vector and spinor indices are raised and lowered with the identity matrix, and we choose $\sigma^1 (\sigma^2)^T \cdots \sigma^7 (\sigma^8)^T = 1$.\footnote{See e.g. \cite{Park:2022pjv} for a discussion of $\g{SO}(8)$ $\sigma$-matrices and the connection to the octonions.} 
Then for 10D Majorana gamma matrices, we have in $8\times 8$ block notation
\begin{align}
(\gamma^i)_{\alpha \beta} &= (\gamma^i)^{\alpha \beta} =
\begin{pmatrix}
0 & \sigma^i\\
(\sigma^i)^T & 0
\end{pmatrix}~,  \quad
(\gamma^0)_{\alpha \beta} = - (\gamma^0)^{\alpha\beta} =
\begin{pmatrix}
1 & 0 \\
0 & 1
\end{pmatrix}~, \quad
(\gamma^9)_{\alpha \beta} = (\gamma^9)^{\alpha\beta} =
\begin{pmatrix}
1 & 0 \\
0 & -1
\end{pmatrix}~.
\end{align}
These 10D $\gamma$-matrices have a natural lift to 11D, where we identify
$\gamma_* = \gamma^{10}$ and $C_{\rm 11D} = C \gamma_*$.

For the right sector of $\g{SO}(1,9)$, we take very similar conventions, with
\begin{align}
\bar\gamma^\rba = \gamma_* \gamma^\ra \, \quad \implies \quad
(\bar\gamma^\rba)_{\ol{\alpha \beta}} = (\gamma^\ra)_{\alpha\beta}~, \qquad
(\bar\gamma^\rba)^{\ol{\alpha \beta}} = -(\gamma^\ra)^{\alpha\beta}~.
\end{align}
When we write $\g{SO}(1,9)$ spinors in Dirac form, we prime the spinors following
\cite{Jeon:2012hp}:
\begin{align}
\psi' =
\begin{pmatrix}
\psi_\balpha \\
\psi^\balpha
\end{pmatrix}~, \qquad
\bar \psi' = \psi'{}^T C =
\begin{pmatrix}
\psi^\balpha & \psi_\balpha
\end{pmatrix}~.
\end{align}

\section{The general decomposition of the supervielbein}
\label{A:SuperV}
In this appendix, we elaborate on how to construct the general form of the supervielbein in type II double field theory in a way that is well-adapted to conventional superspace. This construction in many ways mirrors that of the bosonic case discussed in section \ref{S:TypeIISS.Bosonic}, but involves key additional elements. 

First, we identify some specific components of the generalized vielbein as follows:
\begin{align}
\cV_\alpha{}^M = E_\alpha{}^M~, \qquad
\cV_\balpha{}^M = E_\balpha{}^M~, \qquad
\cV_\ra{}^M = \frac{1}{\sqrt 2} \cE_\ra{}^M~, \qquad
\cV_\rba{}^M = \frac{1}{\sqrt 2} \bar\cE_\rba{}^M~.
\end{align}
In analogy to the two sets of invertible vielbeins $e_\ra{}^m$ and $\bar e_\rba{}^m$ in the bosonic case, we propose the following two sets of invertible vielbeins:
\begin{align}
\cE_A{}^M = 
\begin{pmatrix}
\cE_\ra{}^M \\
E_\alpha{}^M \\
E_\balpha{}^M \\
\end{pmatrix}~, \qquad
\bar \cE_A{}^M = 
\begin{pmatrix}
\cE_\rba{}^M \\
E_\alpha{}^M \\
E_\balpha{}^M
\end{pmatrix}~,
\end{align}
That these should be invertible is a relatively mild assumption; it is true outside of a measure zero set. However, note that both $E_\alpha{}^M$ and $E_\balpha{}^M$ appear in $\cE_A{}^M$ and in $\bar \cE_A{}^M$. Moreover, they $\cE_A{}^M$ and $\bar\cE_A{}^M$ do not transform uniformly under the Lorentz groups.  $\cE_A{}^M$ involves a left vector and both types of spinor, while $\bar \cE_A{}^M$ involves a right vector.

Let their inverses be denoted $\cE_M{}^A$ and $\bar \cE_M{}^A$. We are going to again give special names to some of these components:
\begin{align}
\cE_M{}^A =
\begin{pmatrix}
E_M{}^\ra & \cE_M{}^\alpha & E_M{}^\balpha
\end{pmatrix}~, \quad
\bar \cE_M{}^A =
\begin{pmatrix}
E_M{}^\rba & E_M{}^\alpha & \bar \cE_M{}^\balpha
\end{pmatrix}~.
\end{align}
Following the bosonic case, we identify
a matrix $\Lambda_\rb{}^\rba$ and its inverse $\Lambda_\rba{}^\rb$ via
\begin{align}\label{E:LambdaDefinition}
E_M{}^\rba = E_M{}^\rb \Lambda_\rb{}^\rba~, \qquad
E_M{}^\rb = E_M{}^\rba \Lambda_\rba{}^\rb~.
\end{align}
The fact that such a simple relation exists follows from the fact that $\cE_\alpha{}^M = \bar \cE_\alpha{}^M$ and $\cE_\balpha{}^M = \bar \cE_\balpha{}^M$.
The fact that $\Lambda_\rb{}^\rba$ is again a Lorentz transformation follows from the orthosymplectic structure, which we will show in due course.

The components $E$ that we have identified obey a certain modified orthonormality condition, which we can write as
\begin{align}
\begin{pmatrix}
E_\alpha{}^M & E_\balpha{}^M & E_\ra{}^M & E_\rba{}^M
\end{pmatrix} 
\begin{pmatrix}
E_M{}^\beta \\ E_M{}^\bbeta \\ E_M{}^\rb \\ E_M{}^{\rbb}
\end{pmatrix}
= \begin{pmatrix}
  \delta_\alpha{}^\beta & 0 & 0 & 0\\
  0 & \delta_\balpha{}^\bbeta & 0 & 0\\
  0 & 0 & \delta_\ra{}^\rb & \Lambda_\ra{}^{\rbb} \\
  0 & 0 & \Lambda_\rba{}^\rb & \delta_\rba{}^{\rbb}
  \end{pmatrix}~.
\end{align}
One can think of $(E_\alpha{}^M, E_\balpha{}^M, E_\ra{}^M)$ as a square matrix whose inverse is $(E_M{}^\alpha, E_M{}^{\balpha}, E_M{}^\ra)$. The same statement holds with the vector index $\ra$ replaced with a barred vector index $\rba$, because
\begin{align}\label{E:LambdaDefinition.Inverse}
E_\rba{}^M = \Lambda_\rba{}^\rb E_\rb{}^M~, \qquad
E_\ra{}^M = \Lambda_\ra{}^\rbb E_\rbb{}^M~.
\end{align}
Because of the relations
\eqref{E:LambdaDefinition}
and
\eqref{E:LambdaDefinition.Inverse},
it is straightforward to switch between the barred and unbarred versions of $E_M{}^\ra$ and $E_\ra{}^M$.

We want to re-express $\cE$ and $\bar \cE$  in terms of the superfields $E$.
Start by defining
\begin{align}
S^{\ra \beta} := -\frac{1}{2} \cE^{\ra M} \bar \cE_M{}^\beta~, \qquad
\bar S^{\rba \bbeta} := -\frac{1}{2} \bar \cE^{\rba M} \cE_M{}^\bbeta~.
\end{align}
Then one can work out a complete dictionary as follows:
\begin{subequations}
\begin{alignat}{2}
\cE_M{}^\ra &= E_M{}^{\ra} ~, 
& \qquad
\bar \cE_M{}^\rba &= E_M{}^{\rba} ~, \\
\cE_M{}^\alpha &= E_M{}^\alpha + 2\,E_M{}^\rb S_\rb{}^\alpha~,
& \qquad
\bar \cE_M{}^\alpha &= E_M{}^\alpha~, \\
\cE_M{}^\balpha &= E_M{}^\balpha~, 
& \qquad
\bar \cE_M{}^\balpha &= E_M{}^\balpha + 2\,E_M{}^\rbb S_\rbb{}^\balpha~,
\end{alignat}
\end{subequations}
and inverses
\begin{subequations}
\begin{alignat}{2}
\cE_\alpha{}^M &= E_\alpha{}^M 
& \qquad
\bar\cE_\alpha{}^M &= E_\alpha{}^M ~,\\
\cE_\balpha{}^M &= E_\balpha{}^M 
& \qquad
\bar\cE_\balpha{}^M &= E_\balpha{}^M ~,\\
\cE_\ra{}^M &= E_\ra{}^M -2 S_\ra{}^\beta E_\beta{}^M 
& \qquad
\cE_\rba{}^M &= E_\rba{}^M -2 S_\rba{}^\bbeta E_\bbeta{}^M ~.
\end{alignat}
\end{subequations}
The fermionic superfields $S^{\ra\beta}$ and $S^{\ol{\ra\beta}}$ are the only additional information encoded in $\cE_A{}^M$ and $\bar \cE_A{}^M$ aside from the $E_M{}^A$ vielbeins
(and of course the matrix $\Lambda_\ra{}^\rbb$).

Now, the most general expressions for $D_\alpha$, $D_\balpha$, and $D_{\ha}$ are
\begin{subequations}
\begin{align}
D_\alpha &= E_\alpha{}^M \pa_M + \cV_{\alpha M} \pa^M~, \\
D_\balpha &= E_\balpha{}^M \pa_M + \cV_{\balpha M} \pa^M~, \\
D_\ra &= \frac{1}{\sqrt 2} \cE_\ra{}^M \pa_M + \cV_{\ra M} \pa^M~, \\
D_\rba &= \frac{1}{\sqrt 2} \bar\cE_\rba{}^M \pa_M + \cV_{\rba M} \pa^M~.
\end{align}
\end{subequations}
Let's focus first on $D_\alpha$. Using the property that
$\cV_\alpha{}^\cM \cV_\cM{}_\beta = 0$, one can show that
$E_\alpha{}^M \cV_{M \beta}$ is symmetric in $\alpha$ and $\beta$. This piece can be
identified as part of the Kalb-Ramond two-form, so that
\begin{align}
D_\alpha &= E_\alpha{}^M \Big(\pa_M - B_{MN} \pa^N\, (-)^n \Big)~.
\end{align}
Next using $\cV_\alpha{}^\cM \cV_{\cM \rb} = 0$ and
$\cV_\ra{}^\cM \cV_{\cM \rb} = \eta_{\ra \rb}$, one can show that
\begin{align}
D_\ra &= \frac{1}{\sqrt 2} \cE_\ra{}^M \Big(\pa_M - B_{MN} \pa^N\, (-)^n \Big)
    + \frac{1}{\sqrt 2} (-)^m \cE_M{}_\ra \pa^M
\end{align}
for the same $B$-field. Repeating these conditions with barred indices (as well as
mixed handedness conditions like $\cV_\ra{}^\cM \cV_{\cM \rbb} = 0$), leads to the conclusion
\begin{align}
D_\balpha &= E_\balpha{}^M \Big(\pa_M - B_{MN} \pa^N\, (-)^n \Big)~, \\
D_\rba &= \frac{1}{\sqrt 2} \bar\cE_\rba{}^M \Big(\pa_M - B_{MN} \pa^N\, (-)^n \Big)
    + \frac{1}{\sqrt 2} (-)^m \bar\cE_M{}_\rba \pa^M~,
\end{align}
with the same two-form $B_{MN}$, along 
with the additional requirement that $\Lambda_{\rb \rba} = - \Lambda_{\rba \rb}$. 
This identifies $\Lambda$ as a Lorentz transformation.

This is all quite similar to the bosonic case. The main complications arise when
analyzing $D^\alpha$ and $D^\balpha$. Without loss of generality, $D^\alpha$ can
be written
\begin{align}
D^\alpha
    &=  (\mathscr{Z}^{\alpha \beta} + S^{\rc \alpha} S_\rc{}^\beta) D_\beta
        + \mathscr{Z}^{\alpha \bbeta} D_\bbeta
        + (\mathscr{Z}^{\alpha \rb} + \sqrt{2} S^{\rb \alpha}) D_\rb
        + (\mathscr{Y}_M{}^\alpha + E_M{}^\alpha) \pa^M ~,
\end{align}
in terms of arbitrary $\mathscr{Y}$ and $\mathscr{Z}$ factors. The additional terms constitute shifts in the generic factors, and as we will show, almost all of the $\mathscr{Z}$
and $\mathscr{Y}$ terms vanish. 

For two vectors $X_\cA = X_\cA{}^\cM \pa_\cM$ and $Y_\cA = Y_\cA{}^\cM \pa_\cM$
introduce the inner product notation 
\begin{align}
\langle X_\cA, Y_\cB\rangle = X_\cA{}^\cM Y_\cB{}^\cN \eta_{\cN \cM} (-)^{bm}~.
\end{align}
Then we use
\begin{subequations}
\begin{align}
\delta_\alpha{}^\beta &= \langle D_\alpha, D^\beta\rangle
    = E_\alpha{}^M \mathscr{Y}_M{}^\beta + \delta_\alpha{}^\beta ~, \\
0 &= \langle D_\balpha, D^\beta\rangle
    = E_\balpha{}^M \mathscr{Y}_M{}^\beta~, \\
0 &= \langle D_\ra, D^\beta\rangle
    = \cZ^{\beta}{}_\ra 
    + \frac{1}{\sqrt 2} \cE_\ra{}^M \mathscr{Y}_M{}^\beta ~, \\
0 &= \langle D_\rba, D^\beta\rangle
    =  \frac{1}{\sqrt 2} \bar\cE_\rba{}^M \mathscr{Y}_M{}^\beta
\end{align}
\end{subequations}
to conclude that $\cY_M{}^\alpha$ and $\cZ^{\alpha \rb}$ both vanish.
Next we use
\begin{align}
0 &= \langle D^\alpha, D^\beta \rangle
    = \mathscr{Z}^{\alpha \beta} + S^{\rc \alpha} S_\rc{}^\beta
    - \mathscr{Z}^{\beta \alpha} - S^{\rc \beta} S_\rc{}^\alpha
    - 2 S^{\rb \alpha} S_\rb{}^\beta
    = 2 \mathscr{Z}^{[\alpha \beta]}
\end{align}
to conclude that $\mathscr{Z}^{\alpha \beta}$ is symmetric.
Similar equations hold for $D^\balpha$.
Then we use
\begin{align}
0 &= \langle D^\alpha, D^\bbeta \rangle
    = \mathscr{Z}^{\alpha \bbeta} 
    - \mathscr{Z}^{\bbeta \alpha }
\end{align}
to prove that $\mathscr{Z}^{\alpha \bbeta} = \mathscr{Z}^{\bbeta \alpha }$.
We relabel the remaining $\cZ$ factors as $-S$, giving
\begin{subequations}\label{E:nabla.alpha.up}
\begin{align}
D^\alpha
    &= - (S^{\alpha \beta} - S^{\rc \alpha} S_\rc{}^\beta) D_\beta
        - S^{\alpha \bbeta} D_\bbeta
        + \sqrt{2}\, S^{\rc \alpha} D_\rc
        + \bar \cE_M{}^\alpha \pa^M~, \\
D^\balpha
    &= - (S^{\ol{\alpha \beta}} - S^{\rbc \balpha} S_\rbc{}^\bbeta) D_\bbeta
        - S^{\balpha \beta} D_\beta
        + \sqrt{2}\, S^{\rbc \balpha} D_\rbc
        + \cE_M{}^\balpha \pa^M~.
\end{align}
\end{subequations}

Next, it will be useful to identify how these objects transform under the
$\g{H}_L \times \g{H}_R$ gauge transformations.
Because all superfields transform in a clear way under the double Lorentz group,
we focus on the infinitesimal $\lambda_\rb{}^\alpha$, $\lambda_\rbb{}^\balpha$, $\lambda^{\alpha\beta}$, and $\lambda^{\ol{\alpha\beta}}$ gauge symmetries.
From $\delta \cV_\alpha{}^\cM$,
$\delta \cV_\balpha{}^\cM$,
and $\delta \cV_\ha{}^\cM$, we deduce
\begin{align}
\delta E_\alpha{}^M = \delta E_\balpha{}^M = 0~, \qquad
\delta \cE_\ra{}^M = \sqrt{2} \,\lambda_\ra{}^\beta E_\beta{}^M~, \qquad
\delta \cE_\rba{}^M = \sqrt{2} \,\lambda_\rba{}^\bbeta E_\bbeta{}^M~, \qquad
\delta B_{MN} = 0
\end{align}
and then it follows that
\begin{subequations}
\begin{alignat}{2}
\delta \cE_M{}^\ra &=0~,
& \qquad
\delta \cE_M{}^\rba &=0, \\
\delta \cE_M{}^\alpha &= -\sqrt{2} \,\cE_M{}^\rb \lambda_\rb{}^\alpha~, 
&\qquad
\delta \bar\cE_M{}^\alpha &= 0~, \\
\delta \cE_M{}^\balpha &= 0~,
&\qquad
\delta \bar\cE_M{}^\balpha &= -\sqrt{2} \,\bar\cE_M{}^\rbb \lambda_\rbb{}^\balpha~.
\end{alignat}
\end{subequations}
This can be summed up as follows.
The superfields $S_\ra{}^\alpha$ and $S_\rba{}^\balpha$ simply shift as
\begin{align}
\delta S_\ra{}^\alpha = -\frac{1}{\sqrt2} \lambda_\ra{}^\alpha~, \qquad
\delta S_\rba{}^\balpha = -\frac{1}{\sqrt2} \lambda_\rba{}^\balpha~,
\end{align}
while all the $E_M{}^A$ factors are invariant,
\begin{align}
\delta E_M{}^A = \delta E_A{}^M = 0~.
\end{align}
From $\delta \cV^{\halpha \cM}$ and using the explicit expressions in \eqref{E:nabla.alpha.up}, we determine that
\begin{subequations}
\begin{align}
\delta S^{\alpha \beta} &= -\lambda^{\alpha\beta}
    + \sqrt{2}\, S^{c (\alpha} \lambda_\rc{}^{\beta)}~, \\
\delta S^{\ol{\alpha \beta}} &= -\lambda^{\ol{\alpha\beta}}
    + \sqrt{2}\, S^{\rbc (\balpha} \lambda_\rbc{}^{\bbeta)}~, \\
\delta S^{\alpha \bbeta} &= 0
\end{align}
\end{subequations}

Finally, we can rewrite the entire inverse supervielbein as a product of three factors
\begin{align}
\cV_\cA{}^\cM = 
(\cV_S^{-1})_\cA{}^\cB \times 
(\cV_{E\Lambda})_\cB{}^\cN \times 
(\cV_B^{-1})_\cN{}^\cM 
\end{align}
The third term is built out of the Kalb-Ramond super two-form,
\begin{align}
(\cV_B^{-1})_\cM{}^\cN =
\begin{pmatrix}
\delta_M{}^N & -B_{MN} (-)^n \\
0 & \delta^N{}_M
\end{pmatrix}~, \qquad
\end{align}
The second factor $\cV_{E\Lambda}$ is written, in a chiral decomposition of the indices, as
\begin{align}
(\cV_{E\Lambda})_\cA{}^\cM =
\renewcommand{\arraystretch}{1.5}
\left(\begin{array}{cc}
\frac{1}{\sqrt 2} E_\ra{}^M
    & \frac{1}{\sqrt 2} E_{M\ra} (-)^m \\
E_\alpha{}^M 
    & 0 \\ 
0  
    & E_M{}^\alpha \\ \hline
\frac{1}{\sqrt 2} E_\rba{}^M 
    & \frac{1}{\sqrt 2} E_{M \rba} (-)^m \\
E_\balpha{}^M 
    & 0 \\
0
    & E_M{}^\balpha 
\end{array}\right)~.
\end{align}
The $\cV_S$ factor is given, also in a chiral decomposition, as
\begin{align}
(\cV_S^{-1})_\cA{}^\cB &=
\renewcommand{\arraystretch}{1.5}
\left(\begin{array}{ccc|ccc}
\delta_\ra{}^\rb & -\sqrt{2} S_\ra{}^\beta & 0 
    & 0 & 0 & 0 \\
0 & \delta_\alpha{}^\beta & 0
    & 0 & 0 & 0 \\
\sqrt{2} S^{\rb \alpha} & -S^{\alpha\beta} - S^{\rc\alpha} S_\rc{}^\beta &  \delta^\alpha{}_\beta
    & 0 & -S^{\alpha \bbeta} & 0 \\ \hline
0 & 0 & 0
    &\delta_\rba{}^\rbb & -\sqrt{2} S_\rba{}^\bbeta & 0  \\
0 & 0 & 0
    & 0 & \delta_\balpha{}^\bbeta & 0 \\
0 & -S^{\balpha \beta}  & 0 
    & \sqrt{2} S^{\ol{\rb \alpha}} & -S^{\ol{\alpha\beta}} - S^{\rbc\balpha} S_\rbc{}^\bbeta &  \delta^\balpha{}_\bbeta
\end{array} \right).
\end{align}

\bibliography{library.bib}
\bibliographystyle{utphys_mod_v2}

\end{document}

%% file: sugracom_v3.tex
\newcommand {\cA}{{\cal A}}
\newcommand {\cB}{{\cal B}}
\newcommand {\cC}{{\cal C}}
\newcommand {\cD}{{\cal D}}
\newcommand {\cE}{{\cal E}}
\newcommand {\cF}{{\cal F}}

\newcommand {\cH}{{\cal H}}

\newcommand {\cJ}{{\cal J}}
\newcommand {\cK}{{\cal K}}

\newcommand {\cM}{{\cal M}}
\newcommand {\cN}{{\cal N}}
\newcommand {\cO}{{\cal O}}
\newcommand {\cP}{{\cal P}}

\newcommand {\cR}{{\cal R}}
\newcommand {\cS}{{\cal S}}
\newcommand {\cT}{{\cal T}}

\newcommand {\cV}{{\cal V}}
\newcommand {\cW}{{\cal W}}
\newcommand {\cX}{{\cal X}}
\newcommand {\cY}{{\cal Y}}
\newcommand {\cZ}{{\cal Z}}

\newcommand{\ra}{{\mathrm a}}
\newcommand{\rb}{{\mathrm b}}
\newcommand{\rc}{{\mathrm c}}
\newcommand{\rd}{{\mathrm d}}
\newcommand{\re}{{\mathrm e}}
\newcommand{\rf}{{\mathrm f}}

\newcommand{\bd}{{\dot{\beta}}}
\newcommand{\dalpha}{{\dot{\alpha}}}

\newcommand{\hm}{{\hat{m}}}
\newcommand{\hn}{{\hat{n}}}

\newcommand{\ha}{{\hat{a}}}
\newcommand{\hb}{{\hat{b}}}
\newcommand{\hc}{{\hat{c}}}
\newcommand{\hd}{{\hat{d}}}

\newcommand{\halpha}{{\hat{\alpha}}}
\newcommand{\hbeta}{{\hat{\beta}}}
\newcommand{\hgamma}{{\hat{\gamma}}}
\newcommand{\hdelta}{{\hat{\delta}}}

\newcommand{\veps}{\varepsilon}
\newcommand{\eps}{{\epsilon}}
\newcommand{\eol}{\notag \\}

\newcommand{\ul}{\underline}
\newcommand{\pa}{\partial}


%% file: superDFT2_paper.bbl
\providecommand{\href}[2]{#2}\begingroup\raggedright\begin{thebibliography}{10}

\bibitem{Siegel:1993xq}
W.~Siegel, ``{Two-Vierbein Formalism for String-Inspired Axionic Gravity},''
  {\em Phys. Rev. D} {\bfseries 47} (1993) 5453,
  [\href{http://arxiv.org/abs/hep-th/9302036}{{\ttfamily
  arXiv:hep-th/9302036}}].

\bibitem{Siegel:1993th}
W.~Siegel, ``{Superspace duality in low-energy superstrings},'' {\em Phys. Rev.
  D} {\bfseries 48} (1993) 2826,
  [\href{http://arxiv.org/abs/hep-th/9305073}{{\ttfamily
  arXiv:hep-th/9305073}}].

\bibitem{Hull:2009mi}
C.~M. Hull and B.~Zwiebach, ``{Double field theory},'' {\em JHEP} {\bfseries
  0909} (2009) 099, [\href{http://arxiv.org/abs/0904.4664}{{\ttfamily
  arXiv:0904.4664}}].

\bibitem{Hull:2009zb}
C.~M. Hull and B.~Zwiebach, ``{The gauge algebra of double field theory and
  Courant brackets},'' {\em JHEP} {\bfseries 0909} (2009) 090,
  [\href{http://arxiv.org/abs/0908.1792}{{\ttfamily arXiv:0908.1792}}].

\bibitem{Hohm:2010jy}
O.~Hohm, C.~M. Hull, and B.~Zwiebach, ``{Background independent action for
  double field theory},'' {\em JHEP} {\bfseries 1007} (2010) 016,
  [\href{http://arxiv.org/abs/1003.5027}{{\ttfamily arXiv:1003.5027}}].

\bibitem{Hohm:2010pp}
O.~Hohm, C.~M. Hull, and B.~Zwiebach, ``{Generalized metric formulation of
  double field theory},'' {\em JHEP} {\bfseries 1008} (2010) 008,
  [\href{http://arxiv.org/abs/1006.4823}{{\ttfamily arXiv:1006.4823}}].

\bibitem{Hohm:2011zr}
O.~Hohm, S.~K. Kwak, and B.~Zwiebach, ``{Unification of type-II strings and T
  duality},'' {\em Phys. Rev. Lett.} {\bfseries 107} (2011) ,
  [\href{http://arxiv.org/abs/1106.5452}{{\ttfamily arXiv:1106.5452}}].

\bibitem{Hohm:2011dv}
O.~Hohm, S.~K. Kwak, and B.~Zwiebach, ``{Double field theory of type II
  strings},'' {\em JHEP} {\bfseries 1109} (2011) 013,
  [\href{http://arxiv.org/abs/1107.0008}{{\ttfamily arXiv:1107.0008}}].

\bibitem{Jeon:2012kd}
I.~Jeon, K.~Lee, and J.-H. Park, ``{Ramond-Ramond cohomology and $O(D, D)$
  T-duality},'' {\em JHEP} {\bfseries 1209} (2012) 079,
  [\href{http://arxiv.org/abs/1206.3478}{{\ttfamily arXiv:1206.3478}}].

\bibitem{Jeon:2012hp}
I.~Jeon, K.~Lee, J.-H. Park, and Y.~Suh, ``{Stringy unification of type IIA and
  IIB supergravities under $N=2$ $D=10$ supersymmetric double field theory},''
  {\em Phys. Lett. B} {\bfseries 723} (2013) 245,
  [\href{http://arxiv.org/abs/1210.5078}{{\ttfamily arXiv:1210.5078}}].

\bibitem{West:2010ev}
P.~C. West, ``{$E_{11}$, generalised space-time and IIA string theory},'' {\em
  Phys. Lett. B} {\bfseries 696} (2011) 403,
  [\href{http://arxiv.org/abs/1009.2624}{{\ttfamily arXiv:1009.2624}}].

\bibitem{Rocen:2010bk}
A.~Roc{\'{e}}n and P.~C. West, ``{$E_{11}$, generalised space-time and IIA
  string theory; the $R \otimes R$ sector},'' in {\em Strings, Gauge Fields,
  and the Geometry Behind: The Legacy of Maximilian Kreuzerte}, A.~Rebhan,
  L.~Katzarkov, J.~Knapp, R.~Rashkov, and E.~Scheidegger, eds., pp.~403--412.
\newblock 2012.
\newblock [\href{http://arxiv.org/abs/1012.2744}{{\ttfamily arXiv:1012.2744}}].

\bibitem{Brace:1998xz}
D.~Brace, B.~Morariu, and B.~Zumino, ``{T-duality and Ramond-Ramond backgrounds
  in the Matrix model},'' {\em Nucl. Phys. B} {\bfseries 549} (1999) 181,
  [\href{http://arxiv.org/abs/hep-th/9811213}{{\ttfamily
  arXiv:hep-th/9811213}}].

\bibitem{Fukuma:1999jt}
M.~Fukuma, T.~Oota, and H.~Tanaka, ``{Comments on T-dualities of Ramond-Ramond
  potentials},'' {\em Prog. Theor. Phys.} {\bfseries 103} (2000) 425,
  [\href{http://arxiv.org/abs/hep-th/9907132}{{\ttfamily
  arXiv:hep-th/9907132}}].

\bibitem{Hassan:1999bv}
S.~F. Hassan, ``{T-duality, space-time spinors and R-R fields in curved
  backgrounds},'' {\em Nucl. Phys. B} {\bfseries 568} (2000) 145,
  [\href{http://arxiv.org/abs/hep-th/9907152}{{\ttfamily
  arXiv:hep-th/9907152}}].

\bibitem{Hassan:1999mm}
S.~F. Hassan, ``{$SO(d,d)$ transformations of Ramond-Ramond fields and
  space-time spinors},'' {\em Nucl. Phys. B} {\bfseries 583} (2000) 431,
  [\href{http://arxiv.org/abs/hep-th/9912236}{{\ttfamily
  arXiv:hep-th/9912236}}].

\bibitem{Coimbra:2011nw}
A.~Coimbra, C.~Strickland-Constable, and D.~Waldram, ``{Supergravity as
  generalised geometry I: Type II theories},'' {\em JHEP} {\bfseries 1111}
  (2011) 091, [\href{http://arxiv.org/abs/1107.1733}{{\ttfamily
  arXiv:1107.1733}}].

\bibitem{Coimbra:2012yy}
A.~Coimbra, C.~Strickland-Constable, and D.~Waldram, ``{Generalised geometry
  and type II supergravity},'' {\em Fortschritte der Phys.} {\bfseries 60}
  (2012) 982, [\href{http://arxiv.org/abs/1202.3170}{{\ttfamily
  arXiv:1202.3170}}].

\bibitem{Butter:2022sfh}
D.~Butter, ``{Notes on Ramond-Ramond spinors and bispinors in double field
  theory},'' [\href{http://arxiv.org/abs/2208.11162}{{\ttfamily
  arXiv:2208.11162}}].

\bibitem{Butter:2021dtu}
D.~Butter, ``{Exploring the geometry of supersymmetric double field theory},''
  {\em JHEP} {\bfseries 01} (2022) 152,
  [\href{http://arxiv.org/abs/2101.10328}{{\ttfamily arXiv:2101.10328}}].

\bibitem{Gomis:2017cmt}
J.~Gomis and A.~Kleinschmidt, ``{On free Lie algebras and particles in
  electro-magnetic fields},'' {\em JHEP} {\bfseries 1707} (2017) 085,
  [\href{http://arxiv.org/abs/1705.05854}{{\ttfamily arXiv:1705.05854}}].

\bibitem{Gomis:2018xmo}
J.~Gomis, A.~Kleinschmidt, and J.~Palmkvist, ``{Symmetries of M-theory and free
  Lie superalgebras},'' {\em JHEP} {\bfseries 1903} (2019) 160,
  [\href{http://arxiv.org/abs/1809.09171}{{\ttfamily arXiv:1809.09171}}].

\bibitem{Polacek:2013nla}
M.~Pol{\'{a}}{\v{c}}ek and W.~Siegel, ``{Natural curvature for manifest
  T-duality},'' {\em JHEP} {\bfseries 1401} (2014) 026,
  [\href{http://arxiv.org/abs/1308.6350}{{\ttfamily arXiv:1308.6350}}].

\bibitem{Hohm:2011nu}
O.~Hohm and S.~K. Kwak, ``{$ \mathcal{N} = {1} $ supersymmetric double field
  theory},'' {\em JHEP} {\bfseries 1203} (2012) 080,
  [\href{http://arxiv.org/abs/1111.7293}{{\ttfamily arXiv:1111.7293}}].

\bibitem{Jeon:2011sq}
I.~Jeon, K.~Lee, and J.-H. Park, ``{Supersymmetric double field theory: A
  stringy reformulation of supergravity},'' {\em Phys. Rev. D} {\bfseries 85}
  (2012) 081501, [\href{http://arxiv.org/abs/1112.0069}{{\ttfamily
  arXiv:1112.0069}}].

\bibitem{Hatsuda:2014qqa}
M.~Hatsuda, K.~Kamimura, and W.~Siegel, ``{Superspace with manifest T-duality
  from type II superstring},'' {\em JHEP} {\bfseries 1406} (2014) 039,
  [\href{http://arxiv.org/abs/1403.3887}{{\ttfamily arXiv:1403.3887}}].

\bibitem{Hatsuda:2014aza}
M.~Hatsuda, K.~Kamimura, and W.~Siegel, ``{Ramond-Ramond gauge fields in
  superspace with manifest T-duality},'' {\em JHEP} {\bfseries 1502} (2015)
  134, [\href{http://arxiv.org/abs/1411.2206}{{\ttfamily arXiv:1411.2206}}].

\bibitem{Cederwall:2016ukd}
M.~Cederwall, ``{Double supergeometry},'' {\em JHEP} {\bfseries 1606} (2016)
  155, [\href{http://arxiv.org/abs/1603.04684}{{\ttfamily arXiv:1603.04684}}].

\bibitem{Hull:1998vg}
C.~M. Hull, ``{Timelike T-duality, de Sitter space, large N gauge theories and
  topological field theory},'' {\em JHEP} {\bfseries 07} (1998) 021,
  [\href{http://arxiv.org/abs/hep-th/9806146}{{\ttfamily
  arXiv:hep-th/9806146}}].

\bibitem{Wulff:2016tju}
A.~A. Tseytlin and L.~Wulff, ``{Kappa-symmetry of superstring sigma model and
  generalized 10d supergravity equations},'' {\em JHEP} {\bfseries 06} (2016)
  174, [\href{http://arxiv.org/abs/1605.04884}{{\ttfamily arXiv:1605.04884}}].

\bibitem{DAuria:1982mkx}
R.~D'Auria, P.~Fr{\'{e}}, P.~K. Townsend, and P.~van Nieuwenhuizen,
  ``{Invariance of actions, rheonomy, and the new minimal N = 1 supergravity in
  the group manifold approach},'' {\em Ann. Phys. (N. Y).} {\bfseries 155}
  (1984) 423.

\bibitem{RheonomicBook}
L.~Castellani, R.~D'Auria, and P.~Fr{\'{e}}, {\em {Supergravity and
  superstrings: A Geometric perspective. Vol. 2: Supergravity}}.
\newblock World Scientific, Singapore, 1991.

\bibitem{Gates:1997kr}
S.~J. {Gates, Jr.}, ``{Ectoplasm Has No Topology: The Prelude},'' in {\em
  Supersymmetries and Quantum Symmetries (SQS'97)}, pp.~46--57.
\newblock 1997.
\newblock [\href{http://arxiv.org/abs/hep-th/9709104}{{\ttfamily
  arXiv:hep-th/9709104}}].

\bibitem{Gates:1997ag}
S.~J. {Gates, Jr.}, M.~T. Grisaru, M.~E. Knutt-Wehlau, and W.~Siegel,
  ``{Component actions from curved superspace: Normal coordinates and
  ectoplasm},'' {\em Phys. Lett. B} {\bfseries 421} (1998) 203,
  [\href{http://arxiv.org/abs/hep-th/9711151}{{\ttfamily
  arXiv:hep-th/9711151}}].

\bibitem{Chamseddine:1980cp}
A.~H. Chamseddine, ``{N = 4 supergravity coupled to N = 4 matter and hidden
  symmetries},'' {\em Nucl. Physics, Sect. B} {\bfseries 185} (1981) 403.

\bibitem{Bergshoeff:1981um}
E.~A. Bergshoeff, M.~de~Roo, B.~de~Wit, and P.~{Van Nieuwenhuizen},
  ``{Ten-dimensional Maxwell-Einstein supergravity, its currents, and the issue
  of its auxiliary fields},'' {\em Nucl. Phys. B} {\bfseries 195} (1982) 97.

\bibitem{Cremmer:1978km}
E.~Cremmer, B.~Julia, and J.~Scherk, ``{Supergravity in theory in 11
  dimensions},'' {\em Phys. Lett. B} {\bfseries 76} (1978) 409.

\bibitem{Romans:1985tz}
L.~J. Romans, ``{Massive N = 2a supergravity in ten dimensions},'' {\em Phys.
  Lett. B} {\bfseries 169} (1986) 374.

\bibitem{Schwarz:1983wa}
J.~H. Schwarz and P.~C. West, ``{Symmetries and transformations of chiral N =
  2, D = 10 supergravity},'' {\em Phys. Lett. B} {\bfseries 126} (1983) 301.

\bibitem{Howe:1983sra}
P.~S. Howe and P.~C. West, ``{The complete $N = 2$, $d = 10$ supergravity},''
  {\em Nucl. Phys. B} {\bfseries 238} (1984) 181.

\bibitem{Nilsson:1981bn}
B.~E. Nilsson, ``{Simple 10-dimensional supergravity in superspace},'' {\em
  Nucl. Phys. B} {\bfseries 188} (1981) 176.

\bibitem{Carr:1986tk}
J.~L. Carr, S.~J. {Gates, Jr.}, and R.~N. Oerter, ``{D = 10, N = 2a
  supergravity in superspace},'' {\em Phys. Lett. B} {\bfseries 189} (1987) 68.

\bibitem{Wulff:2013kga}
L.~Wulff, ``{The type II superstring to order $\theta^4$},'' {\em JHEP}
  {\bfseries 07} (2013) 123, [\href{http://arxiv.org/abs/1304.6422}{{\ttfamily
  arXiv:1304.6422}}].

\bibitem{Bergshoeff:2001pv}
E.~A. Bergshoeff, R.~Kallosh, T.~Ort{\'{i}}n, D.~Roest, and A.~{Van Proeyen},
  ``{New formulations of D = 10 supersymmetry and D8-O8 domain walls},'' {\em
  Class. Quantum Gravity} {\bfseries 18} (2001) 3359,
  [\href{http://arxiv.org/abs/hep-th/0103233}{{\ttfamily
  arXiv:hep-th/0103233}}].

\bibitem{Howe:2015hpa}
P.~S. Howe and J.~Palmkvist, ``{Forms and algebras in (half-)maximal
  supergravity theories},'' {\em JHEP} {\bfseries 1505} (2015) 032,
  [\href{http://arxiv.org/abs/1503.00015}{{\ttfamily arXiv:1503.00015}}].

\bibitem{Bergshoeff:2005ac}
E.~A. Bergshoeff, M.~de~Roo, S.~F. Kerstan, and F.~Riccioni, ``{IIB
  supergravity revisited},'' {\em JHEP} {\bfseries 08} (2005) 098,
  [\href{http://arxiv.org/abs/hep-th/0506013}{{\ttfamily
  arXiv:hep-th/0506013}}].

\bibitem{Bergshoeff:2007ma}
E.~A. Bergshoeff, P.~S. Howe, S.~Kerstan, and L.~Wulff, ``{Kappa-symmetric
  SL(2, $\mathbb R$) covariant D-brane actions},'' {\em JHEP} {\bfseries 10}
  (2007) 050, [\href{http://arxiv.org/abs/0708.2722}{{\ttfamily
  arXiv:0708.2722}}].

\bibitem{Kleinschmidt:2003mf}
A.~Kleinschmidt, I.~Schnakenburg, and P.~C. West, ``{Very extended Kac-Moody
  algebras and their interpretation at low levels},'' {\em Class. Quantum
  Gravity} {\bfseries 21} (2004) 2493,
  [\href{http://arxiv.org/abs/hep-th/0309198}{{\ttfamily
  arXiv:hep-th/0309198}}].

\bibitem{Arutyunov:2015mqj}
G.~Arutyunov, S.~Frolov, B.~Hoare, R.~Roiban, and A.~A. Tseytlin, ``{Scale
  invariance of the $\eta$-deformed $AdS_5 \times S^5$ superstring, T-duality
  and modified type II equations},'' {\em Nucl. Phys. B} {\bfseries 903} (2016)
  262, [\href{http://arxiv.org/abs/1511.05795}{{\ttfamily arXiv:1511.05795}}].

\bibitem{Sakatani:2016fvh}
Y.~Sakatani, S.~Uehara, and K.~Yoshida, ``{Generalized gravity from modified
  DFT},'' {\em JHEP} {\bfseries 04} (2017) 123,
  [\href{http://arxiv.org/abs/1611.05856}{{\ttfamily arXiv:1611.05856}}].

\bibitem{Park:2016sbw}
J.-H. Park, ``{Green-Schwarz superstring on doubled-yet-gauged spacetime},''
  {\em JHEP} {\bfseries 1611} (2016) 005,
  [\href{http://arxiv.org/abs/1609.04265}{{\ttfamily arXiv:1609.04265}}].

\bibitem{Bandos:2015cha}
I.~A. Bandos, ``{Superstring in doubled superspace},'' {\em Phys. Lett. B}
  {\bfseries 751} (2015) 408,
  [\href{http://arxiv.org/abs/1507.07779}{{\ttfamily arXiv:1507.07779}}].

\bibitem{Sakamoto:2018krs}
J.-i. Sakamoto and Y.~Sakatani, ``{Local $\beta$-deformations and Yang-Baxter
  sigma model},'' {\em JHEP} {\bfseries 06} (2018) 147,
  [\href{http://arxiv.org/abs/1803.05903}{{\ttfamily arXiv:1803.05903}}].

\bibitem{Morand:2017fnv}
K.~Morand and J.-H. Park, ``{Classification of non-Riemannian
  doubled-yet-gauged spacetime},'' {\em Eur. Phys. J. C} {\bfseries 77} (2017)
  , [\href{http://arxiv.org/abs/1707.03713}{{\ttfamily arXiv:1707.03713}}].

\bibitem{Cho:2019ofr}
K.~Cho and J.-H. Park, ``{Remarks on the non-Riemannian sector in Double Field
  Theory},'' {\em Eur. Phys. J. C} {\bfseries 80} (2020) 0,
  [\href{http://arxiv.org/abs/1909.10711}{{\ttfamily arXiv:1909.10711}}].

\bibitem{Park:2020ixf}
J.-H. Park and S.~Sugimoto, ``{String Theory and Non-Riemannian Geometry},''
  {\em Phys. Rev. Lett.} {\bfseries 125} (2020) 1,
  [\href{http://arxiv.org/abs/2008.03084}{{\ttfamily arXiv:2008.03084}}].

\bibitem{Berman:2019izh}
D.~S. Berman, C.~D.~A. Blair, and R.~Otsuki, ``{Non-Riemannian geometry of
  M-theory},'' {\em JHEP} {\bfseries 07} (2019) 175,
  [\href{http://arxiv.org/abs/1902.01867}{{\ttfamily arXiv:1902.01867}}].

\bibitem{Gomis:2000bd}
J.~Gomis and H.~Ooguri, ``{Nonrelativistic closed string theory},'' {\em J.
  Math. Phys.} {\bfseries 42} (2001) 3127,
  [\href{http://arxiv.org/abs/hep-th/0009181}{{\ttfamily
  arXiv:hep-th/0009181}}].

\bibitem{West:2003fc}
P.~C. West, ``{$E_{11}$, SL(32) and central charges},'' {\em Phys. Lett. B}
  {\bfseries 575} (2003) 333,
  [\href{http://arxiv.org/abs/hep-th/0307098}{{\ttfamily
  arXiv:hep-th/0307098}}].

\bibitem{Bossard:2017wxl}
G.~Bossard, A.~Kleinschmidt, J.~Palmkvist, C.~N. Pope, and E.~Sezgin, ``{Beyond
  $E_{11}$},'' {\em JHEP} {\bfseries 1705} (2017) 020,
  [\href{http://arxiv.org/abs/1703.01305}{{\ttfamily arXiv:1703.01305}}].

\bibitem{Bossard:2021ebg}
G.~Bossard, A.~Kleinschmidt, and E.~Sezgin, ``{A master exceptional field
  theory},'' {\em JHEP} {\bfseries 06} (2021) 185,
  [\href{http://arxiv.org/abs/2103.13411}{{\ttfamily arXiv:2103.13411}}].

\bibitem{Bossard:2019ksx}
G.~Bossard, A.~Kleinschmidt, and E.~Sezgin, ``{On supersymmetric $E_{11}$
  exceptional field theory},'' {\em JHEP} {\bfseries 1910} (2019) 165,
  [\href{http://arxiv.org/abs/1907.02080}{{\ttfamily arXiv:1907.02080}}].

\bibitem{Park:2022pjv}
J.-H. Park, ``{Lecture note on Clifford algebra},'' {\em J. Korean Phys. Soc.}
  {\bfseries 81} (2022) 1, [\href{http://arxiv.org/abs/2205.09509}{{\ttfamily
  arXiv:2205.09509}}].

\end{thebibliography}\endgroup
